\documentclass[aps,pre,reprint,groupedaddress]{revtex4-2}

\usepackage{graphicx}  
\usepackage{dcolumn}   
\usepackage{bm}        
\usepackage{amssymb}   
\usepackage{amsmath}
\usepackage{comment}
\usepackage[utf8]{inputenc}
\usepackage{float}
\usepackage[dvipsnames]{xcolor}
\usepackage{graphicx}
\usepackage{comment}
\usepackage{bigints}
\usepackage{tabularx}
\usepackage{pbox}
\usepackage{tabu}

\newcommand{\Go}{G_0}
\newcommand{\dgamma}{\dot{\gamma}}
\newcommand{\tY}{\tilde{Y}}
\newcommand{\chip}{\chi^{\rm{peak}}}
\newcommand{\gc}{\gamma_c}
\newcommand{\md}{\mathrm{d}}

\newcommand{\xmin}{x_{\mathrm{min}}}
\newcommand{\fext}{T}
\newcommand{\gammap}{\gamma_p}
\newcommand{\chid}{\chi_{\mathrm{dis}}}
\newcommand{\fs}{\kappa}

\newcommand{\pltime}{\tau_{\rm{pl}}}
\newcommand{\avtau}{\tilde{t}}

\usepackage{mathtools}
\usepackage{makecell,tabularx}

\usepackage{caption}
\usepackage{subcaption}
\setcellgapes{3pt}

\newcolumntype{b}{X}
\newcolumntype{s}{>{\hsize=.65\hsize}X}

\usepackage{subcaption}

\begin{document}


\title{Ductile and brittle yielding of athermal amorphous solids: a mean-field paradigm beyond the random field Ising model}



\author{Jack T. Parley}
\email[Author to whom correspondence should be addressed: ]{jack.parley@uni-goettingen.de}
\affiliation{Institut f{\"u}r Theoretische Physik, University of G{\"o}ttingen,
Friedrich-Hund-Platz 1, 37077 G{\"o}ttingen, Germany}

\author{Peter Sollich}
\affiliation{Institut f{\"u}r Theoretische Physik, University of G{\"o}ttingen,
Friedrich-Hund-Platz 1, 37077 G{\"o}ttingen, Germany}
\affiliation{Department of Mathematics, King’s College London, London WC2R 2LS, UK}


\date{\today}

\begin{abstract}

Amorphous solids can yield in either a {\it ductile} or {\it brittle} manner under strain: plastic deformation can set in gradually, or abruptly through a macroscopic stress drop. Developing a unified theory describing both ductile and brittle yielding constitutes a fundamental challenge of non-equilibrium statistical physics. Recently, it has been proposed that, in the absence of thermal effects, the nature of the yielding transition is controlled by physics akin to that of the quasistatically driven Random field Ising model (RFIM), which has served as the paradigm for understanding the effect of quenched disorder in slowly driven systems with short-ranged interactions. However, this theoretical picture neglects both the dynamics of, and the elasticity-induced long-ranged interactions between, the mesoscopic material constituents. Here, we address these two aspects and provide a unified theory building on the Hébraud-Lequeux elastoplastic description. The first aspect is crucial to understanding the competition between the imposed deformation rate and the finite timescale of plastic rearrangements: we provide a dynamical description of the macroscopic stress drop, as well as predictions for the shifting of the brittle yield strain and the scaling of the peak susceptibility with inverse shear rate. The second is essential in order to capture properly the behaviour in the limit of quasistatic driving, where avalanches of plasticity
diverge with system size at any value of the strain. We fully characterise the avalanche behaviour, which is radically different to that of the RFIM. In the quasistatic, infinite size limit, we find that both models have mean-field Landau exponents, obscuring the effect of the interactions. We show, however, that the latter profoundly affect the behaviour of finite systems approaching the spinodal-like brittle yield point, where we recover qualitatively the finite-size trends found in particle simulations. The interactions also modify the nature of the random critical point separating ductile and brittle yielding, where we predict critical behaviour on top of the marginality present at any value of the strain. We finally discuss how all our predictions can be directly tested against particle simulations and eventually experiments, and make first steps in this direction.

\end{abstract}


\maketitle

\section{Introduction}\label{sec:intro}

Amorphous solids are all around us: from golf clubs made up of metallic glass, to colloidal gels such as toothpaste or emulsions such as mayonnaise, all the way to granular packings such as sand heaps~\cite{nicolas_deformation_2018,berthier_theoretical_2011}. Despite the very wide range of particle size and energy scales involved, such systems, be they made up of individual molecules, colloids, droplets, or bubbles, display a common phenomenology. Under a strain deformation, they initially present an elastic (i.e.\ reversible) solid-like response. At higher deformations, the solid-like elastic response eventually gives way to an irreversible plastic deformation, a process referred to as {\it{yielding}}. As we know from everyday life, yielding can be {\it{ductile}} or {\it{brittle}}, depending on whether plastic (irretrievable) deformation sets in gradually or abruptly (a more precise definition will be given below). The ultimate fate of the material after yielding also varies, of course: it may enter a stationary regime of plastic flow, as is the case for most soft amorphous solids, or exhibit instead macroscopic fracture and material failure, as is the case for the silica glass in our windows.

From a statistical physics perspective, an open challenge is to develop a satisfactory and comprehensive theory which can at least {\it aspire} to capture the universal features of ductile and brittle yielding across a wide range of materials. In the spirit of statistical physics, such a general theory must necessarily be simple to capture only the key universal ingredients of the physics, but can then serve as a building block for including material-specific peculiarities and parameters. 

The challenge to develop such a theory lies at two of the frontiers of modern physics, namely {\it{non-equilibrium}} and {\it{disorder}}. On the one hand, the application of a finite shear rate clearly drives the system out of equilibrium; in many cases such as colloidal glasses, the material may in fact be aging and hence already out of equilibrium before shear is applied. On the other hand, by definition amorphous solids lack any clear regular structure, and the corresponding arrangements of particles look (structurally) as disordered as a liquid. Indeed, unlike crystals, the specific sample structure and properties depend on the preparation: one therefore has to talk of {\it ensembles} of sample realisations given a preparation protocol. 

In the hope of capturing universal features of amorphous solids, the community has invested large efforts into computer-based particle simulations of model glasses, on the one hand, and mesoscopic lattice-based elastoplastic models, on the other. Turning to the first, simulation advances in recent years~\cite{ninarello_models_2017} now allow to generate computer glass samples over an unprecedented range of annealing or equivalently stability. The initial level of annealing in these simulations is determined by the preparation temperature $T_{\rm ini}$ at which liquid configurations are equilibrated before quenching to $T=0$. In the work of Ozawa et al.~\cite{ozawa_random_2018}, the range of $T_{\rm ini}$ considered was argued to be sufficient to describe systems ranging from poorly annealed glasses (e.g.\ wet foams~\cite{lauridsen_shear-induced_2002}) and colloidal systems~\cite{schall_structural_2007,bonn_yield_2017,siebenburger_creep_2012}, to well-annealed glasses (e.g.\ metallic glass experiments~\cite{hufnagel_deformation_2016,greer_shear_2013,gu_ductile--brittle_2009,su_atomic_2022}) and ultrastable glasses~\cite{fullerton_density_2017} \cite{ozawa_random_2018}. Indeed, for the first of these one expects ductile behaviour, while the latter are typically brittle.

Ozawa et al.~\cite{ozawa_random_2018} proposed a fascinating theoretical picture for this change in the nature of yielding. Applying shear deformation in the athermal ($T=0$) and quasistatic (zero strain-rate) limits, the annealing-dependent regimes of ductile and brittle yielding could be clearly identified. It was argued that ductile and brittle yielding are separated by a critical value of the initial disorder or annealing, which further corresponds to a random critical point as it appears in the random-field Ising model (RFIM, defined below). From a theoretical standpoint, arguably the main appeal of this 
picture is that the random critical point of the athermal quasistatically driven RFIM is well understood: at least at a mean-field level, its critical behaviour corresponds to that of the equilibrium (thermal) RFIM at the critical disorder~\cite{dahmen_hysteresis_1996}. Ultimately, therefore, the original non-equilibrium problem with disorder is mapped to an existing paradigm of {\it equilibrium} statistical physics in the presence of disorder. 

Before we return to this point, we note that this proposed picture sparked a discussion in the community regarding the existence of a random critical point in finite dimensions. Barlow et al.~\cite{barlow_ductile_2020,pollard_yielding_2022} proposed instead that, in the athermal limit, any system displaying an overshoot, however small, will undergo a macroscopic stress drop and hence be brittle. Only extremely poorly annealed samples with no overshoot can then display ductile yielding. The particle simulations of Richard et al.~\cite{richard_finite-size_2021} pointed to the fact that, if a critical disorder value (with finite associated overshoot) exists, its determination is rendered challenging by very strong finite-size effects. 

Strong support for a random critical point, however, has been recently obtained from lattice elastoplastic models (EPMs)~\cite{rossi_finite-disorder_2022}. These are mesoscopic models that attempt to capture the basic ingredients of deformation and flow in amorphous solids. Based on substantial numerical and experimental observations from a wide range of systems~\cite{nicolas_deformation_2018,barrat_heterogeneities_2011,argon_plastic_1979,desmond_measurement_2015,maloney_amorphous_2006,tanguy_plastic_2006,puosi_time-dependent_2014}, they propose the simple picture that deformation is mediated by localised plastic events, where a mesoscopic block of material locally fluidises and deforms plastically. This localised plastic strain in turn induces a stress redistribution in the surrounding medium, which is typically modelled by the Eshelby stress propagator~\cite{picard_elastic_2004} (see below for more details). Using lattice EPMs in $d=2$ and $d=3$, the authors of \cite{rossi_finite-disorder_2022} showed, by accessing much larger system sizes than in particle simulations and performing finite-size scaling, that at least for $d=3$ (and within the elastoplastic framework) a finite disorder critical point persists in the infinite size limit. 

Returning to the theoretical picture developed in \cite{ozawa_random_2018} for the random critical point and brittle yielding, there are at least two important directions in which it is incomplete, and which we will address in this work: the first is related to {\it dynamics}, while the second has to do with {\it interactions}.

Turning to the first, it is clear that, in any physical system, the localised plastic events carry an associated finite timescale, which must compete with the applied driving rate. This point has been studied in detail with particle simulations~\cite{singh_brittle_2020} in the overdamped athermal setting, showing that the sharp brittle yielding transition is smeared out at any finite shear rate. In finite dimension, the macroscopic stress drop is due to the formation of a shear band (we will discuss its mean-field representation below), with finite timescales associated with both the destabilisation of individual mesoscopic blocks and the elastic stress propagation. Considering the macroscopic stress versus strain curve in the large system size limit, this will only present a sharp discontinuity when the shear band timescale is negligible with respect to the timescale set by the external shear~\cite{singh_brittle_2020}.

Turning to {\it interactions}, the theoretical description in \cite{ozawa_random_2018} neglects the long-range, sign-varying nature of the elastic interactions associated with stress redistribution due to plastic events. These contrast with the (standard) RFIM, where interactions are strictly positive and short-range. It is a priori not at all clear how this difference modifies the standard RFIM picture. One thing that is well known is that the long-range sign-varying interactions induce {\it marginal stability} of the amorphous solid~\cite{muller_marginal_2015,lin_scaling_2014}. This profoundly affects the avalanche behaviour in the quasistatic limit, and causes avalanches of plasticity to be system-spanning 
and to scale sub-extensively with system size, as studied in detail in particle simulations~\cite{maloney_subextensive_2004,shang_elastic_2020}. In the athermal quasistatic limit, where avalanches are sharply defined, brittle yielding will therefore have to arise on top of this underlying marginality of the amorphous solid. 

To address these issues, we will turn to an elastoplastic description of the amorphous solid. In recent work, the basic postulates of EPMs have been greatly strengthened through empirical measurements of local residual stresses in particle systems~\cite{barbot_local_2018}, which have been shown to constitute the best-performing structural predictors of plasticity~\cite{richard_predicting_2020}. By properly calibrating the parameters, EPMs can further be used to directly match both the transient and steady state behaviour of disordered particle systems~\cite{liu_elastoplastic_2021}. Here, we will consider the simplest possible EPM, the H\'{e}braud-Lequeux (HL) mean-field elastoplastic model.

As detailed in Sec.~\ref{sec:mf_theory}, in the HL model the interactions between mesoscopic blocks are approximated as Gaussian mechanical noise. Thanks to this simplification, the model is exactly solvable, at least in steady state, where in particular one may analytically obtain an expansion of the solution up to any order in shear rate $\dgamma$~\cite{agoritsas_relevance_2015,puosi_probing_2015} (these solutions are provided for completeness in App.~\ref{app:details}). However, despite its simplicity, we believe that HL
can play a paradigmatic role
in this context, as it arguably contains the minimal necessary physical ingredients. We mention here just some of the applications and successes of the model: it recovers the widespread Herschel-Bulkley rheological law in steady state~\cite{hebraud_mode-coupling_1998}, i.e. a flow curve exponent $\beta=2$ defined via $\dgamma \sim (\Sigma-\Sigma_y)^{\beta}$, where $\Sigma_y$ is a dynamical yield stress~\footnote{One must note that in both two- and three-dimensional lattice elastoplastic models one in fact measures $\beta \approx 1.5$ ($\beta \approx 2$ for the Herschel-Bulkley law), where the flow curve exponent is defined by $\dgamma \sim (\Sigma-\Sigma_c)^{\beta}$~\cite{liu_driving_2016}. Importantly, it was shown~\cite{liu_driving_2016} that the measured flow exponent depends crucially on the shear rate window considered, with $\beta \approx 2$ for larger shear rates where the mechanical noise crosses over to a Gaussian spectrum as in the HL model. We note also the coarse-graining idea of \cite{fernandez_aguirre_critical_2018} where the value $\beta \approx 1.5$ measured in 2$d$ elastoplastic models is argued to derive from an effective mechanical noise exponent $\mu \approx 3/2\neq 1$ (see meaning of $\mu$ in Sec.~\ref{sec:mf_theory}).}.
With the addition of heterogeneity in the yield thresholds, it further gives a qualitatively complete description of the complex dependence of yielding under oscillatory shear on the initial level of annealing of the glass~\cite{parley_mean-field_2022}. Finally, it is possible (in the finite driving regime) to calibrate it directly against particle simulations~\cite{puosi_probing_2015}.

Before we go on, it is important to note that, at least in the limit of quasistatic shear, the HL model is not expected to be a ``true'' mean-field model in the sense that it becomes an exact description of the system above some upper critical dimension. Away from the quasistatic limit, in finite-dimensional lattice elastoplastic models, where one can measure directly the mechanical noise on a given site, one finds that the noise indeed crosses over to a Gaussian distribution as in the HL model~\cite{liu_driving_2016}. In the quasistatic limit, however, replacing the Gaussian noise of the HL model by power-law distributed noise, as shown by Lin \& Wyart~\cite{lin_mean-field_2016} (see also Sec.~\ref{sec:mf_theory}), provides improved predictions for the pseudogap exponent, which describes the density of mesoscopic blocks close to a plastic instability. Their results showed that the mean-field prediction becomes accurate above four dimensions, suggesting this value as an upper critical dimension (though without any supporting Ginzburg-type arguments)~\cite{lin_mean-field_2016}. Alternatively, Ferrero \& Jagla have argued~\cite{ferrero_properties_2021} that, by appropriately redefining the mechanical noise, a mean-field description of the flowing state in the quasistatic limit (at least as regards the scaling of the strain steps between avalanches) can be extended down to dimensions $d=2,3$. Throughout this paper, we will provide a complete mean-field picture of the ductile and brittle yielding transitions within the HL model, which has the important advantage of analytical tractability, while also paving the way towards refined descriptions using power-law noise or dimensionality-dependent effective noise distributions. 

The paper is structured as follows: in Sec.~\ref{sec:mf_theory}, we present the HL mean-field elastoplastic approach, and discuss the relation to the paradigmatic RFIM. 
In Secs.~\ref{sec:first_part} and \ref{sec:second_part}, we then tackle the yielding of amorphous solids distinguishing the two limits $N \to \infty$ and $\dgamma \to 0$. In the first approach, Sec.~\ref{sec:first_part}, we will fix $N=\infty$ and study the behaviour for low but nonzero shear rates. For such infinitely large systems, a master equation description holds, allowing for analytical progress. In the second approach, Sec.~\ref{sec:second_part}, we will instead fix a quasistatic shear rate $\dgamma=0^{+}$ and study the {\it avalanche} behaviour for large finite system sizes. The key to tackling this problem will be a mapping to a first-passage problem first introduced by Jagla~\cite{jagla_avalanche-size_2015}. 
Throughout this Sec.~\ref{sec:second_part}, we will perform a detailed comparison against the corresponding behaviour of the athermal, quasistatically driven RFIM. This will include, besides the avalanche statistics, both the behaviour in the infinite size limit and the finite-size behaviour approaching the brittle yield point and around the random critical point. Finally, in Sec.~\ref{sec:outlook} we offer a summary and discussion of the results and an outlook towards future research.


\begin{figure}
\centering
\begin{subfigure}{0.4\textwidth}
   \includegraphics[width=1\linewidth]{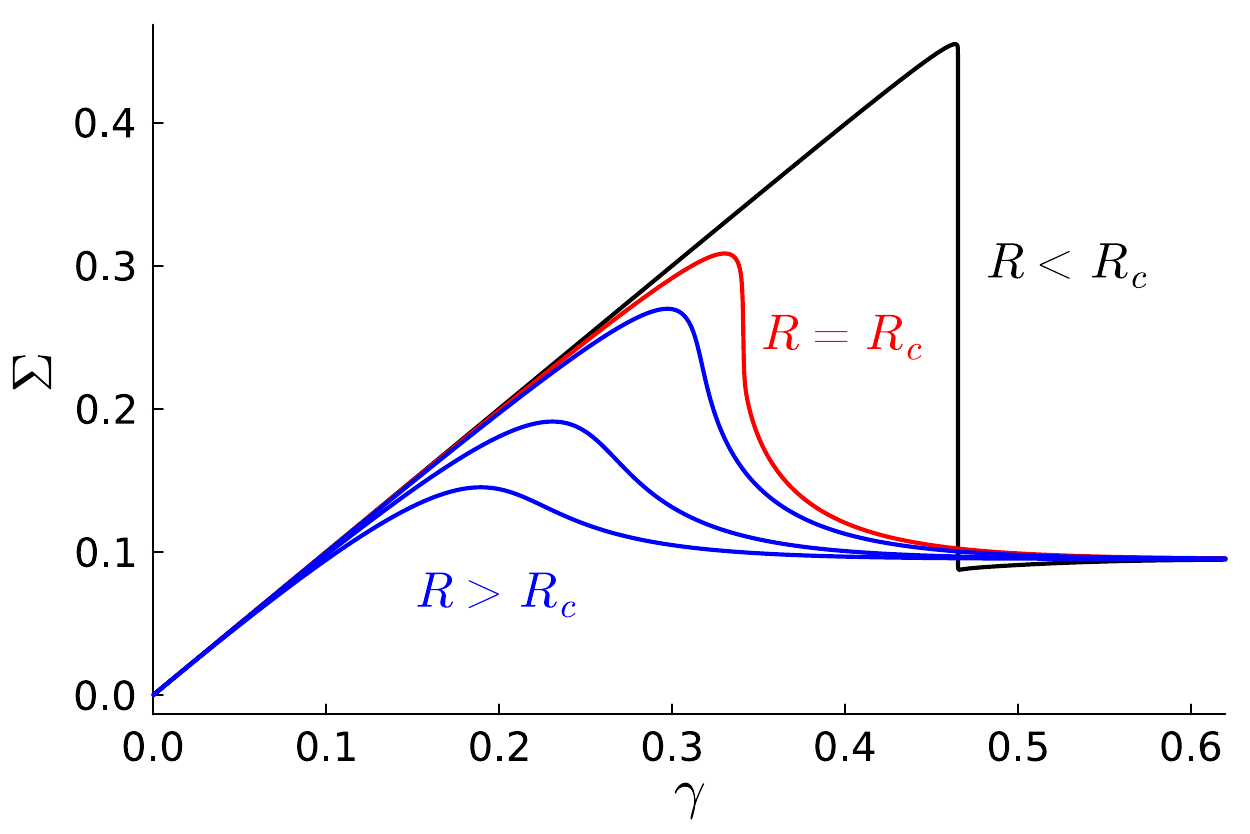}
   \caption{HL model (amorphous solid): evolution of macroscopic stress $\Sigma(\gamma)$ for different values of initial disorder $R$ under a very slow shear rate $\dgamma \pltime=10^{-7}$ and for coupling fixed to $\alpha=0.45$.}
   \label{fig:Ng1} 
\end{subfigure}

\begin{subfigure}{0.4\textwidth}
   \includegraphics[width=1\linewidth]{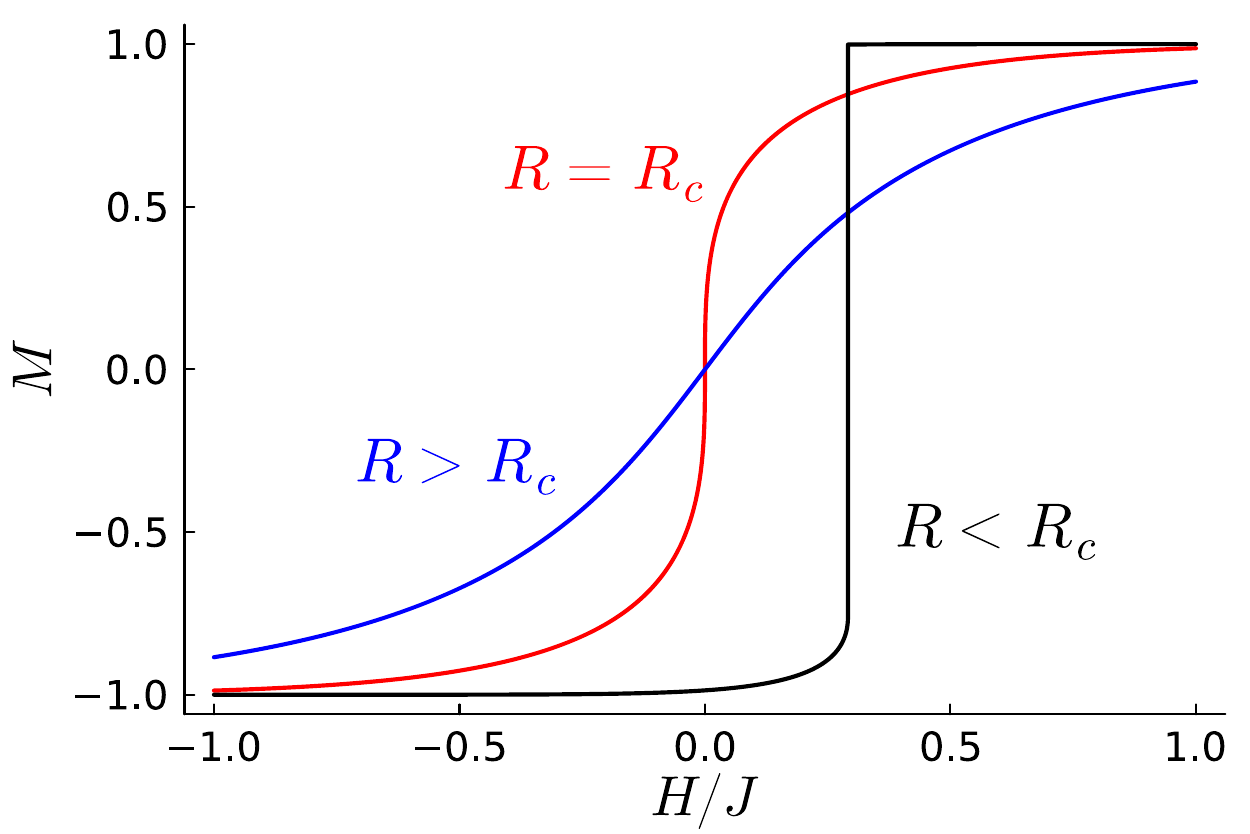}
   \caption{Magnetization of the random field Ising model at zero temperature in the mean-field limit, driven by an external field $H$ increasing quasistatically from $H=-\infty$.}
   \label{fig:Ng2}
\end{subfigure}

\caption{\label{fig:comparison}$\Sigma(\gamma)$ in yielding amorphous solids vs $M(H)$ in disordered magnets. The level of disorder $R$, corresponding to the standard deviation of a Gaussian, refers respectively to the initial local stress distribution and the distribution of random local fields. In both cases a discontinuity emerges for $R<R_c$ in the infinite size limit where, at a critical value of imposed strain $\gamma_c$ (or external field $H_c$), a finite fraction of the mesoscopic blocks (spins) rearrange (flip) in a so-called $\infty$-avalanche. At $R=R_c$, the macroscopic stress (magnetisation) curve develops an infinite slope.}
\end{figure}



\section{\label{sec:mf_theory}Mean-field theory}

The HL model proposes a mesoscopic elastoplastic description of the amorphous solid, in terms of $N$ mesoscopic blocks, on the scale of localised plastic events, but large enough so that a local shear stress $\sigma_i$ can be defined. One obtains the {\it macroscopic stress} by averaging over the local elements, i.e. $\Sigma=\sum_i \sigma_i /N$. Typically, each block is assumed to occupy a site $i$ on a regular lattice. The HL dynamics can then be defined in discrete time, by introducing a coupled set of stochastic update rules for each block~\footnote{We note that this is not how the HL model was derived originally in~\cite{hebraud_mode-coupling_1998}, where the master equation was introduced directly.}. Considering a time step $\Delta t \ll 1$, small enough so that there is at most one yield event per time interval, and labeling the site where this event takes place by $l$, the yielding rule is given by
\begin{equation}
\sigma_l(t+\Delta t)=0 \quad \text{with probability} \  \frac{\Delta t }{\pltime} \quad \text{if}\  |\sigma_l(t)|>\sigma_c 
\label{sigma_l_update}
\end{equation}
where $\pltime$ is the timescale associated with plastic rearrangements and $\sigma_c$ is the local yield threshold (considered uniform throughout this paper). The yielding rule models the relaxation of the local stress (which is reset to 0~\footnote{In~\cite{agoritsas_non-trivial_2017}, the effect of an incomplete stress relaxation was shown not to be qualitatively important, at least for the rheology.}) after a plastic rearrangement. 

The stresses at all other sites $\{\sigma_i\}$, $i \neq l$, evolve as
\begin{equation}
  	\sigma_i(t+\Delta t)=G_0\dot{\gamma}\Delta t+\sigma_i(t)+\delta \sigma_i
  	\label{sigma_i_update}
\end{equation}
Here, $\dgamma$ is the external shear rate, $G_0$ is a local shear modulus, and $\delta \sigma_i$ is a stress ``kick'' that models the stress propagation (from site $l$, with propagation taken as instantaneous for simplicity) due to the sign-varying and long-range Eshelby stress propagator.

\begin{figure}
\centering
\begin{subfigure}{0.35\textwidth}
   \includegraphics[width=1\linewidth]{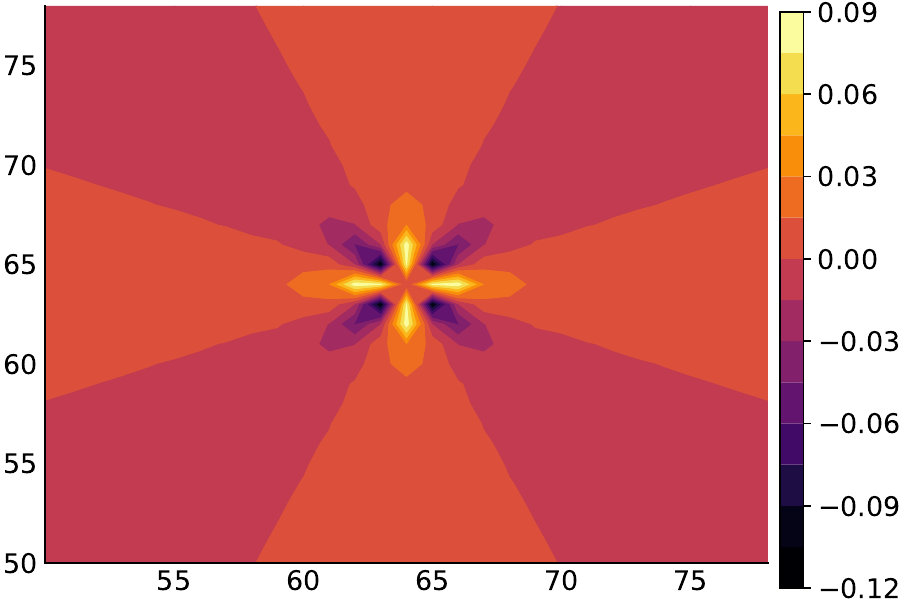}
   \label{fig:noises_1} 
\end{subfigure}

\begin{subfigure}{0.4\textwidth}
   \includegraphics[width=1\linewidth]{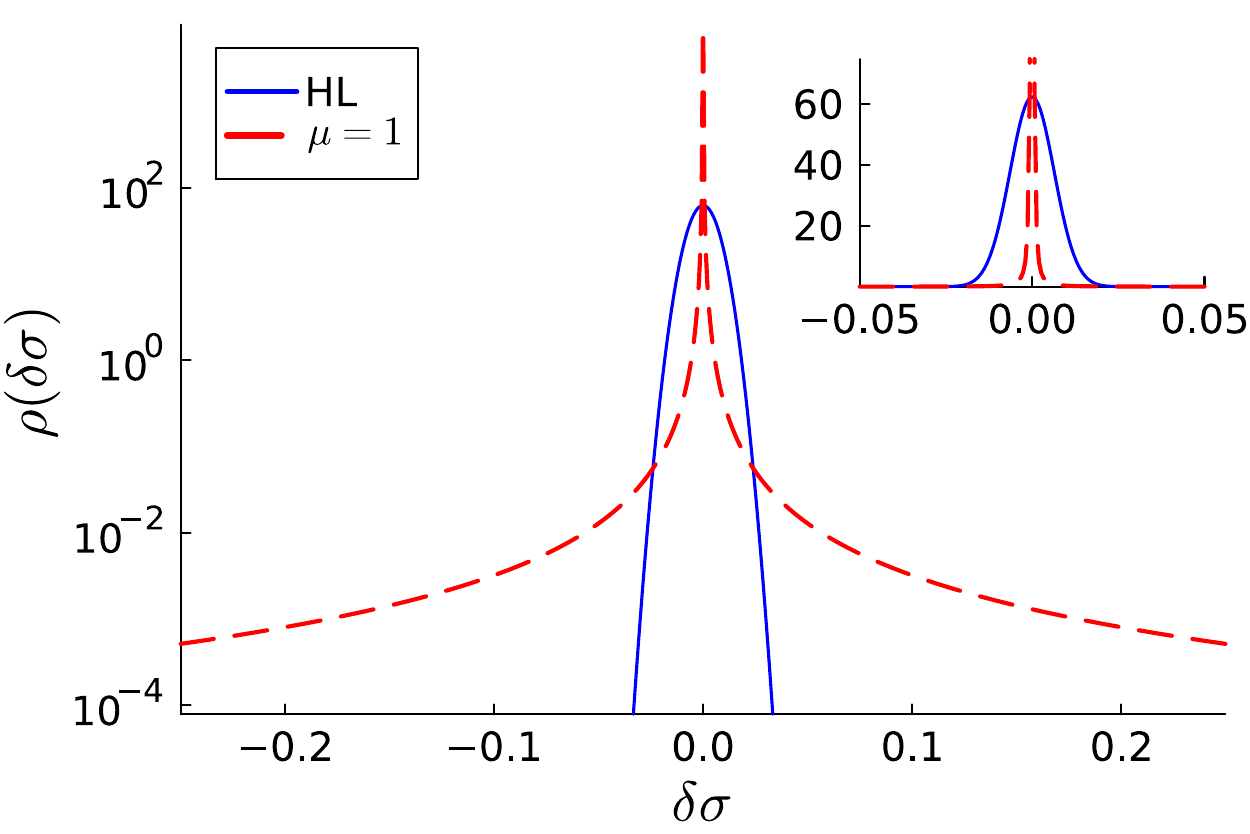}
   \label{fig:noises_2}
\end{subfigure}

\caption{\label{fig:noises}Top: (shear) stress propagator elements resulting from a localised plastic relaxation (unit stress drop) occurring at the center of a $128 \times 128$ square lattice, reproduced from~\cite{parley_mean-field_2022-1}. Bottom: comparison of the noise kick distribution $\rho(\delta \sigma)$ for $\mu=1$ (\ref{eq:power_law_noise}) with $A=0.32$ and $N=10^{4}$, against the HL model with $N=10^{4}$ and $\alpha \simeq 0.2$ chosen to have the same root mean square stress kick. The $\mu=1$ case, which corresponds to the distribution of stress kick values felt at a given site due to a yield event located uniformly at random in the surrounding lattice,
is mainly made up of tiny $\mathcal{O}(N^{-1})$ kicks (see inset on linear scale), but attains the same variance thanks to its power-law tails (main plot, on semi-log scale).}
\end{figure}

The stress ``kick'' felt at a site $i$ due to a yield event at site $l$ should of course depend on the relative position $\mathbf{r}_{il}$ of the two sites, through the complex spatial structure of the propagator, e.g.\ displaying a quadrupolar form in two dimensions~\cite{picard_elastic_2004} (see Fig.~\ref{fig:noises} top). The presence of special directions on which the interaction is purely constructive is indeed responsible for the appearance of {\it shear bands} in finite-dimensional systems. In a mean-field approach, one can get rid of the spatial structure while retaining the two main features of the stress propagation: its long-range and sign-varying nature, which are captured as a {\it mechanical noise} acting on each local stress. Such a mean-field approach to the yielding problem can be justified, at least within a lattice elastoplastic framework, along the lines of~\cite{ferrero_elastic_2019}, who showed that one can interpolate smoothly between the critical exponents of mean-field depinning and yielding in finite dimension. The intuition is that, for the long-range Eshelby interaction decaying as $r_{il}^{-d}$, where $d$ is the dimension, the effect of a single site onto another site is negligible compared to the combined effect of all other sites on the lattice~\cite{ferrero_elastic_2019}.

In fact, considering the sites as being placed on a lattice, with a long-range interaction decaying as a power-law, {\it randomising} the spatial structure leads to a family of mean-field models defined by a symmetric distribution of noise kicks~\cite{lemaitre_plastic_2007,lin_mean-field_2016}
\begin{equation}\label{eq:power_law_noise}
    \rho(\delta \sigma)\sim A \ N^{-1}|\delta \sigma|^{-\mu-1}
\end{equation}
with the {\it mechanical noise exponent} $0< \mu \leq 2$, and a coupling constant $A$. For the elastic propagator decaying as $r^{-d}$ one expects $\mu=1$, as studied in the model by Lin and Wyart~\cite{lin_mean-field_2016}. The distribution has a lower cutoff vanishing with system size as $N^{-1/\mu}$, and a system-size independent upper cutoff $\delta \sigma_u$ (corresponding to the largest stress kick felt in the system). 
Due to the upper cutoff, in all cases the root mean square (rms) stress kick scales as
\begin{equation}\label{eq:rms}
    \sqrt{\langle \delta \sigma^2 \rangle}\propto A N^{-1/2}
\end{equation}
where the proportionality involves geometrical factors~\cite{parley_aging_2020}. Although the rms stress kick scales in the same way, the overall kick distribution is of course radically different for different values of the noise exponent $\mu$ (Fig.~\ref{fig:noises}). In the absence of shear, decreasing $\mu$ can be seen as progressively leading to the dominance of ``near-field'' events (large kicks), which completely dominate the physics for $\mu<1$~\cite{parley_aging_2020}. Different values of $\mu$ can also be contrasted by observing the scaling of the lower cutoff $N^{-1/\mu}$, corresponding to the scale of stress kicks from far away events, which make up the bulk of the noise distribution. While for $\mu=2$ these typical kicks scale like the rms (\ref{eq:rms}), for $\mu=1$ they are of order $N^{-1}$, reflecting the power-law decay $L^{-d}$ of the Eshelby propagator at large lenghtscales. This scaling $N^{-1}$ for the physical case $\mu=1$ lies at the root of the smooth interpolation discussed above between mean-field depinning (where, as in a mean-field ferromagnet, the ``kick'' in the local field due to a spin flip on any other site scales as $N^{-1}$) and the Eshelby propagator in finite dimension, as it implies that both interactions are ``equally long-range''~\cite{ferrero_elastic_2019}.

In the limit $N\rightarrow \infty$, where the local dynamics at each site decouple and depend only on the current {\it average} number of unstable elements, one can derive~\cite{parley_aging_2020} a master equation describing the dynamical evolution of the local stress distribution in the amorphous solid under any arbitrary imposed shear rate $\dgamma$. For the general form of mechanical noise (\ref{eq:power_law_noise}) with exponent $0<\mu<2$, this leads to a family of mean-field models incorporating L\'{e}vy noise~\cite{parley_aging_2020}. In the special limiting case of $\mu \rightarrow 2^{-}$, the L\'{e}vy noise simplifies to standard diffusion and one obtains the HL model~\cite{parley_aging_2020}. HL is clearly dominated by far-field events (small kicks): the noise kicks are simply independent Gaussian random variables, with standard deviation $(2 \alpha /N)^{1/2}$, which defines the HL {\it coupling constant} $\alpha$. The regime of physical interest of the model is for small coupling $\alpha<\alpha_c=\sigma_c^2/2$, where HL describes a jammed solid with finite yield stress and Herschel-Bulkley rheology~\cite{hebraud_mode-coupling_1998}. The coupling constant $\alpha$ of the HL model can in principle be calibrated against particle simulations~\cite{puosi_probing_2015}, while within the kinetic elastoplastic approach of~\cite{bocquet_kinetic_2009} it can be related explicitly to the values of the propagator elements, as considered in~\cite{ozawa_elasticity_2023}. We note that, on the other hand, for $\mu=1$ the coupling $A$ can be chosen to precisely match the {\it{shuffled}} finite-dimensional lattice propagator~\cite{lin_mean-field_2016,parley_aging_2020}~\footnote{We mention also the view of~\cite{fernandez_aguirre_critical_2018,ferrero_criticality_2019}, which suggest values in the whole range $1<\mu<2$ are physically significant and effectively arise from the coarse-graining of plastic events to consider the aggregate effect of avalanches.}. The HL master equation for the local stress distribution reads:
\begin{multline}\label{eq:hl_first}
    \partial_t P(\sigma,t)=-\Go \dgamma \partial_{\sigma} P(\sigma,t)+\alpha Y(t)\partial_{\sigma}^2 P(\sigma,t)\\-\frac{1}{\pltime} \theta \left (|\sigma|-\sigma_c\right)P(\sigma,t)+Y(t)\delta(\sigma)
\end{multline}
with the noise kicks of standard deviation $(2 \alpha /N)^{1/2}$ turning into the HL diffusion coefficient $\alpha Y(t)$ in the $N \rightarrow \infty$ limit. The {\it yield rate} is defined self-consistently as 
\begin{equation}
    Y(t)=\frac{1}{\pltime}\int\mathrm{d}\sigma \ \theta \left (|\sigma|-\sigma_c\right)P(\sigma,t) 
\end{equation}
so that the noise in the system is indeed proportional to the average fraction of unstable elements. The {\it macroscopic stress} at any point in time is, as in the discrete model, given by the average
\begin{equation}\label{eq:macro_stress}
    \Sigma (t)=\int \mathrm{d}\sigma \ \sigma  P(\sigma,t) 
\end{equation}
In the following, we will consider the HL model for simplicity, as it is more amenable to an analytical approach. The impact of considering instead models with $\mu <2$ will be discussed further below (Sec.~\ref{subsec:extension_PL}).


We discuss now the relation to the Random field Ising model (RFIM), a fascinating connection first discussed in~\cite{ozawa_random_2018}. There, a mean-field elastoplastic model is introduced, and studied in the quasistatic ($\dgamma=0^{+}$) limit. Importantly, however, this model contains strictly {\it positive} interactions, in the sense that the de-stabilisation of any element will always 
de-stabilise any other component of the system. The model in this sense is reminiscent of the mean-field RFIM driven by a slowly varying external field at zero temperature, and the authors therefore consider the very interesting analogy between the non-equilibrium RFIM and yielding under quasistatic shear. 

Indeed, the RFIM arguably constitutes the paradigmatic model for understanding the de-stabilisation of a metastable state in an athermal system with quenched diosorder and strictly positive (ferromagnetic) interactions~\cite{sethna_crackling_2001}. The RFIM has also played an important role in socioeconomic physics, providing a theoretical picture of e.g.\ collective opinion shifts~\cite{michard_theory_2005}. From a theoretical perspective, the appealing feature of the RFIM is that it is amenable to systematic renormalisation group techniques~\cite{dahmen_hysteresis_1996}. Crucially, it turns out that the critical exponents describing the non-equilibrium behaviour at zero temperature are the same as those of the equilibrium model~\cite{dahmen_hysteresis_1996}. 
Following the interpolation argument~\cite{ferrero_elastic_2019} mentioned above, within the general view that yielding is a mean-field transition, we will restrict our discussion throughout this paper to the {\it mean-field} RFIM (and refer to this simply as the RFIM from now on). The mean-field elastoplastic model of \cite{ozawa_random_2018} essentially recovers the same behaviour as the RFIM; in the following, we will therefore contrast our results to the RFIM, while bearing in mind that this also applies to the elastoplastic model with strictly positive interactions.

We briefly summarise the main features of the RFIM. This is described by a Hamiltonian
\begin{equation}
    \mathcal{H}=-\frac{1}{2}\sum_{i j}J_{ij}s_i s_j -\sum_{i}(H+f_i)s_i
\end{equation}
Here the $J_{ij}>0$ are translationally invariant ferromagnetic interactions, $H$ is an external field and $f$ is a random local field extracted independently for each spin $i$ from a distribution $\rho(f)$. To connect to our approach, we will consider a Gaussian with standard deviation $R$: $\rho(f)\sim e^{-f^2/(2R^2)}$. To obtain the mean-field (fully connected) limit, one replaces the short range interactions by a scaled all-to-all interaction $J=J_{ij}/N$. 
One may then regard the effect of a single spin flip on the local effective field of another spin as a (strictly positive) ``kick'' of order $N^{-1}$, to be compared to the order $N^{-1/2}$ rms (sign-varying) stress kick (\ref{eq:rms}) due to mechanical noise. Introducing the magnetisation $M=\sum_i s_i /N$, the mean-field RFIM reduces to a collection of spins each feeling an effective local field given by
\begin{equation}
    h_i^{\rm eff}=JM+H+f_i
\end{equation}

Consider now the following scenario. Starting from a configuration $\forall s_i =-1$, and $H=-\infty$, slowly increase the external field until all spins have flipped sign (at zero temperature, the spin flip occurs when $h_i^{\rm eff}$ changes sign). Performed quasistatically, this process proceeds via {\it avalanches} of spin flips, so that $M(H)$ is made up of flat segments interrupted by sudden jumps in magnetisation. In the infinite size limit, where $M(H)$ is smooth, there is a critical value of the initial disorder $R_c$ separating two fundamentally different behaviours.

For $R<R_c$, one eventually triggers the so-called infinite ($\infty$-) avalanche, where a finite fraction of the spins flips and $M(H)$ jumps discontinuously. At $R=R_c$, $M(H)$ is continuous, but develops an infinite slope. As we will see below, and as discussed in \cite{ozawa_random_2018}, the quasistatic behaviour of the macroscopic stress $\Sigma (\gamma)$ in an amorphous solid has features similar to $M(H)$ (see Fig.~\ref{fig:comparison}). However, it is clear that, due to the sign-varying and long-range nature of the interactions in an amorphous solid, one expects, at least for the avalanche behaviour, significant differences in the mean-field predictions.

Some partial answers as to what these differences may be were provided in the second work we now discuss~\cite{popovic_elastoplastic_2018}. This must be considered as the starting point of our approach, as it was the first to show the presence of brittle yielding in the HL model. However, it was restricted to the quasistatic limit, and limited to showing the possibility of macroscopic failure for well-annealed initial conditions (the precise relation to our results will be clarified in Sec.~\ref{subsec:TD_limit}). Indeed, the study~\cite{popovic_elastoplastic_2018} left many open questions. How does brittle yielding appear in the HL model as $\dgamma \rightarrow 0$? What do the avalanche {\it distributions} 
look like, and what are they controlled by? What happens around the critical disorder $R_c$ of the HL model? We will address these and other questions below.

\section{\label{sec:first_part}
Part 1: $N=\infty$, $\dot{\gamma}\ll 1$}
For $N=\infty$, the HL dynamics is described exactly by its master equation (\ref{eq:hl_first}). Before we turn to the description of brittle yielding, we will firstly lay out the analytical framework needed to study this master equation. This includes the so-called {\it boundary layer expansion}, which will be necessary to understand the behaviour in the limit of very slow shear rates, $\dgamma \ll 1$, where the sharp brittle yielding transition arises.

\subsection{\label{sec:background}Analytical framework}

Without loss of generality, it is firstly convenient to write the HL equation (\ref{eq:hl_first}) in dimensionless variables, with $\sigma_c$ providing the stress scale and $\pltime$ the timescale ($\Go$ can be absorbed into the definition of shear rate). The dimensionless HL equation is then
\begin{multline}\label{eq:hl_adim}
    \partial_t P(\sigma,t)=- \dgamma \partial_{\sigma} P(\sigma,t)+\alpha Y(t)\partial_{\sigma}^2 P(\sigma,t)\\- \theta \left (|\sigma|-1\right)P(\sigma,t)+Y(t)\delta(\sigma)
\end{multline}
with 
\begin{equation}
    Y(t)=\int \mathrm{d}\sigma \ \theta \left (|\sigma|-1\right)P(\sigma,t) 
    \label{Y_exact}
\end{equation}
Throughout this paper, we will consider a strain-controlled protocol, that is, we will study the dynamics subject to a fixed external shear rate $\dgamma$. It is therefore convenient to write (\ref{eq:hl_adim}) as a function of total strain $\gamma=\dgamma t$ instead of time
\begin{multline}\label{eq:hl_strain}
     \partial_{\gamma} P(\sigma,\gamma)=- \partial_{\sigma} P(\sigma,\gamma)+\alpha \tY (\gamma)\partial_{\sigma}^2 P(\sigma,\gamma)\\- \frac{1}{\dgamma}\theta \left (|\sigma|-1\right)P(\sigma,\gamma)+\tY (\gamma)\delta(\sigma)
\end{multline}
where we have defined the re-scaled yield rate $\tY(\gamma)=Y(t)/\dgamma$. Important for the following will be the equation of motion of the macroscopic stress, which can be obtained by inserting the definition (\ref{eq:macro_stress}) into (\ref{eq:hl_strain}):
\begin{equation}\label{eq:stress_eom}
    \frac{\partial \Sigma}{\partial \gamma}\equiv -\chi (\gamma)=1-\tY (\gamma) \langle \sigma_u \rangle
\end{equation}
where we have defined the susceptibility $\chi (\gamma)$, which measures the (negative) stress response to an infinitesimal increment in strain. $\langle \sigma_u \rangle$ is the average unstable stress, given by
\begin{equation}
    \langle \sigma_u \rangle=\frac{\int \mathrm{d}\sigma \  \sigma \ \theta \left (|\sigma|-1\right)P}{\int \mathrm{d}\sigma \ \theta \left (|\sigma|-1\right)P}
\end{equation}
The physical meaning of (\ref{eq:stress_eom}), which in the original variables reads
\begin{equation}
    \dgamma=\frac{\partial_t \Sigma}{\Go}+Y \frac{\langle \sigma_u \rangle}{\Go}
\end{equation}
is simply the decomposition of the total strain increment (which is fixed from outside) into elastic strain increment (measured by the elastic stress change) and plastic strain increment, to which each local rearrangement contributes an amount $\sigma_u /\Go$.

As noted in the introduction, the HL equation (\ref{eq:hl_strain}) at any fixed shear rate $\dgamma$ can be solved exactly in steady state; during the transient evolution, however, this is generally not the case. To gain analytical insight into the yielding behaviour for $\dgamma \ll 1$, we therefore turn now to a {\it boundary layer expansion} of the master equation (\ref{eq:hl_strain}) in shear rate, following~\cite{agoritsas_non-trivial_2017,agoritsas_relevance_2015,sollich_aging_2017}. 

\begin{figure}
\includegraphics[width=0.45\textwidth]{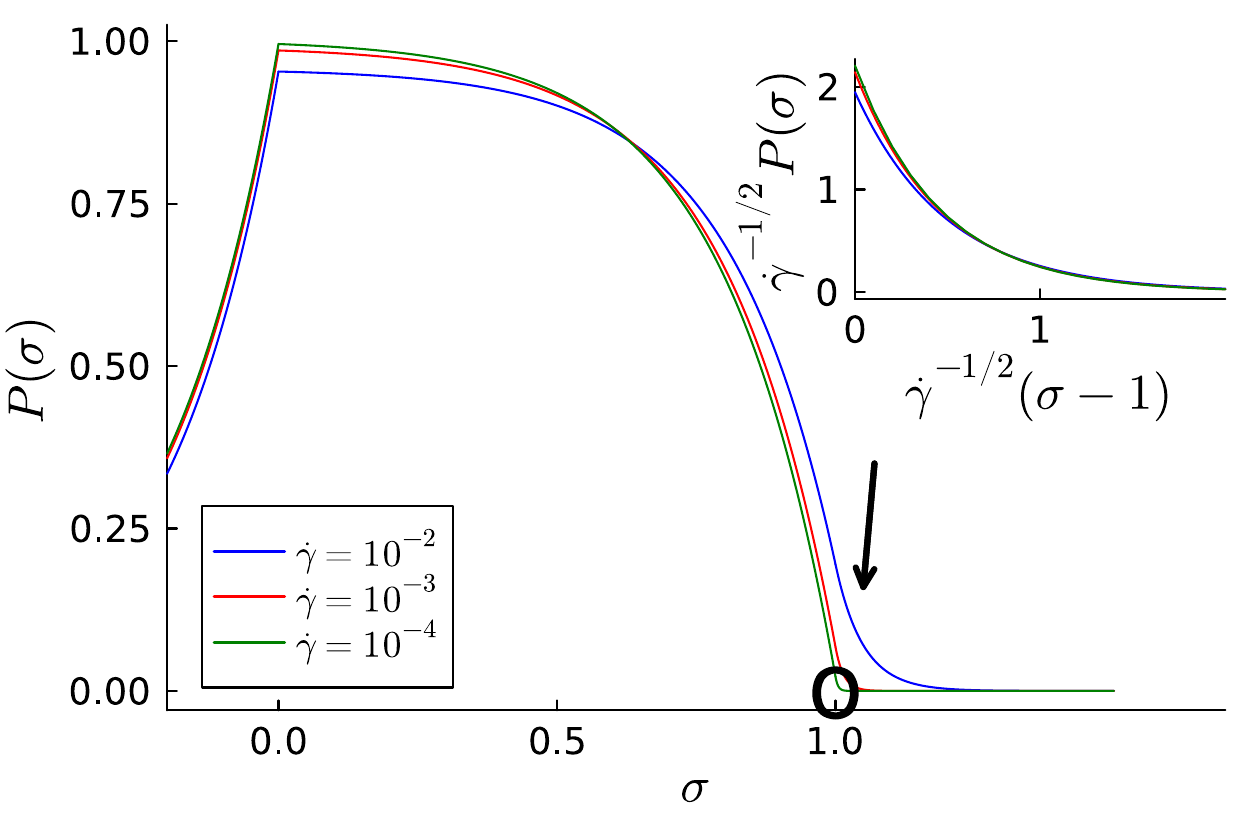}
\caption{\label{fig:BL_illustration}Illustration of the notion of a stress boundary layer. Main panel: exact solution for the local stress distribution in the yielded steady state ($\alpha=0.2$) for a range of $\dgamma$ (see App.~\ref{app:details} for the limiting distribution $Q_0$ when $\dgamma\to 0$). Inset: as explained in the text, for $\dgamma\ll 1$ local stress values beyond the threshold $\sigma_c=1$ are confined to a boundary layer of width $\mathcal{O}(\dgamma^{1/2})$.
}
\end{figure}

The physical motivation of this expansion is simple. For $R>R_c$, where there exists a smooth solution for $\forall \gamma$ in the limit $\dgamma \rightarrow 0$, one necessarily has $\tY=\mathcal{O}(1)$ or equivalently $Y=\mathcal{O}(\dgamma)$ for $\dgamma \ll 1$. Therefore, the probability {\it weight} of the (stress) boundary layers around the local thresholds $\sigma=\pm 1$, as quantified by (\ref{Y_exact}), will be $\mathcal{O}(\dgamma)$. To estimate the width of the layers, one considers that, once an element becomes unstable, it takes a time of $\mathcal{O}(1)$ ($\mathcal{O}(\pltime)$ in the original units) to undergo a rearrangement, during which time it will receive $\mathcal{O}(N Y)=\mathcal{O}(N\dgamma)$ stress kicks. Given that the stress kicks add up with random signs, this will lead to a stress change of $\mathcal{O}(\dgamma^{1/2})$, 
which 
sets the width of the boundary layers. Their height, finally, must also be $\mathcal{O}(\dgamma^{1/2})$ to recover the total boundary layer weight of $\mathcal{O}(\dgamma)$. The concept of a boundary layer is illustrated in Fig.~\ref{fig:BL_illustration}, where we take as example the yielded steady state.

All in all, one may therefore write the following expansion, distinguishing between the interior ($|\sigma|<1$) and exterior ($|\sigma|>1$) stress regions:
\begin{equation}\label{eq:expa}
    P(\sigma,\gamma)=Q_0(\sigma,\gamma)+\dgamma^{1/2}Q_1(\sigma,\gamma)+\mathcal{O}(\dgamma), \quad |\sigma|<1
\end{equation}
\begin{multline}
    P(\sigma,\gamma)=
    \dgamma^{1/2}\fext_1^{\pm}\left(\dgamma^{-1/2}(\pm \sigma-1),\gamma\right)+\mathcal{O}(\dgamma), \quad |\sigma|>1
\end{multline}
with boundary conditions (up to $\mathcal{O}(\dgamma^{1/2})$):
\begin{eqnarray}
    Q_0(\pm 1,\gamma)&=&0\\
    Q_0'(\pm 1,\gamma)&=&(\fext_1^{\pm})^{'}(0,\gamma)\\
    Q_1(\pm 1,\gamma)&=&\fext_1^{\pm}(0,\gamma)
\end{eqnarray}
Here and in the following, primes as in $Q_0'$ denote a derivative with respect to local stress, $\partial_{\sigma}Q_0$. The exterior functions $\fext_1^{\pm}$, where the $\pm$ distinguishes between the positive (at $\sigma=1$) and negative ($\sigma=-1$) boundary layers, decay exponentially on a scale set by the scaling variable $\dgamma^{-1/2}(\pm \sigma-1)$. Defining the boundary layer {\it weights} $c_0$ and $c_1$ respectively as the sum of the integrals (with a factor of $\alpha$) over the exterior functions $\fext_1^{\pm}$ and the next-order correction $\fext_2^{\pm}$, we have the following expansion for the rescaled yield rate
\begin{equation}\label{eq:expa_yield_rate}
    \tY (\gamma)=\frac{c_0(\gamma)}{\alpha}+\frac{c_1(\gamma)}{\alpha}\dgamma^{1/2}+\mathcal{O}(\dgamma^{3/2})
\end{equation}
while for the macroscopic stress
\begin{eqnarray}
    \Sigma(\gamma)&=&\int \md \sigma\  \sigma Q_0(\sigma,\gamma)+\int \md \sigma\  \sigma Q_1(\sigma,\gamma) \ \dgamma^{1/2}+\mathcal{O}(\dgamma) \nonumber\\
    &\equiv& \Sigma_0(\gamma)+\Sigma_1(\gamma)\dgamma^{1/2}+\mathcal{O}(\dgamma)
\end{eqnarray}
Inserting the expansion (\ref{eq:expa}) into the master equation one finally obtains the equations of motion of $Q_0$ and $Q_1$, at $\mathcal{O}(1)$ and $\mathcal{O}(\dgamma^{1/2})$ respectively. The first of these reads
\begin{equation}\label{eq:Q0_eom}
    \partial_{\gamma}Q_0=-\partial_{\sigma}Q_0+c_0(\gamma)\partial_{\sigma}^2 Q_0 +\frac{c_0(\gamma)}{\alpha}\delta(\sigma) 
\end{equation}
Normalisation requires $\int \mathrm{d}\sigma \ Q_0 (\sigma,\gamma)=1$ $\forall \gamma$, 
leading to the quasistatic loading condition
\begin{equation}\label{eq:bc_dQ0}
    |Q_0'(1,\gamma)|+Q_0'(-1,\gamma)=\frac{1}{\alpha},\quad \forall \gamma
\end{equation}
where we have used the fact that, on physical grounds, $Q_0'(1,\gamma)<0$ and $Q_0'(-1,\gamma)>0$ (the diffusive flux due to mechanical noise is always de-stabilising). Note that, for an arbitrary initial condition, the quasistatic loading condition (\ref{eq:bc_dQ0}) can clearly not hold immediately after starting the dynamics at $\gamma=0$. In practice, one finds that the transient regime where (\ref{eq:bc_dQ0}) does not hold is short~\footnote{This of course does not hold for initial conditions with a finite support, which are perfectly elastic during an initial transient where the distribution is simply advected by shear.}.

In steady state, by solving in parallel for the exterior functions, 
the leading order equations can be solved analytically to obtain $c_0$, $c_1$, $Q_0$ and $Q_1$ for any $\alpha$~\cite{puosi_probing_2015} (see also App.~\ref{app:details}). In fact, one can in principle solve analytically for the $\{Q\}$ and $\{\fext\}$ functions up to any arbitrary order. Outside of steady state, although the analytical solution is not known in closed form, we have checked numerically that for $R>R_c$ the expansion also holds for $\forall \gamma$. As we will see below, brittle yielding for $R<R_c$, and the singular behaviour at $R=R_c$, must be understood as {\it breakdowns} of this expansion, signalled by divergences of $c_0(\gamma)$ and $c_1(\gamma)$.

Given that, for low values of $\dgamma$, the boundary layer stress scale of $\mathcal{O}(\dgamma^{1/2})$ is challenging to resolve numerically~\footnote{Given a discretisation in stress $\md \sigma \sim M^{-1}$, resolving the boundary layers requires $M\sim \dgamma^{-1/2}$. However, due to the diffusive term one needs the strain step to satisfy $\md \gamma \sim (\md \sigma)^2$, so that $\md \gamma \sim \dgamma$}, an appealing alternative is to study directly the quasistatic limit by evolving only the interior functions. Note, however, that Eq. (\ref{eq:Q0_eom}) is not closed, as it depends on the leading order boundary layer weight $c_0$. 
One can instead introduce a {\it self-consistent} dynamics, where the interior functions are evolved while inferring the boundary layer weights from the dynamics of the macroscopic stress. This effectively amounts to {\it integrating out} the boundary layers (see App.~\ref{app:details} for details).

The leading order self-consistency equation, which we will refer to in the following as the {\it 0th order self-consistent dynamics}, is given by
\begin{equation}\label{eq:self_consistent}
    c_0(\gamma)={\left (|Q_0'(1,\gamma)|-Q_0'(-1,\gamma)\right)}^{-1}\left (1-\frac{\partial \Sigma_0}{\partial \gamma}\right)
\end{equation}
which, together with the equation of motion for $Q_0$ (\ref{eq:Q0_eom}), fully defines the dynamics solely in terms of properties of $Q_0$ (specifically its derivatives at the boundaries, and its average via $\Sigma_0$). This is (up to the approximation $Q_0'(-1,\gamma)\approx 0$, which we do not make here), equivalent to the quasistatic HL equation derived in \cite{popovic_elastoplastic_2018}. 

This self-consistent approach can in principle be extended up to any order. 
Indeed, one can write down the next order equation determining $c_1(\gamma)$ self-consistently from $Q_0$, $Q_1$ and $\Sigma_1$ (see App.~\ref{app:details}). However, these higher order self-consistency equations become rather cumbersome, and we have not found a numerically stable way of integrating the 
equations 
for $c_1$ and $Q_1$. To study the quasistatic limit we will therefore restrict ourselves to the 0th order self-consistent dynamics.

\subsection{Results}
We now consider a fixed coupling constant $\alpha$~\footnote{Note that for the numerics in Sec.~\ref{sec:first_part} we fix $\alpha=0.45$. This rather large (but within the physical jammed regime of the model $\alpha<\alpha_c=1/2$) value is chosen simply for numerical convenience, but does not affect any of the scalings; $R_c$ decreases for smaller $\alpha$, so that the numerical solution of the master equation becomes more challenging. Note that with this choice of $\alpha$ one can still neglect yield events at the negative threshold $\sigma=-1$ during the transient (i.e.\ $Q_0'(-1,\gamma)\ll 1$), but this would not be true in the yielded steady state. To study the avalanche dynamics in Sec.~\ref{subsec:avalanches} we will therefore switch to a lower value $\alpha=0.2$.} and study the dependence on the initial annealing by tuning the amount of initial disorder quantified by $R$. Considering the same ansatz as in~\cite{rossi_finite-disorder_2022}, where brittle yielding was studied in lattice elastoplastic models, the initial stress distribution is modelled as a Gaussian of width $R$
\begin{equation}\label{eq:annealing}
    P(\sigma,\gamma=0)\propto (1-\sigma^2)e^{-\frac{\sigma^2}{2R^2}}
\end{equation}
where the parabolic prefactor is added to have a small but non-zero yield rate during the initial transient: this is useful for numerical purposes, but not qualitatively relevant. Although the Gaussian shape is clearly a simplification with respect to real particle systems, the annealing ansatz (\ref{eq:annealing}) captures qualitatively the strong depletion of the tail of soft elements (i.e.\ sites with $1-|\sigma_i|\ll 1$) for well-annealed samples. This is indeed what is found for brittle samples by empirical measurements of local residual stresses in particle simulations~\cite{richard_predicting_2020}. We note that the ansatz (\ref{eq:annealing}) additionally implies that the overall distribution becomes narrower with increasing annealing, a trend not seen in particle simulations~\cite{barbot_local_2018}. In Sec.~\ref{subsec:hetero}, we will discuss a physically better justified mechanism inducing brittle yielding, within an augmented version of the HL model accounting for yield stress heterogeneity. This is not expected, however, to cause qualitative changes to the scalings we derive (see Sec.~\ref{subsec:hetero}), so that we will consider the original HL model and ansatz (\ref{eq:annealing}) throughout, which will also facilitate the comparison to the RFIM.

\begin{figure}
\includegraphics[width=0.45\textwidth]{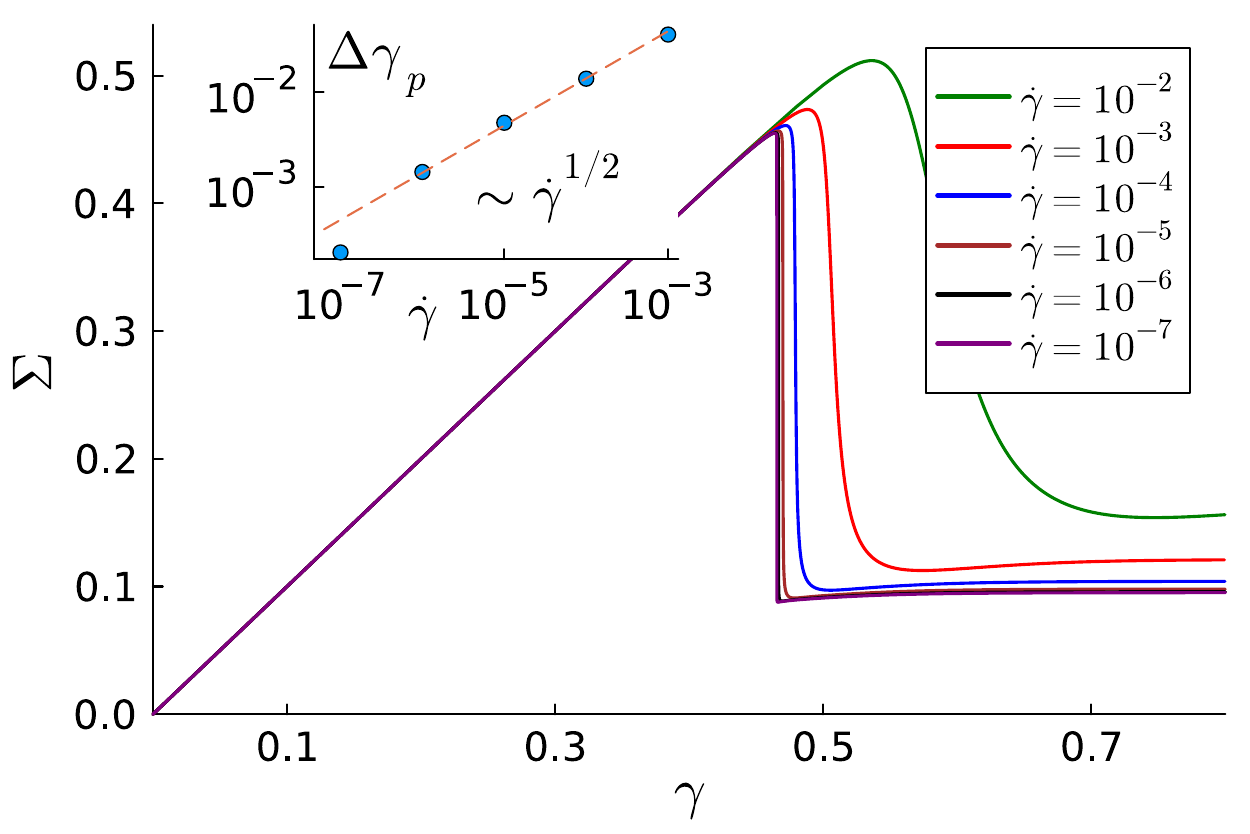}
\caption{\label{fig:brittle}Macroscopic stress $\Sigma(\gamma)$ against strain, obtained from the numerical solution of the HL master equation (\ref{eq:hl_adim}) for different shear rates. The initial disorder is fixed to $R=0.2<R_c$ ($R_c\approx 0.305$ for $\alpha=0.45$, see Sec.~\ref{subsec:TD_limit}): in this brittle regime, the stress versus strain curve develops a sharp discontinuity for $\dgamma \to 0$ as found in particle simulations~\cite{singh_brittle_2020}. Inset: approach of the yield rate peak position $\gamma_p$ to the quasistatic yield point $\gamma_c$ as $\dgamma \to 0$, consistent with $\Delta \gamma_p\equiv(\gamma_p(\dgamma)-\gamma_c)\sim \dgamma^{1/2}$.} 
\end{figure}

\begin{figure}
\includegraphics[width=0.45\textwidth]{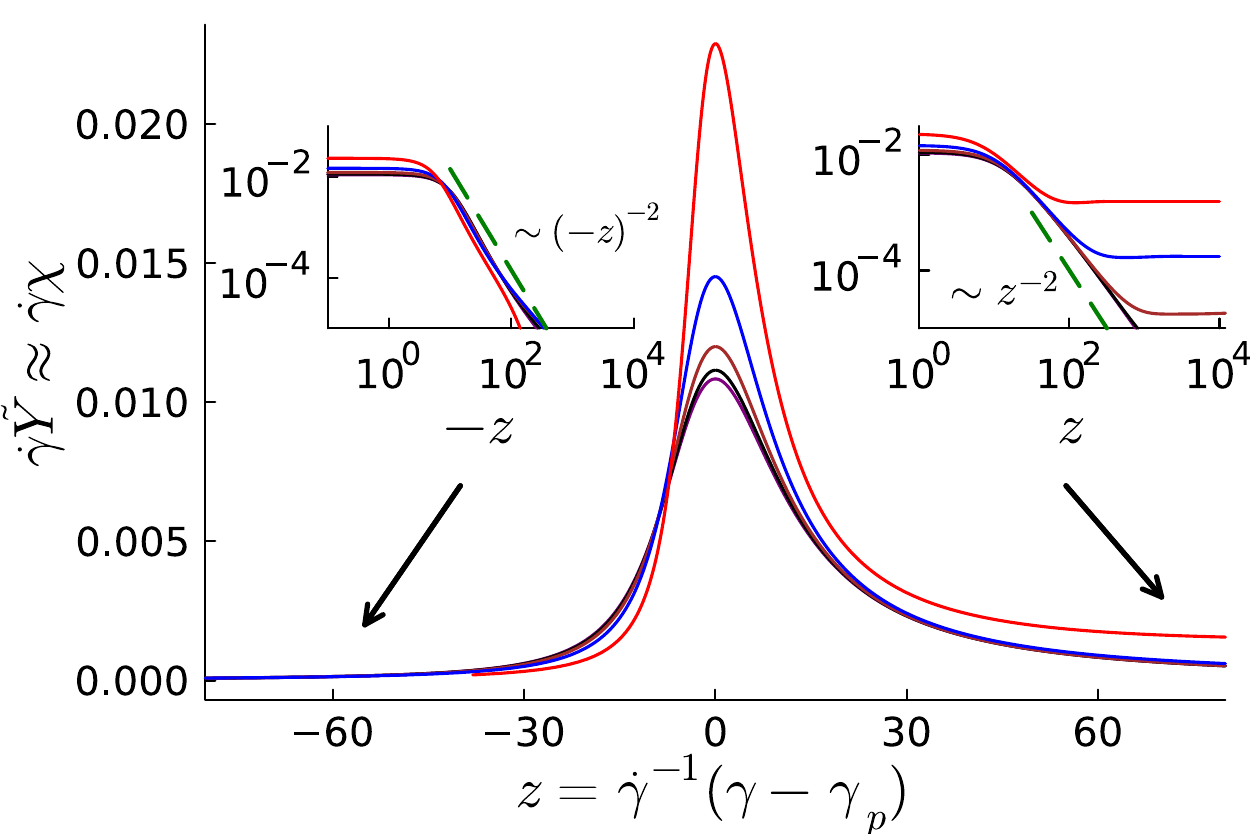}
\caption{\label{fig:yield_rate}Resolving the $\infty-$avalanche: for the same runs (except $\dgamma=10^{-2}$) shown in Fig.~\ref{fig:brittle}, we ``zoom in'' on the macroscopic stress drop by considering the time variable $\dgamma^{-1} (\gamma-\gamma_p)$ centered at the peak position for each shear rate $\dgamma$. The rescaled yield rate curves $\tY(\gamma)$ collapse following (\ref{eq:ansatz}), reflecting the expected delta-like behaviour of the  susceptibility $\chi (\gamma)$ ($\tY\approx \chi$, see text). Insets: positive and negative tails of the scaling function $f(z)$, $f_{+}$ and $f_{-}$, which display the universal power-law $f_{\pm}\sim (\pm z)^{-2}$ independent of the initial condition (see App.~\ref{app:universality}). 
}
\end{figure}

\subsubsection{Resolving the $\infty$-avalanche/ macroscopic stress drop}
Focussing now on the brittle regime $R<R_c$, our first contribution in this paper is to provide a scaling description of the discontinuous stress drop. As one may see in Fig.~\ref{fig:brittle}, where we show numerical results for different shear rates $\dgamma$, one indeed finds a sharp transition for $\dgamma \rightarrow 0$. Borrowing from the terminology of the RFIM, we will refer to this stress drop as the {\it infinite} ($\infty$-)avalanche. This is to distinguish it from the precursor avalanches that are all $\ll N$ and hence not visible in the stress-strain curve in the infinite size limit.

Given that $\Sigma (\gamma)$ has a finite discontinuity for $\dgamma \rightarrow 0$, this implies that its derivative (with respect to strain) must become {\it delta-like}. In Eq.~(\ref{eq:stress_eom}), given that $\tY\gg 1$ in the relevant regime, one can neglect the first term; on the other hand, around the brittle yield point one may safely assume that $\langle \sigma_u \rangle \approx 1$, simply reflecting the fact that elements can effectively only become unstable in the orientation of the imposed shear, and in the quasistatic limit do not become significantly stressed beyond $\sigma=1$.

Overall one therefore has the relation $\chi \approx \tY$  for the slope of the stress-strain curve. Given that during the infinite avalanche the total number of rearrangements must integrate to a finite number per mesoscopic element, one requires $\int \md t'\, Y(t') =\int \md \gamma\, \tY(\gamma) =\mathcal{O}(1)$ across the stress drop. Within the $\infty$-avalanche, one may therefore write down an ansatz of the form
\begin{equation}\label{eq:ansatz}
    \tY(\gamma)\sim \dgamma^{-b} \, f\!\left (\dgamma^{-b}(\gamma-\gammap)\right), \quad \mathrm{for}\quad \dgamma \ll 1
\end{equation}
where $\gammap$ is defined via the position of the peak in $\tY(\gamma)$. While the above ansatz holds within the peak region, to recover the steady state value $c_0^{ss}/\alpha$ after the peak an additional correction would be required, which we have omitted in (\ref{eq:ansatz}).

Now, because the $\infty$-avalanche must involve a macroscopic fraction of the entire system, 
it requires
$Y=\mathcal{O}(1)$ or equivalently $\tY=\mathcal{O}(\dgamma^{-1})$, implying $b=1$. In Fig.~\ref{fig:yield_rate}, where we show the yield rate rescaled according to (\ref{eq:ansatz}), we indeed find a good collapse for $\dgamma \rightarrow 0$. The scaling variable $z$ within the peak region therefore turns out to be $z=\dgamma ^{-1}(\gamma-\gammap)$. On reflection, this is simply the (centered) {\it time} variable. This result can be straightforwardly understood as follows. Once the boundary layer expansion has broken down (we will see below how this breakdown occurs) and the system has entered the ``peak region'' or $\infty-$avalanche, then in the quasistatic limit one expects the equation of motion for the local stresses to become purely diffusive, that is
\begin{equation}\label{eq:diffusive_hl}
    \partial_t P(\sigma,t)=\alpha Y(t)\partial_{\sigma}^2 P-\theta \left(|\sigma|-1\right) P +Y(t)\delta (\sigma)
\end{equation}
with $Y=\mathcal{O}(1)$. The HL dynamics as a function of {\it time} is indeed perfectly smooth for any $R>0$. The presence of purely diffusive dynamics over a finite timescale, on the other hand, leads to a discontinuity when considering dynamics in terms of {\it strain} in the $\dgamma \rightarrow 0$ limit.

The strictly diffusive dynamics (\ref{eq:diffusive_hl}), starting from an initial condition with $Y=\mathcal{O}(1)$, was studied in the context of {\it athermal aging} in~\cite{sollich_aging_2017,parley_aging_2020}. It was found that, asymptotically, the aging behaviour is given by $Y(t)\sim t^{-2}$, stemming from the boundary layer scalings which lead to the equation $\partial_t Y \sim Y^{3/2}$ governing the depletion of the boundary layer. Here, we correspondingly find that, for $z\gg 1$, $f_{+}(z)\sim B_{+}z^{-2}$ so that the final asymptotic approach to steady state is governed by the same athermal aging exponent. (For large $z$ the approach to the steady state leads to a correction $(c_0^{\rm ss}/\alpha) \,\dgamma$, with the height of this plateau vanishing as $\dgamma\to 0$.)  
Interestingly, the same scaling also determines the negative tail so that for $|z|\gg 1$, $z<0$, we find $f_{-}(z)\sim B_{-}\left(-z\right)^{-2}$ as shown in the inset of Fig.~\ref{fig:yield_rate}. Thus the onset of the infinite avalanche, corresponding to the diffusive {\it buildup} of the boundary layer, is described by the same exponent as the aging dynamics. One may think of this as the mean-field analogue of the initial growth of a shear band in a finite-dimensional system, where a finite fraction of the bulk of the stress distribution crosses the yield threshold due to the diffusive noise propagation. The scaling $f_{-}(z)\sim 
\left(-z\right)^{-2}$ implies that the onset of the macroscopic stress drop scales as $\Sigma \sim \Sigma_c-C|z|^{-1}$, where $\Sigma_c$ is the limiting value before entering the peak region and $C$ is a constant (see App.~\ref{app:collapse_stress} for a plot of the stress).

Importantly, the tail behaviours of $\tY$ are {\it universal}, in the sense that they do not depend on the precise form of the initial condition. We check this in App.~\ref{app:universality}, where we consider an initial condition with roughly the same $R$ but possessing power-law instead of Gaussian tails.

\subsubsection{\label{sec:breakdown}Breakdown of boundary layer expansion}
Remaining within the brittle regime $R<R_c$, we now address the question of how the expansion in shear rate (\ref{eq:expa}) breaks down, which {\it must} occur so that the system can pass from the strain-dependent evolution to the $\infty-$avalanche regime, which is described by time-dependent purely diffusive dynamics. Associated with this question is the limiting $\dgamma=0^{+}$ behaviour, given by $c_0(\gamma)$ and $\Sigma_0 (\gamma)$, on approaching the macroscopic stress drop. 

\begin{figure}
\centering

\begin{subfigure}[b]{0.4\textwidth}
   \includegraphics[width=1\linewidth]{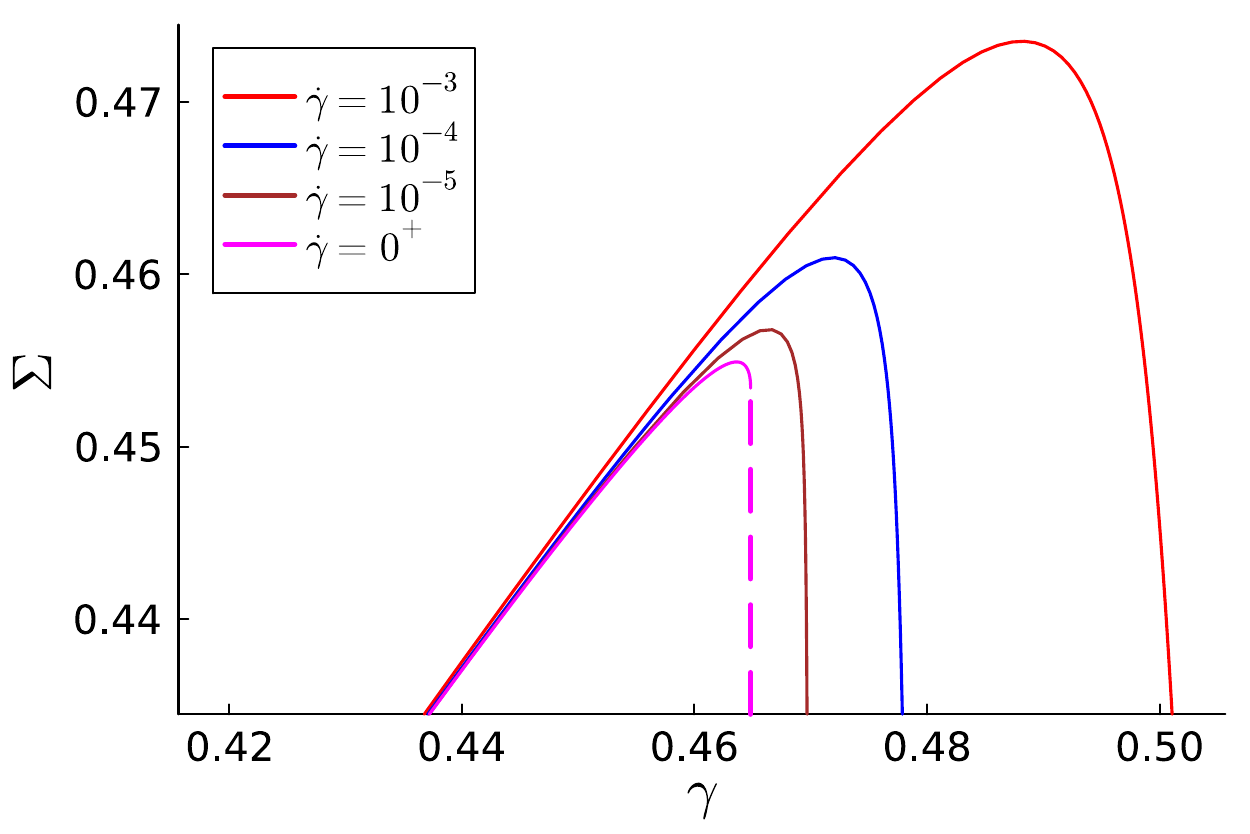}
   \label{fig:spinodal_stress}
\end{subfigure}

\begin{subfigure}[b]{0.4\textwidth}
   \includegraphics[width=1\linewidth]{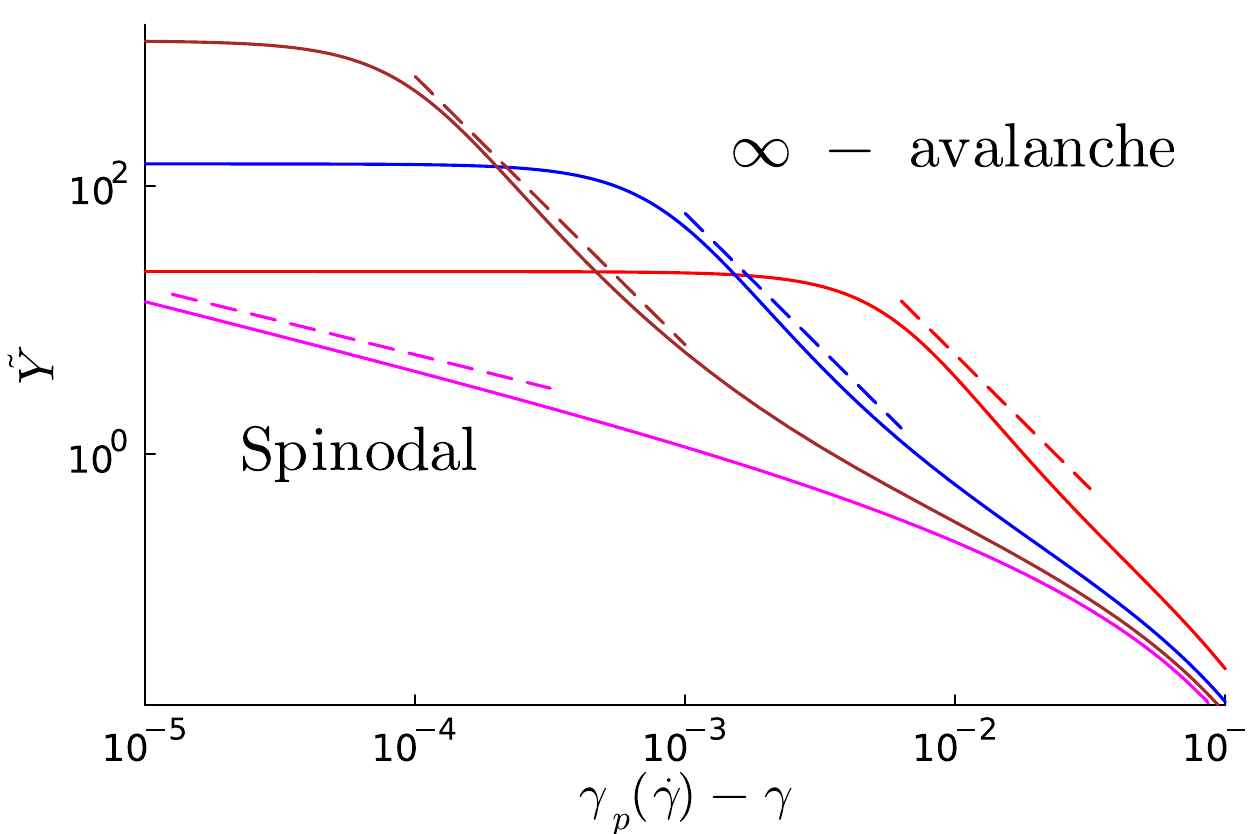}
   \label{fig:spinodal_ty} 
\end{subfigure}

\caption{\label{fig:spinodal}Breakdown of the boundary layer expansion. Top: ``zooming in'' on the macroscopic stress $\Sigma(\gamma)$ around the brittle yield point $\gamma_c$ for a subset of the runs shown in Fig.~\ref{fig:brittle}. We add the limiting quasistatic solution $\Sigma_0(\gamma)$, which ends in a square-root singularity. Bottom: corresponding behaviour of the rescaled yield rate $\tY$: for finite shear rates, the system enters the $\infty$-avalanche, indicated by the power law tail $(\gamma_p(\dgamma)-\gamma)^{-2}$ (dashed lines) reflecting the diffusive buildup of the boundary layer, while for $\dgamma=0^{+}$ one finds the spinodal divergence $c_0(\gamma)\sim (\gamma_c-\gamma)^{-1/2}$ (magenta dashed line).}
\end{figure}

Turning first to the second question, it will be useful to relate $c_0(\gamma)$ to a property of the local stress distribution, namely its curvature at the yield threshold $\sigma=1$. To see this relation, consider the equation of motion for $Q_0$ (\ref{eq:Q0_eom}). Due to the boundary condition $Q_0(1,\gamma)=0$ $\forall \gamma$, and because on approaching the yield point $Q_0'(-1,\gamma)\ll 1$ so from (\ref{eq:bc_dQ0}) $|Q_0'(1,\gamma)|\approx 1/\alpha$, one finds $Q_0''(1,\gamma)\approx -(\alpha c_0(\gamma))^{-1}$. That is, the curvature is negative, and behaves as the inverse of $c_0$. If we then write down the equation for $Q_0''(1,\gamma)$ by taking the second derivative of (\ref{eq:Q0_eom}), and insert the last result for $c_0$, we have
\begin{equation}\label{eq:curvature_eom}
    \partial_{\gamma}Q_0''(1,\gamma)=-\partial_{\sigma}^3 Q_0(1,\gamma)-\frac{1}{\alpha Q_0''(1,\gamma)}\partial_{\sigma}^4 Q_0(1,\gamma)
\end{equation}
Now, assuming (we will come back to this below) that the fourth derivative does not vanish, we have that $\partial_{\gamma}Q_0''(1,\gamma)\sim -(Q_0''(1,\gamma))^{-1}$. Positing a power-law ansatz $Q_0''(1,\gamma)\sim (\gamma_c-\gamma)^{\beta}$ this implies $\beta=1/2$ and hence 
\begin{equation}
    c_0(\gamma)\sim (\gamma_c-\gamma)^{-1/2}
\end{equation}
and from Eq.~(\ref{eq:stress_eom})
\begin{equation}
    \Sigma_0(\gamma)\sim \Sigma_c+A (\gamma_c-\gamma)^{1/2}
\end{equation}
We therefore find that the $\dgamma=0^{+}$ solution ends in a {\it spinodal} with an associated square root singularity in the macroscopic stress. This is simply due to the fact that, in the brittle regime $R<R_c$, only the curvature of the local stress distribution $Q_0$ at the yield threshold vanishes, while higher order derivatives remain finite. This recovers the behaviour of the magnetisation in the RFIM $M_c-M\sim (H_c-H)^{1/2} $; in Sec.~\ref{subsec:TD_limit} below, we will argue this result is implied by the smoothness assumption of~\cite{popovic_elastoplastic_2018}. To confirm the spinodal behaviour, we numerically solve the 0th order self-consistent dynamics, as shown in Fig.~\ref{fig:spinodal}. This also gives a numerical estimate for the brittle yield point $\gamma_c \equiv \gammap (\dgamma=0^{+})$ in the quasistatic limit.

For finite $\dgamma$, the solution must instead leave the spinodal before reaching $\gamma_c$, in order to enter the $\infty-$ avalanche regime (Fig.~\ref{fig:spinodal}). This is signalled by the breakdown of the low shear rate expansion. Considering the finite $\dgamma$ behaviour of $\tY(\gamma)$ against $\gamma_c-\gamma$, we can investigate how this breakdown occurs. We find (see App.~\ref{app:details} for details) that the finite shear rate curves ``come off'' the quasistatic limiting solution at a value $\gamma_{\rm sp}$ following $\gamma_c-\gamma_{\rm sp}\sim \mathcal{O}(\dgamma^{1/2})$. 
Finally, we find numerically that the same scaling also controls the shifting of the brittle yield point with shear rate, that is $\Delta \gammap \equiv \gammap(\dgamma)-\gamma_c\sim \dgamma^{1/2}$, as shown in the inset of Fig.~\ref{fig:brittle}.

\subsubsection{Scaling at the critical disorder}

\begin{figure}
\includegraphics[width=0.45\textwidth]{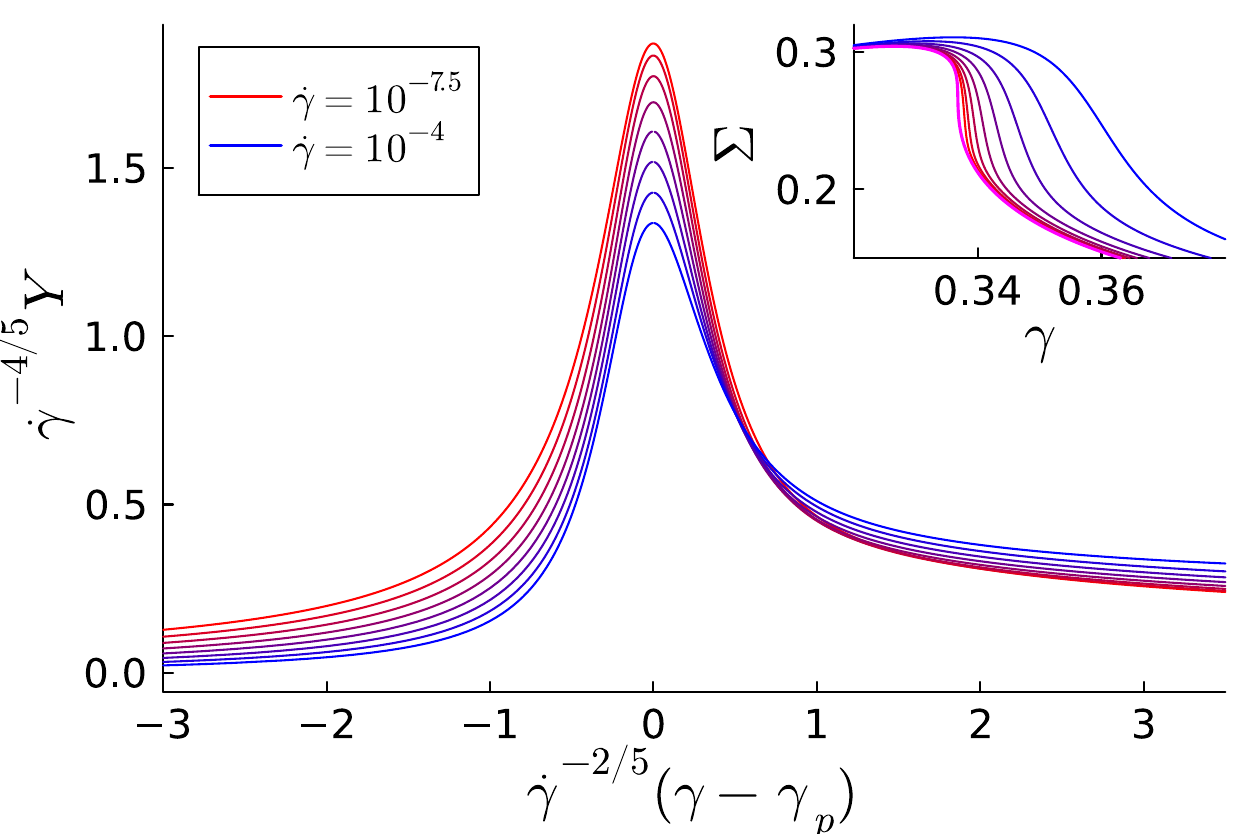}
\caption{\label{fig:Rc_scaling}Scaling plot of the yield rate close to the critical disorder ($R=0.306 \gtrapprox R_c$): shown are $8$ shear rates logarithmically spaced between $\dgamma=10^{-4}$ (blue, bottom) and $\dgamma=10^{-7.5}$ (red, top). As in Fig.~\ref{fig:yield_rate}, we ``zoom in'' on the stress drop, but now considering the scaling variable $\dgamma^{-2/5}(\gamma-\gamma_p)$ centered at the peak position for each shear rate $\dgamma$. In contrast to the brittle case (Fig.~\ref{fig:yield_rate}), note that the number of rearrangements across the peak, and the associated stress drop, does not integrate to a finite amount in the $\dgamma=0^{+}$ limit, but rather vanishes slowly as $\sim \dgamma^{1/5
}$, reflecting the fact that the quasistatic solution is continuous. Inset: stress-strain curves for the same runs, where we also include the limiting curve in the $\dgamma=0^{+}$ limit (magenta line) obtained with the self-consistent dynamics.} 
\end{figure}

\begin{table*}
   \begin{tabular}{|c|c|c|c|}
      \hline 
   & Scaling variable & Peak yield rate  & Peak susceptibility  \\
   & &  (fraction of unstable elements) &  (max. slope of $\Sigma(\gamma)$)  \\
         \hline
     $R>R_c$ \newline (Ductile) & $\gamma-\gamma_p$ (Strain) & $\mathcal{O}(\dgamma)$ & $\chip \sim \mathcal{O}(1)$ \\ \hline
     $R=R_c$ \newline (Critical) & $\dgamma^{-2/5}(\gamma-\gamma_p)$ & $\mathcal{O}(\dgamma^{4/5})$ & $\chip \sim \dgamma^{-1/5}$ \\ \hline
     $R<R_c$ \newline (Brittle) & $\dgamma^{-1}(\gamma-\gamma_p)$ (Time) & $\mathcal{O}(1)$ & $\chip \sim \dgamma^{-1}$\\ \hline
\end{tabular}
\caption{\label{table:chip} Summary of scaling behaviours of the HL model studied in the limits $N=\infty$, $\dgamma \ll 1$ (Sec.~\ref{sec:first_part}). For each regime we give the appropriate scaling variable to collapse the yield rate function around its (shear-rate dependent) peak value at $\gamma_p$, as well as the scaling with shear rate of this peak value. Note that, in time units of $\pltime$, the yield rate is nothing but the fraction of unstable sites, which must become macroscopic in the brittle regime. The corresponding scalings of the peak susceptibility follow from (\ref{eq:stress_eom}).}
    \end{table*}
We have thus far seen that, in the brittle regime $R<R_c$, the yield rate attains a peak value of $\mathcal{O}(1)$ even in the limit $\dgamma \to 0^{+}$ of quasistatic shear; this simply reflects the fact that, within the $\infty-$avalanche, a macroscopic fraction of the system becomes unstable. During this $\infty-$avalanche, the material displays a self-sustaining macroscopic cascade of plastic events {\it without} any further application of shear, and accordingly the scaling variable is simply given by the physical time $\dgamma^{-1}(\gamma-\gamma_p)$ (see Fig.~\ref{fig:yield_rate}) associated with this process. On the other hand, for ductile samples $R>R_c$ displaying an overshoot, the peak value of the yield rate in the quasistatic limit is always $\mathcal{O}(\dgamma)$ (see Sec.~\ref{sec:background}), and the macroscopic stress in the quasistatic limit is a smooth function of the applied strain. 

These behaviours can also be recast in terms of the peak susceptibility $\chip$. This quantity, which can be directly measured in particle simulations along the lines of \cite{singh_brittle_2020}, is nothing but the maximum slope of the macroscopic stress-strain curve. From the relation (\ref{eq:stress_eom}) and the above discussion, we know that it should diverge as $\chip \sim \dgamma^{-1}$ for $R<R_c$, while it will attain a finite value $\chip\sim \mathcal{O}(1)$ for $R>R_c$ (see Table~\ref{table:chip}). At the critical disorder $R=R_c$, where the macroscopic stress in the quasistatic limit is continuous but develops an infinite slope (see inset of Fig.~\ref{fig:Rc_scaling}), one naturally expects an intermediate power-law $\chip \sim \dgamma^{-b}$ (with $0<b<1$). This will correspond to an abnormally large peak yield rate $Y^{\rm peak}=\mathcal{O}(\dgamma^{1-b})$, which has to vanish in the quasistatic limit (as there is no $\infty-$avalanche), but with an exponent $1-b<1$.

In the following, we will attempt to gain insight into the change in the (transient) yielding behaviour upon crossing $R_c$, from the behaviour in the yielded steady state upon crossing the critical coupling $\alpha_c$. For $R<R_c$, due to the well-annealed initial condition, what we have seen is that the quasistatic loading puts the system on the verge of a macroscopic self-sustaining plastic activity without any further application of shear, instead maintained purely by the mechanical noise. This transition to self-sustaining activity is reminiscent of the behaviour under steady shear as the coupling parameter $\alpha$ is varied. The HL model reproduces a ``jamming'' transition as the coupling is decreased below a critical value $\alpha_c$~\cite{hebraud_mode-coupling_1998}. For $\alpha<\alpha_c$, the system is ``jammed'' and displays a yield stress as $\dgamma \to 0$, with the yield rate scaling as $\mathcal{O}(\dgamma)$. For $\alpha>\alpha_c$, the coupling is instead strong enough to sustain a $\mathcal{O}(1)$ steady plastic activity in the absence of shearing, and the model behaves as a Newtonian fluid under shear. $\alpha_c$ therefore represents the critical coupling at which plastic rearrangements just {\it barely} become self-sustaining. The self-sustaining ``fluid'' regime of the HL model for $\alpha>\alpha_c$ has been referred to as being unphysical, as it describes a steady state with dissipation in the absence of external energy input~\cite{agoritsas_relevance_2015}. Here we are working strictly in the jammed regime $\alpha<\alpha_c$ and the self-sustaining $\infty-$avalanche arises instead due to loading of a well-annealed initial local stress distribution. 

The critical behaviour of the HL model under steady shear has been fully characterised within the boundary layer ansatz framework through so-called matched
asymptotic expansions~\cite{olivier_glass_2011,agoritsas_relevance_2015}. Here we will summarise the main features, leaving the detailed equations to App.~\ref{app:scaling_Rc}. Although a rigorous extension of the scaling results for criticality under steady shear~\cite{olivier_glass_2011} to the transient regime at the critical disorder is beyond the scope of this work, we will give an intuition of why these results are expected to apply also in the latter case, and provide supporting numerical evidence. 

The main special feature of the shear rate expansion at criticality is that the leading order correction in the bulk ($|\sigma|<\sigma_c$) decouples from the leading order exterior ($|\sigma|>\sigma_c$) tail. Due to the diffusive scaling, one expects the width of the boundary layer to always scale as $Y^{1/2}$, implying also the same scaling for the height of the leading order tail $T_1$ (see Sec.~\ref{sec:background}). This then matches the amplitude of the leading order bulk correction $Q_1$. At criticality, $Y\sim \dgamma^{4/5}$, implying a boundary layer width $\sim \dgamma^{2/5}$. The leading order interior correction however scales instead as $\sim \dgamma^{1/5}$, which is only possible if $Q_1$ vanishes at $\pm \sigma_c$ to satisfy the boundary conditions. This provides a loose intuition of why the exponent denominator $s=5$ arises in the matched asymptotic expansion (see App.~\ref{app:scaling_Rc}), as it is the minimum value capable of accommodating the three different scalings. 

Within the peak regime, one expects the evolution to occur on the strain scale set by the boundary layer, implying the scaling variable $\dgamma^{-2/5}(\gamma-\gamma_p)$. Importantly, this ensures (see App.~\ref{app:scaling_Rc}) that the 0th order bulk distribution $Q_0$ remains fixed to the quasistatic limit throughout the peak regime, which is necessary given that there should be no finite jump in $Q_0(\sigma,\gamma)$ at $R=R_c$. The corresponding scaling plot of the yield rate is shown in Fig.~\ref{fig:Rc_scaling}, while in App.~\ref{app:scaling_Rc} we provide numerical data consistent with the $\dgamma^{1/5}$ scaling in the interior and the $\dgamma^{2/5}$ scaling of the boundary layer. We note that, as visible in Fig.~\ref{fig:Rc_scaling}, a collapse only appears to set in for the lowest shear rates studied; this reflects the numerical challenge of accessing the asymptotic regime at $R=R_c$ due to the slowly decaying ($\sim \dgamma^{2/5}$) width of the boundary layer.

Summing up, we have argued that from a {\it dynamical} perspective the critical disorder $R_c$ corresponds to an anomalous scaling of the rate of plastic events with shear rate. This leads to a critical divergence of the peak susceptibility with inverse shear rate, which can be directly compared to particle simulations along the lines of~\cite{singh_brittle_2020}. There, a critical scaling with inverse $\dgamma$ around the critical preparation temperature $T_{\rm ini,c}$ (equivalent to $R_c$ here) was indeed suggested by the data~\footnote{We note that in \cite{singh_brittle_2020} the authors consider the scaling of a typical lengthscale between shear bands, and not $\chip$ (see Fig.~7 there); we expect the latter however to be straightforwardly measurable in the simulations (in fact, it is shown already in Fig.~4 for $R<R_c$).}. In the second part of this paper, we will investigate the nature of $R_c$ from the complementary perspective of avalanches, where it will correspond to a random critical point marking the onset of an extensive macroscopic event, the $\infty-$avalanche. This $\infty-$avalanche, the shape of which in low dimensions will of course be dictated by the ``preferential'' directions of the elastic propagator, will be distinguished from the system-spanning but subextensive, or ``empty'' (in the sense that the density of plastic rearrangements within them vanishes as the avalanche linear extension grows~\cite{popovic_scaling_2022,lin_scaling_2014,nicolas_deformation_2018}) avalanches in amorphous solids occurring at any value of the strain.

\section{\label{sec:second_part} Part 2: $\dot{\gamma}=0^{+}$, $N \gg 1$; comparison to the Ising model in a random field}

In this second part we will now turn to the complementary limit of $\dgamma=0^{+}$, $N \gg 1$. That is, we will fix the shear rate to the quasistatic limit and study the behaviour of large but finite system sizes. Considering the $\dgamma=0^{+}$ limit will allow us in particular to compare to the behaviour of the quasistatically driven RFIM at zero temperature.

This second part is structured as follows: in \ref{subsec:avalanches} we will firstly fully characterise the avalanche distribution in the flowing steady state, and during the transient for ductile samples displaying a mild stress overshoot, contrasting the results for both cases with the avalanche statistics of the RFIM. In \ref{subsec:TD_limit}, we will compare the behaviour of both models in the infinite size limit, showing that Landau exponents result in both cases but with starkly different underlying signatures. In \ref{subsec:precursors}, we then consider the brittle yielding of finite-size systems. 
We show that the way the spinodal limit controls the approach to failure is qualitatively different in the two models, recovering for the HL model the qualitative trends found in particle simulations~\cite{ozawa_role_2020}, and we consider the role of precursors and the possibility of predicting failure. 
Finally, in \ref{subsec:criticality} we consider finite-size samples prepared at the critical disorder. As in the RFIM, the avalanche distribution is controlled by a random critical point. Importantly, however, this criticality arises on top of the underlying marginality, and hence possesses scale-invariance properties different  from those of the RFIM critical point. 

\subsection{\label{subsec:avalanches}Avalanches} 
With the aim of studying avalanches of plastic rearrangements we turn again to the discrete block version of the dimensionless HL model (\ref{eq:hl_adim}), introduced in Sec.~\ref{sec:mf_theory}, and study it directly in the quasistatic limit. Consider $N$ mesoscopic blocks, each one of which is assigned a shear stress $\sigma_i$. As is typically done, it will be useful to introduce also the local {\it stability} $x_i$, defined as $1-\sigma_i$, which simply measures the strain increment (recall $\Go=1$) needed to make the block unstable.

Initially, we assume $x_i >0$ (and $<2$) for all blocks $i$. We then consider the following update rules. We first look for the element closest to instability, which defines $\xmin\equiv \min_i \{ x_i\}$, and strain all elements by this amount. We then set the stress of the unstable element to $0$ and apply stress kicks to all the rest as detailed in Sec.~\ref{sec:mf_theory}. We finally build a list of all unstable elements ($|\sigma_i|>1$) after the stress kicks, the length of which we refer to as $N_u$. 

This constitutes the first step of the avalanche. Subsequent steps are performed iteratively as follows: 
\begin{enumerate}
\item Pick an element at random from the unstable list, and reset its stress to $0$.
\item Apply independently drawn Gaussian stress kicks to all other elements, with standard deviation $(2 \alpha/N)^{1/2}$.
\item Update the list of unstable elements. If $N_u>0$, return to step 1.
\end{enumerate}
These update rules precisely match the quasistatic limit of the HL master equation (\ref{eq:hl_adim}): once an element is unstable, the resetting is a Poisson stochastic process with rate one. Importantly, an unstable element can re-stabilise before it resets by yielding. In the quasistatic limit, where any finite timescale becomes instantaneous relative to the timescale of the shear, this reduces to the algorithm detailed above.

Once the avalanche has terminated, we can define the {\it avalanche size} $S$ as the total number of steps taken in the above algorithm, i.e.\ the number of rearrangements that have taken place. In the regime of physical interest of the model, where one can neglect the yield events occurring in the orientation opposite to the imposed shear, the avalanche size is related to the associated stress drop as $\Delta \Sigma=S/N$.

\begin{figure}
\includegraphics[width=0.45\textwidth]{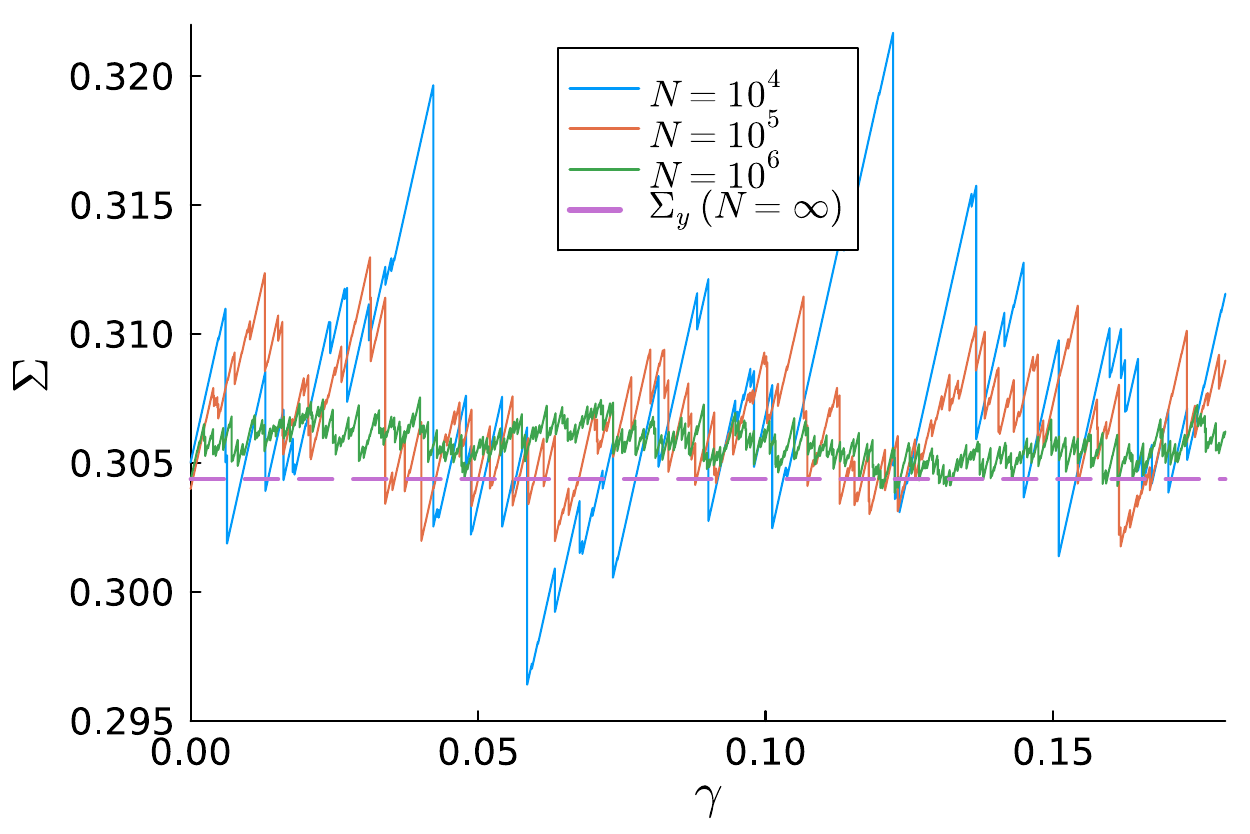}
\caption{\label{fig:ava_ss}Macroscopic stress versus strain in the flowing steady state of the HL model (with coupling $\alpha=0.2$) under quasistatic driving for different system sizes $N$. The dynamics consists of periods of elastic loading interrupted by stress drops $\Delta \Sigma$, related to the avalanche size $S$ as $\Delta \Sigma=S/N$. As in real amorphous solids, avalanche sizes depend on system size, with the average stress drop scaling as $\langle \Delta \Sigma \rangle \sim N^{-\tilde{\alpha}}$ with $\tilde{\alpha}=1/2$. We also show (dashed line) the analytically derived yield stress for $N=\infty$, which is indeed approached as the system size is increased.} 
\end{figure}

The steady state stress-strain behaviour of the HL model obtained from the above quasistatic dynamics is shown in Fig.~\ref{fig:ava_ss}. The HL model is capable of reproducing what is arguably the key feature of quasistatically deformed amorphous solids~\cite{maloney_subextensive_2004,shang_elastic_2020}, namely the {\it sub-extensive} scaling of the stress drops $\langle \Delta \Sigma \rangle \sim N^{-\tilde{\alpha}}$, with $\tilde{\alpha}<1$ ($\tilde{\alpha}=1/2$ in HL, see below). Before we even turn to the avalanche distribution, we may note already that such a scaling with system size is in contrast to the RFIM, where the avalanche sizes are always $\mathcal{O}(1)$ and do not grow with system size. 

To provide a theoretical understanding of the avalanche behaviour, our main source of inspiration will be the pioneering ideas of Jagla~\cite{jagla_avalanche-size_2015} (which were also partially applied in~\cite{popovic_elastoplastic_2018}). It is important to note, however, that the analysis of~\cite{jagla_avalanche-size_2015} does not apply directly to the HL model. Importantly, for the dynamics studied in~\cite{jagla_avalanche-size_2015}, once a site crosses the threshold, its fate is ``sealed'' in the sense that it cannot re-stabilise by receiving a 
stress kick of negative sign~\footnote{See footnote 38 in \cite{jagla_avalanche-size_2015}.}. The dynamics can therefore be thought of as corresponding to an infinitesimal $\pltime$; this should lead to important qualitative differences: e.g.\ with $\pltime=0^+$, even the Herschel-Bulkley law is not recovered. Further quantitative differences are expected due to the different form of the local stress relaxation in the model studied by Jagla~\cite{jagla_avalanche-size_2015}, where the stress release is exponentially distributed. We will nonetheless see below that the basic ideas of~\cite{jagla_avalanche-size_2015} can in fact be applied to the HL model, both in the transient and in steady state, {\it provided} one accounts for a fundamental qualitative difference in the shape of $P(x)$. Our main novel result will be to show that, although the avalanche exponent $\tau$ can {\it in principle} take non-universal values, the finite-size HL dynamics in fact universally imposes $\tau \approx 1$.

The powerful idea of Jagla~\cite{jagla_avalanche-size_2015} is to consider the following mapping, of the evolution of $N_u$ during an avalanche, to an effective random walk; {\it three} terms will arise in this mapping. Consider the discrete steps performed during the avalanche iteration (see algorithm above), which we label by $j=1,2,\ldots$
and the current number of unstable elements as $N_{u,j}$. At each step $j\rightarrow j+1$, we know $N_u$ experiences a deterministic reduction by unity; the number of elements that re- or de-stabilise due to stress kicks, however, and hence the overall increment $N_{u,j+1}-N_{u,j}$, is a random variable. As the baseline for the following discussion consider the situation where, on average, each yield event leads to one additional unstable site, so that $N_u$ (on average) remains constant. This implies a {\it strictly} linear form for the distribution $P(x)$ of the $x_i$ close to the boundary at $x=0$, in particular $x/\alpha$, which is the steady state solution in front of an absorbing boundary at $x=0$ (the diffusive flux across the boundary $\alpha Y\partial_x P$ exactly balances the re-injection rate $Y$, and is in addition constant across $x>0$ implying steady state). Under these conditions, the number of new unstable elements generated at each step will be a Poisson random variable with mean one (given that the values $x_i$ are uncorrelated~\cite{jagla_avalanche-size_2015}) , so that overall the increment will correspond to $N_{u,j+1}-N_{u,j}\sim \mathrm{Poiss} (\lambda=1)-1$, which is a random variable with mean zero and unit variance. 

In the limit of many steps, we can switch to a continuous time description $j \rightarrow \avtau$. Note that $\avtau$ here is unrelated to the physical time in the model, and just replaces the discrete counter. From the above, $N_u(\avtau)$ converges to an unbiased random walk, with $\langle N_u^2(\avtau)\rangle=\avtau$. The avalanche {\it stops} when $N_u$ returns to $0$, hence the statistics of $S$ will be related to a first-passage problem.

The other two terms that appear in the mapping are effectively {\it corrections} to the $P(x)=x/\alpha$ behaviour, and hence to the unbiased random walk. The first of these accounts for corrections to the slope; for $N \gg 1$, this correction will be accounted for by the (negative) curvature at the threshold $P''(0)<0$. To estimate the contribution from this correction, we can proceed following~\cite{popovic_elastoplastic_2018}. Up to the avalanche ``time'' $\avtau$, the stress kicks generated lead to a stress scale $\sim (2\alpha \avtau/N)^{1/2}$, due to the standard deviation $(2 \alpha /N)^{1/2}$ of each kick. The accumulated number of elements that have {\it incorrectly} been accounted for in the unbiased random walk can then be estimated as
\begin{equation}\label{eq:g_of_z}
    N \int_{0}^{(2\alpha \avtau/N)^{1/2}} \left (\frac{x}{\alpha}-P(x)\right) \md x \equiv N g\left(\sqrt{\frac{2\alpha \avtau}{N}}\right)
\end{equation}
where one assumes a constant fixed shape for $P(x)$ during the avalanche dynamics. For $z \ll 1$, $g(z)$ can be expanded as $\sim |P''(0)|z^3/6$, so that the contribution up to ``time'' $\avtau$ may be approximated by
\begin{equation}
    \approx N^{-1/2}\frac{|P''(0)|}{6}(2 \alpha)^{3/2}\avtau^{3/2}
\end{equation}
which constitutes a {\it negative} accumulated drift term in the random walk mapping, and will be responsible for the appearance of a finite cutoff (on the scale $N^{1/2}$) in the avalanche distribution.

The final term to consider will instead constitute a {\it positive} accumulated drift. We will refer to this as the {\it plateau source}. The original idea in~\cite{jagla_avalanche-size_2015} was that, after straining by $\xmin$ to trigger the next avalanche, effectively a plateau is generated at the yield threshold, of height $\sim \xmin/\alpha$. However, we find instead that, for any finite system size $N$, where the stress kicks have a finite typical size $\Delta (N)\propto (2\alpha/N)^{1/2}$, $P(x)$ in fact {\it always} has a plateau on the finite stress scale $x \sim \Delta (N)$, even before straining the distribution by $\xmin$.
\begin{figure}
\includegraphics[width=0.45\textwidth]{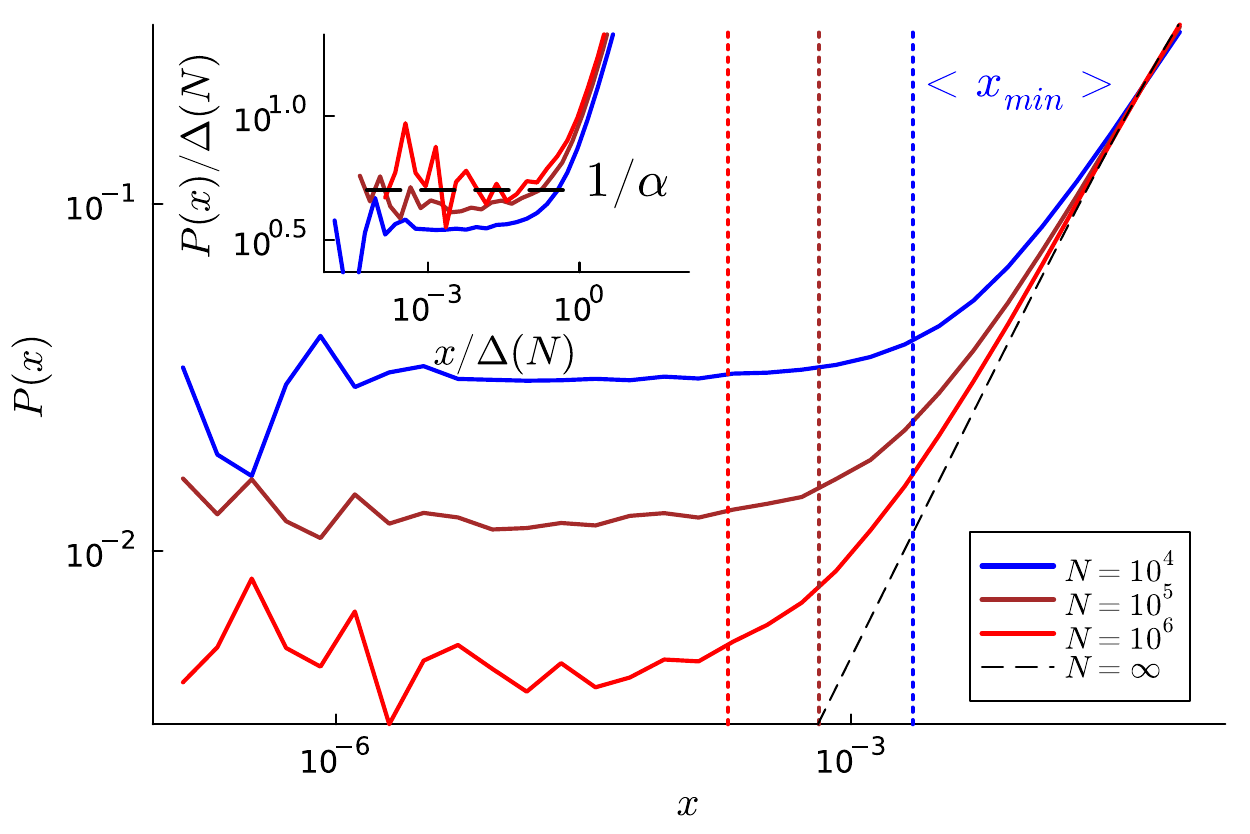}
\caption{\label{fig:plateau}Histogram of local stabilities $\{x_i \}$ after an avalanche in the steady state of the HL model (Fig.~\ref{fig:ava_ss}), obtained by averaging over many avalanches. In the limit $N=\infty$, $P(x)$ is simply linear $P(x)\approx x/\alpha$ for $x \ll 1$ (dashed line). For any finite $N$, $P(x)$ instead develops a finite plateau below a crossover scale $\Delta (N)\sim N^{-1/2}$. We indicate with dotted lines the location of the average $\langle x_{\rm min}\rangle$ for each system size $N$, which always lies at the edge of the plateau. Inset: collapse of the plateau region by rescaling both axes by $\Delta (N)= c(2\alpha/N)^{1/2}$, with a numerical prefactor $c \approx \sqrt{2}$. With this prefactor, the plateau height (see text) follows $P_0(N)/\Delta (N)\approx 1/\alpha $, as indicated by the dashed line.} 
\end{figure}

This is illustrated in Fig.~\ref{fig:plateau}, where we measure the statistics of $\{x_i\}$ after each avalanche in steady state. $\Delta(N)$ marks a crossover scale, below which $P(x)$ tends to the constant value $P_0(N)$. Finding analytically the value of $P_0$ would require the knowledge of the full analytical form of $P(x)$ for finite but large $N$. A naive estimate is to consider $P_0(N)=\alpha^{-1}\,\Delta(N)$ by intersecting the linear behaviour $x/\alpha$ with the typical scale $\Delta (N)=c(2\alpha/N)^{1/2}$; we find the unknown prefactor in this proportionality to be numerically close to $c=\sqrt{2}$, so that $P_0(N)\approx 2\alpha^{-1/2}  N^{-1/2}$.

The presence of a plateau in $P(x)$ within elastoplastic models is not new: such behaviour was described also in~\cite{ferrero_properties_2021,ruscher_residual_2020}. In fact, the HL plateau scalings we find here correspond, as one would expect, to those found for a quenched random interaction kernel on a finite-dimensional lattice~\cite{ferrero_properties_2021}. The appearance of the plateau fundamentally alters the form of $P(\xmin)$. Whereas a linear $P(x)$ leads to a (one-sided) Gaussian $P(\xmin)$ with linear prefactor~\cite{jagla_avalanche-size_2015}, the plateau implies that $P(\xmin=0)\neq 0$, and the distribution is instead initially exponential, $P(\xmin)\sim e^{-N P_0(N) \xmin}$. In Fig.~\ref{fig:plateau} we indicate the location of the average value $\langle \xmin \rangle$. This lies at the edge of the plateau for each of the three system sizes studied, so that typical values of $\xmin$ lie within the plateau. This means that, typically, straining by $\xmin$ leads to the same plateau value $P_0$. Sampled values of $\xmin$ that lie in the linear range of $P(\xmin)$ beyond the plateau can only lead to a higher plateau after straining by $\xmin$, so $P_0(N)$ may also be thought of as a {\it lower bound} on the generated plateau height.

We now need to find the correction to the unbiased random walk $N_u(\avtau)$ from this plateau source. As in~\cite{jagla_avalanche-size_2015}, we estimate this via the problem of diffusion from a finite plateau of height $P_0$ for $x>0$ across an absorbing boundary at $x=0$. At long times, the accumulated diffusive flux from this process is given by~\footnote{For the problem of diffusion from a plateau of height $P_0$, with diffusion constant $D$ across an absorbing boundary at $x=0$, the time-dependent outward flux is given by $P_0/\sqrt{\pi D \avtau}$. Integrating this up to avalanche time $\avtau$, and using that the diffusion constant in avalanche time is given by $D=\sigma^2/2$, where $\sigma=(2 \alpha/N)^{1/2}$ is the standard deviation of the kicks, one recovers the result (\ref{eq:plateau_source}).}
\begin{equation}\label{eq:plateau_source}
    2 \sqrt{\frac{\alpha}{\pi}} B \avtau^{1/2}\equiv \left(A \avtau\right)^{1/2}
\end{equation}
Here we have defined $B$ as the prefactor in $P_0(N)=B N^{-1/2}$ and the constant $A$ via $A^{1/2}=2 (\alpha/\pi)^{1/2} B$, which after inserting the numerical estimate for $B$ gives \begin{equation}
    A^{1/2}\approx 4/\sqrt{\pi}
    \label{eq:A}
\end{equation}

\begin{figure}
\includegraphics[width=0.45\textwidth]{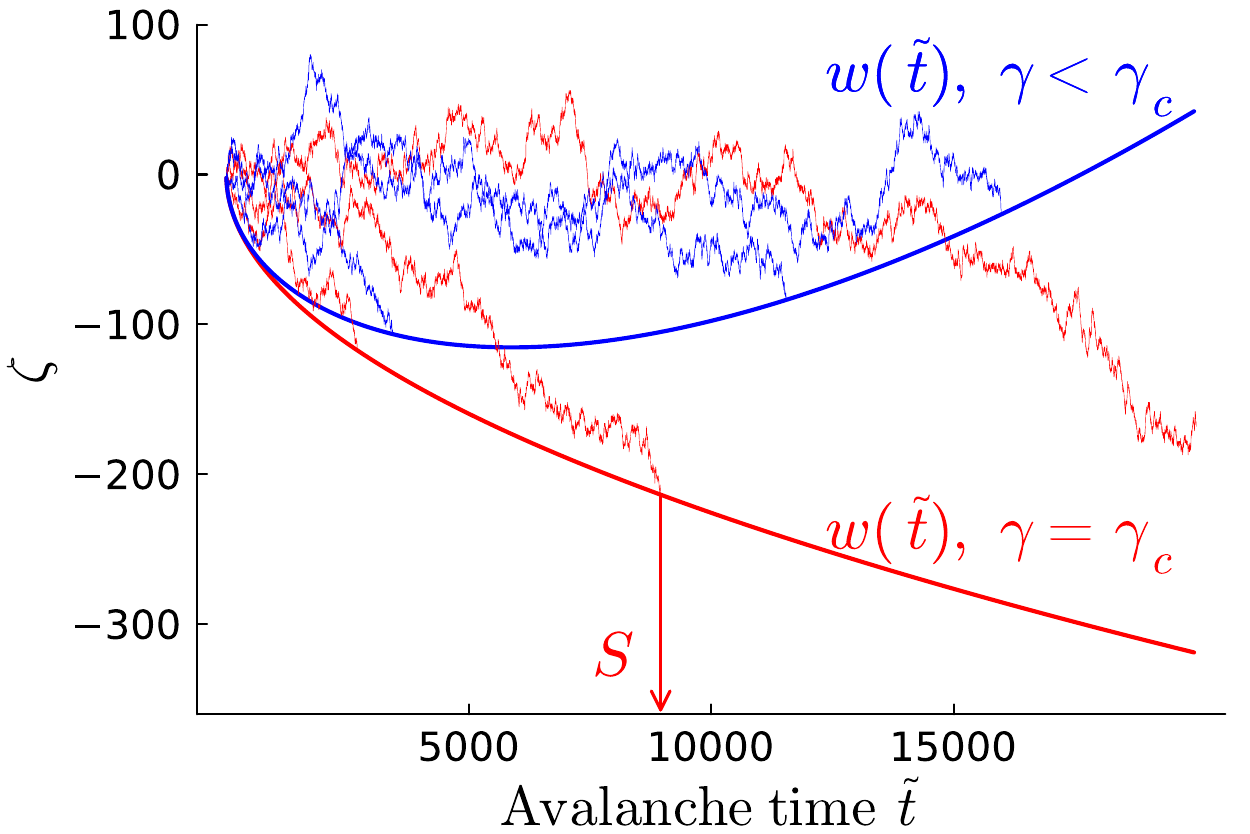}
\caption{\label{fig:mapping}Graphical illustration of the mapping introduced by Jagla~\cite{jagla_avalanche-size_2015}. Instead of the original first-passage problem for $N_u$ with additional drifts, one considers an unbiased random walker $\zeta(\avtau)$ in the presence of a moving absorbing boundary $w(\avtau)$ given by (\ref{eq:wall}). The avalanche size $S$ is then determined by the first passage across this wall (arrow). For the plot we consider $N=10^{6}$, typical values for $\alpha$ and $P''(0)$ and $A_{\mathrm{min}}^{1/2}\approx 4/\sqrt{\pi}$ for $A$ (see text). At $\gamma=\gamma_c$, $P''(0)=0$ and there is no finite cutoff size, as illustrated by the random walk that has not been absorbed until the longest avalanche time shown.} 
\end{figure}

Because the statistics of $S$ are determined by the first passage problem for $N_u(\avtau)$, it is useful to recast the accumulated drift terms as a {\it moving} absorbing wall~\cite{jagla_avalanche-size_2015}. 
That is, instead of a random walk with non-trivial drift terms, one may instead think of an unbiased random walker (with unit variance), whose position at avalanche time $\avtau$ we denote by $\zeta(\avtau)$, and its first-passage problem across an absorbing wall $w(\avtau)$ moving as (see Fig.~\ref{fig:mapping})
\begin{equation}\label{eq:wall}
    w(\avtau)\approx N^{-1/2} \frac{|P''(0)|}{6}(2\alpha)^{3/2}\avtau^{3/2}-(A\avtau)^{1/2}
\end{equation}
Although the full analytical solution to this first passage problem is not known, one may derive~\cite{jagla_avalanche-size_2015} a heuristic expression for the form of the cutoff in the avalanche (first passage time) distribution, by considering that the avalanche stops when $\zeta(\avtau)\approx w(\avtau)$. This leads to~\footnote{Specifically, (\ref{eq:p_of_S}) is obtained by considering the asymptotic form expected for the random walk $\zeta(\avtau)\sim e^{-\zeta^2/(2\avtau)}$, inserting $w$ in the place of $\zeta$ and replacing $\avtau$ by $S$ (as the avalanche size $S$ is determined by the time at which the wall is hit $\zeta (S)=w(S)$).}
\begin{equation}\label{eq:p_of_S}
    P(S)\sim S^{-\tau}\exp{\left(-\left(\frac{S}{S_c}\right)^{2}+a \frac{S}{S_c} \right)}
\end{equation}
where instead of the typical exponential cutoff (see (\ref{eq:p_of_S_RFIM}) below) one obtains instead a Gaussian tail, along with an overshoot term ($a>0$). We will now show how the random walk mapping correctly accounts for the measured values of the avalanche exponent $\tau$ and the cutoff $S_c(N)$, which we define through Eq. (\ref{eq:p_of_S}).

\begin{figure}
\includegraphics[width=0.45\textwidth]{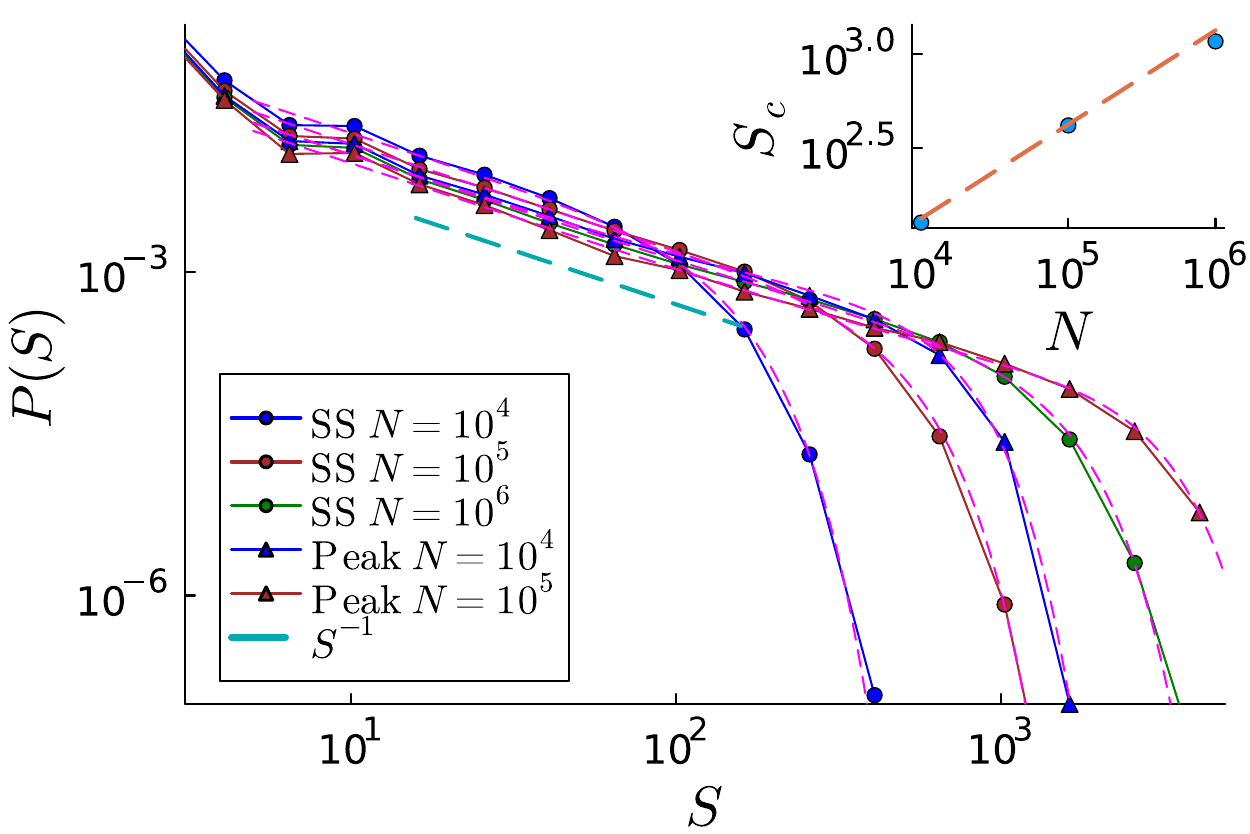}
\caption{\label{fig:ava_non_collapsed}Avalanche distribution $P(S)$ obtained for different system sizes, in steady state (SS) and around the peak of the susceptibility (Peak) for each finite $N$ (see Fig.~\ref{fig:suscep_N}). Magenta dashed lines correspond to fits of the form (\ref{eq:p_of_S}). Inset: cutoff sizes $S_c(N)$ obtained from the fit (\ref{eq:p_of_S}) in steady state; we find good agreement with the analytical prediction $S_c(N)\approx 3 \alpha^{1/2}N^{1/2}$ (dashed line).} 
\end{figure}

Turning first to the cutoff $S_c(N)$, from the heuristic derivation~\cite{jagla_avalanche-size_2015} leading to (\ref{eq:p_of_S}) one has that $S_c \approx 3 \alpha^{-3/2}|P''(0)|^{-1}N^{1/2}$. In steady state, we can resort to the $N\to\infty$ limit to estimate the curvature. Given that $\tY\approx 1$ from Eq.~(\ref{eq:stress_eom}), $\partial_{\gamma}Q_0(1,\gamma)=0$ implies that $|Q_0''(1)|\approx \alpha^{-2}$. This leads to the following cutoff in the steady state of the HL model
\begin{equation}\label{eq:SS_Sc}
    S_c(N)\approx 3 \alpha^{1/2}N^{1/2}
\end{equation}
which holds well numerically (see inset of Fig.~\ref{fig:ava_non_collapsed}).

We now turn to the value of the avalanche exponent $\tau$. If the curvature vanishes, one is left with the problem of an unbiased random walk 
in the presence of a boundary receding as $(A \avtau)^{1/2}$. As pointed out in~\cite{jagla_avalanche-size_2015}, this corresponds to the problem of a one-dimensional random walker in a cage limited by $(-w_0(\avtau),\infty)$, where the left boundary is expanding as $w_0(\avtau)=(A \avtau)^{1/2}$. It turns out that, of all possible functional forms of $w_0(\avtau)$, $w_0(\avtau)\sim \avtau^{1/2}$ is the most interesting, as it competes with the scaling of the random walk. Indeed, this is the {\it marginal} case, and the exponent $\sim \avtau^{-\beta}$ describing the decay of the survival probability becomes non-universal~\cite{krapivsky_life_1996,bray_persistence_2013,redner_first_2022}, as it depends on the value of $A$~\footnote{In general, it depends on the ratio $A/D$ where $D$ is the diffusion coefficient associated to the random walk. For the present case of a random walk with unit variance, $D=1/2$.}: for $A \ll 1$, $\beta \rightarrow 1/2$ from below, while for $A\gg 1 $ $\beta \rightarrow 0$. The exponent $\beta+1$ of the first passage time distribution, defined as the first time the random walker hits the boundary of the cage, therefore varies between $3/2$ and $1$: we recall that the first passage sets the size of the avalanche, so that $\beta+1$ is also the avalanche exponent $\tau$.

In the HL model, therefore, as in the model studied by Jagla~\cite{jagla_avalanche-size_2015}, one in principle ends up with a mixture of power laws making up the avalanche distribution, as each value of $\xmin$ leads to a different value of $A$ and hence $\tau$. Indeed, Jagla reported exponents roughly in the range $\tau=1.1\ldots 1.2$. We now show, however, that due to the appearance of a plateau in $P(x)$, $\tau \approx 1$ in the HL model:  although the exponent $\tau$ is {\it in principle} non-universal, it effectively saturates its lower bound.

To see this, we need to turn again to the form of $P(x)$ shown in Fig.~\ref{fig:plateau}. As already noted above, whatever the value of $\xmin$ that is sampled, the plateau height that is generated is always bounded from below by $P_0=P(x=0)$. Accordingly $A$ is lower bounded by the value $A_{\mathrm{min}}^{1/2}\approx 4/\sqrt{\pi}$ given in (\ref{eq:A}), which is {\it independent} of $\alpha$.  
This lower 
bound on $A$ gives 
an {\it upper bound} on $\tau$. 
Numerically, $A_{\rm min} \approx 5$ 
turns out to be well within the asymptotic regime $A \gg 1$. In fact, although the full problem of finding $\tau\left(A\right)$ corresponds to finding the ground state of a quantum harmonic oscillator with an infinite wall on one side~\cite{jagla_avalanche-size_2015}, for $A\gg 1$ one can safely approximate this by~\cite{redner_first_2022,krapivsky_life_1996}~\footnote{Specifically, we use Eq.(9.10) in \cite{redner_first_2022}: this so-called ``free'' approximation consists in assuming the probability density profile of a free random walker at long times, and self-consistently relating the decay of the survival probability to the flux axross the expanding boundary. Here, only one of the boundaries is expanding, leading to an overall factor of $1/2$ with respect to (9.10) (and we recall $D=1/2$, see footnote above).}
\begin{equation}
    \tau \simeq 1+\frac{1}{4\sqrt{\pi}}\sqrt{2A}e^{-\frac{A}{2}}
\end{equation}
Plugging in the value of $A_{\mathrm{min}}$,this gives $\tau_{\mathrm{max}}\approx 1.03$. Therefore, regardless of the precise prefactor in $A_{\mathrm{min}}$, one expects that universally $\tau \approx 1$ in the HL model.

\begin{figure}
\includegraphics[width=0.45\textwidth]{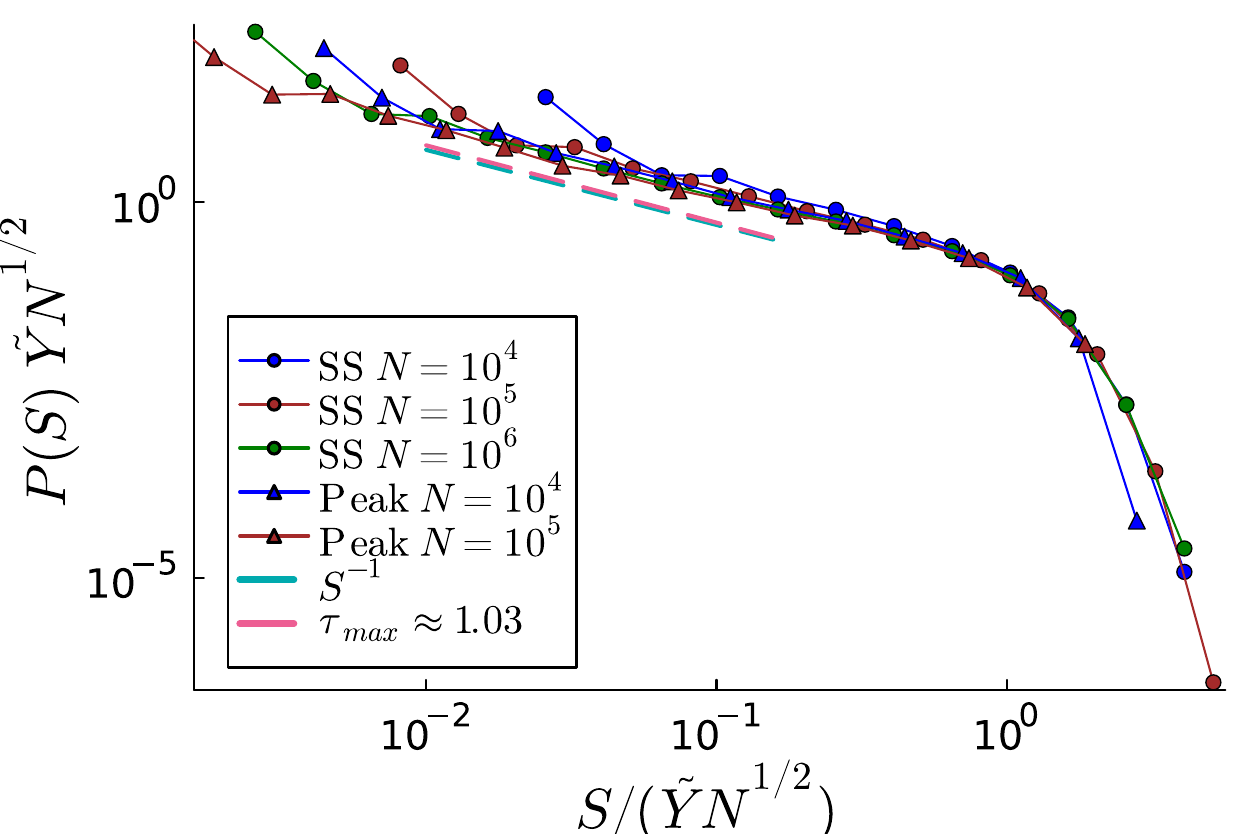}
\caption{\label{fig:ava_collapsed}Avalanche distributions from Fig.~\ref{fig:ava_non_collapsed} collapsed onto a single master curve, by scaling the cutoff sizes as $S_c\sim \tilde{Y} N^{1/2}$. In steady state $\tilde{Y}\approx 1$, whereas for the peak distributions we use the strain-averaged $\tY$ across a narrow interval (see Fig.~\ref{fig:suscep_N}). We additionally show the power law $S^{-\tau_{\rm max}}$ with $\tau_{\rm max}\approx 1.03$ calculated in the text; this is of course numerically indistinguishable from $S^{-1}$.} 
\end{figure}

Given that $\tau \approx 1$, to fit the avalanche distributions we simply use the expression (\ref{eq:p_of_S}) with $\tau=1$. As can be seen in Fig.~\ref{fig:ava_non_collapsed} for the distributions in steady state, the fit is excellent. Due to the scaling $S_c\sim N^{1/2}$ of the cutoff, one can also collapse the distributions onto a master curve (Fig.~\ref{fig:ava_collapsed}).

\begin{figure}
\includegraphics[width=0.45\textwidth]{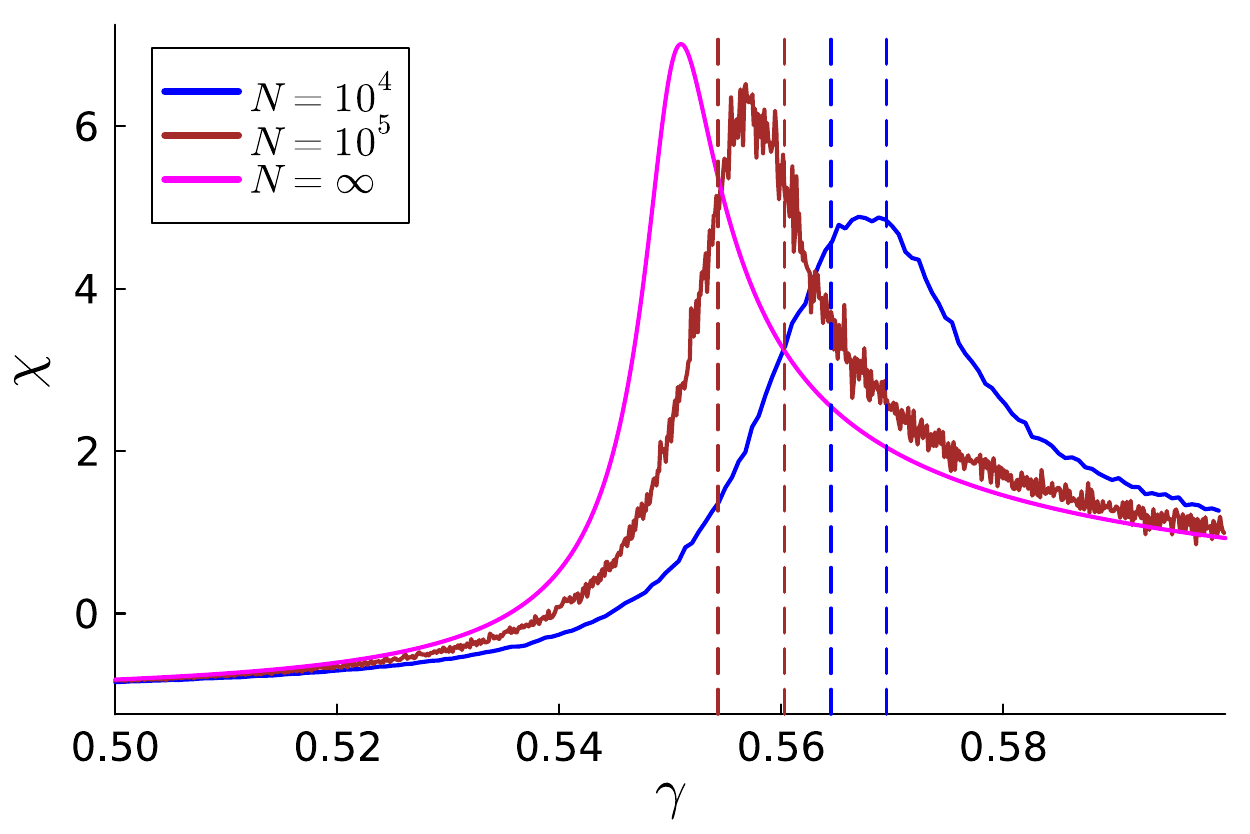}
\caption{\label{fig:suscep_N}Susceptibility curve $\chi(\gamma)$ for finite $N$, obtained from the slope of the {\it averaged} stress-strain curve. The averaging is over realisations of both the initial condition and the stochastic dynamics. We show also the analytical form obtained from the quasistatic solution $\Sigma_0(\gamma)$, which is approached for $N\to\infty$. Dashed lines indicate the narrow intervals around the peaks over which we record the ``peak'' avalanche statistics shown in Figs.~\ref{fig:ava_non_collapsed} and \ref{fig:ava_collapsed}.} 
\end{figure}

We finally show that the above analysis holds also during the transient. Choosing a value $R>R_c$ but with a significant peak susceptibility $\chip=\mathcal{O}(10)$ in the $N\rightarrow\infty$ limit, we run the full transient dynamics for finite $N$. Averaging over realisations~\footnote{We point out here an important difference to the RFIM: in the RFIM, once the initial sample is drawn, the dynamics is purely deterministic, while in the HL model the dynamics itself is by definition stochastic.}, one can find the average stress strain curve, $\langle\Sigma (\gamma)\rangle$ for finite $N$, where the average is over both realisations of the initial condition and of the loading dynamics. From here, one can obtain the susceptibility $\chi(\gamma)$ for finite $N$
\begin{equation}\label{eq:chi_con}
    \chi(\gamma)=-\frac{\partial \langle \Sigma \rangle}{\partial \gamma}
\end{equation}
which corresponds to the {\it connected susceptibility} considered in~\cite{ozawa_random_2018,ozawa_role_2020}. The $\chi(\gamma)$ curves for $N=10^{4}$ and $N=10^{5}$ are shown in Fig.~\ref{fig:suscep_N}. We recall that the rescaled yield rate is related to $\chi(\gamma)$ as $\tY(\gamma)\approx 1+\chi(\gamma)$, while, in the quasistatic limit, from the boundary condition on $Q_0$ one has $|Q_0''(1)|\sim \tY^{-1}$. Given the scaling of $S_c$ described above, one therefore expects the cutoff at any strain to scale as $S_c\sim \tY(\gamma)N^{1/2}\approx (1+\chi(\gamma))N^{1/2}$. 

Numerically, avalanches must of course be recorded over a finite strain interval: if this interval is made too narrow, obtaining good statistics is challenging. To check the scaling of the cutoff, we therefore use a finite interval around the peak of $\chi(\gamma)$, where $\tY$ is roughly constant. More concretely, we consider the strain-averaged $\tY$ within this finite interval. 

In Figs.~\ref{fig:ava_non_collapsed} and \ref{fig:ava_collapsed}, we include the avalanche distributions obtained around the peak for $N=10^{4}$ and $N=10^{5}$. We see clearly that indeed the same scaling holds as in steady state, and the ``peak'' distributions can be collapsed onto the same master curve. One expects that, also in the transient, the linear $Q_0(x)\approx x/\alpha$ behaviour at the boundary is always masked by a plateau, as this is simply a generic feature of a random walk with finite step size in the presence of a boundary~\cite{ferrero_properties_2021}.

Before continuing we compare the avalanche behaviour we have found to the well-known results for the RFIM~\cite{dahmen_hysteresis_1996,ozawa_random_2018}. Here the avalanche size distribution can be worked out exactly and depends solely on the parameter $\overline{n}=2 J P(x)$ (see Table~\ref{T1} for definition of $x$) quantifying the average number of secondary flips triggered by one spin flip:
\begin{equation}\label{eq:p_of_S_RFIM}
    P(S)=\frac{(\overline{n} S)^{S-1}}{S!}e^{-\overline{n} S}\approx \frac{1}{\overline{n}\sqrt{2\pi S^3}} e^{-(1-\overline{n})^2S/2}
\end{equation}
where the approximate form holds for $S\gg 1$ and near criticality $\overline{n}\to 1^{-}$. This can be written as $P(S)\sim S^{-\tau}e^{-S/S_c}$, with $\tau=3/2$ and a cutoff $S_c\propto 1/(1-\overline{n})^2 \propto \chi^2$~\footnote{More precisely, the susceptibility $\chi=\partial M/\partial H$ in the RFIM is related to $\overline{n}$ as $\chi
=2 P(x)/(1-\overline{n})$.}. Importantly, the cutoff $S_c$ does not scale with system size. It is interesting to note that the form (\ref{eq:p_of_S_RFIM}) precisely matches the distribution obtained in another well-known problem in complex systems, namely the distribution of sizes $S$ of small components in a Poisson random graph~\cite{newman_networks_2018}, where $\overline{n}$ is replaced by the connectivity $c$. Criticality is there approached as $c\to 1^{-}$, where the giant component percolates. The fact that this analogy to a problem of ``static'' networks with fixed connectivity exists also highlights the difference between avalanche criticality in the RFIM and in HL: in the HL model, the state of the system at the beginning of each avalanche is such that it would give an infinite avalanche size; the cutoff due to finite system size then only arises as the avalanche progresses.


The key feature we have stressed of the HL avalanche behaviour is that avalanches become ``scale-free'' both for $N \to \infty$ (growing system size) and for $\tY\to\infty$ or equivalently $\chi \to \infty$ (growing transient susceptibility). In the next subsection, we will turn to the infinite size limit to derive analytically the divergence of $\chi$, both at the critical disorder and in the brittle regime. We will see, however, that care must be exercised when applying the infinite size behaviour to predict e.g. the growth of precursor avalanches approaching the brittle yield point, as will be  discussed later in Sec.~\ref{subsec:precursors}.

\subsection{\label{subsec:TD_limit}The infinite size limit: analyticity implies Landau theory with non-trivial signatures}
We study now the quasistatic behaviour in the infinite size limit, in particular to elucidate the behaviour at the critical disorder $R=R_c$. To do this, we will consider the 0th order self-consistent dynamics introduced in Sec.~\ref{sec:background}.

By numerically solving the self-consistent equations (see App.~\ref{app:plastic} for details), we study firstly the divergence of $c_0(\gamma)=\alpha \lim_{\dgamma \to 0} \tY$ (see Eq.~\ref{eq:expa_yield_rate}), recalling that when $\tY \gg 1$ this quantity also corresponds to the susceptibility $\chi$ in the quastistatic limit (\ref{eq:stress_eom}). To study this divergence, it is useful to introduce an effective exponent as 
\begin{equation}\label{eq:eff_expo}
    b(\gamma)=\frac{\partial \log{c_0(\gamma)}}{\partial \log{ \Delta \gamma}}
\end{equation}
From the discussion in Sec.~\ref{sec:breakdown}, we already expect that, for $R<R_c$, $b \to -1/2$ when $\Delta \gamma \to 0$, where $\Delta \gamma=\gamma_c-\gamma$. On the other hand, if for $R>R_c$ we define $\Delta \gamma=\gamma_{p}-\gamma$ as the strain deviation from the peak of $\chi$, we should have that $b \to 0$ for $\Delta \gamma \to 0$ given that $\chi$ eventually flattens out at the peak. 

\begin{figure}
\includegraphics[width=0.45\textwidth]{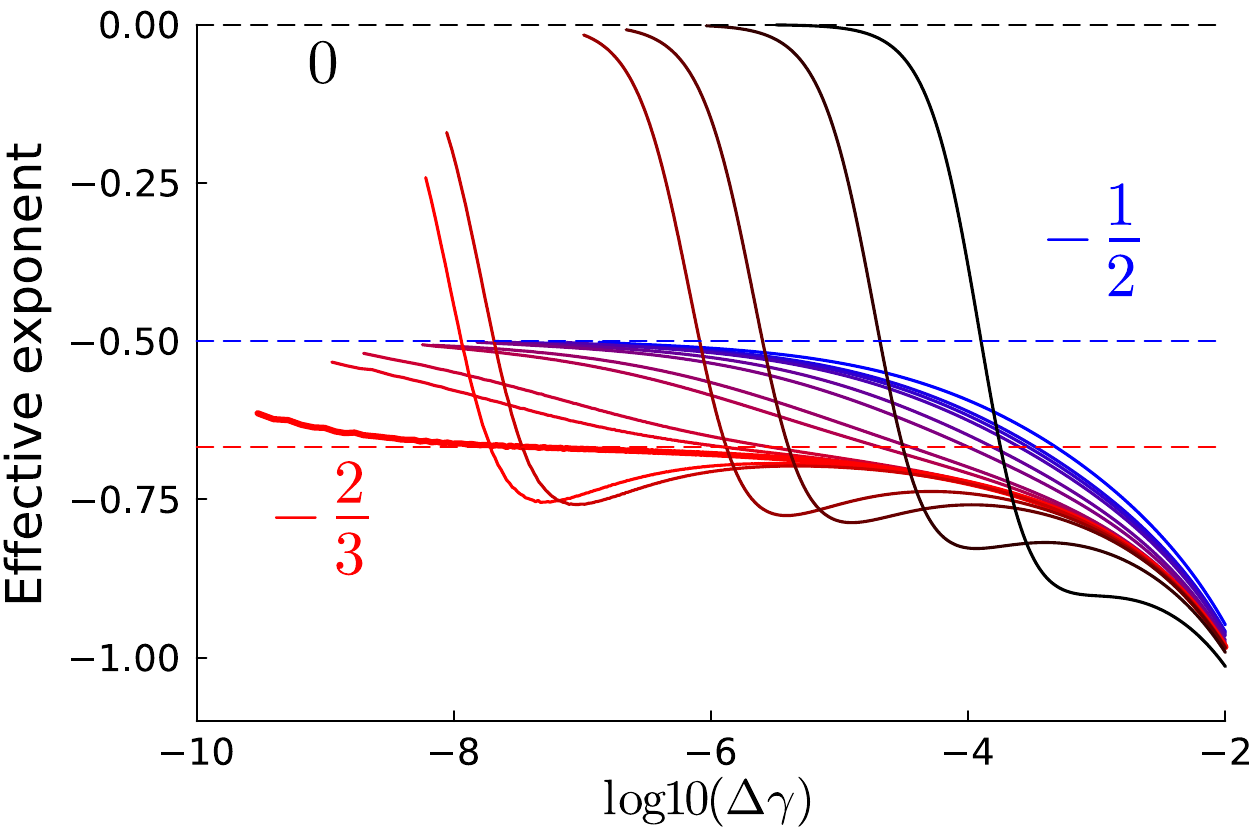}
\caption{\label{fig:effective_exponent}Effective exponent defined by (\ref{eq:eff_expo}), plotted against the strain deviation, $\Delta \gamma$, defined as $\gamma_c-\gamma$ for $R<R_c$ and $\gamma_p-\gamma$ for $R>R_c$. Blue to red (thick) curves show the approach $R\to R_c^{-}$, with the red (thick) curve corresponding to $R_c\approx 0.305215$. Red to black curves are for increasing $R$ values above $R_c$ (see App.~\ref{app:plastic} for precise $R$ values). We notice the limiting exponents $b=-1/2$, $b=-2/3$ and $b=0$, for $R<R_c$, $R=R_c$ and $R>R_c$ respectively.} 
\end{figure}

We check these behaviours numerically in Fig.~\ref{fig:effective_exponent}, where we show $b(\gamma)$ against $\Delta \gamma$ for $R$ values straddling the critical disorder $R_c$ for $\alpha=0.45$. Around $R=R_c$, we notice the appearance of a critical exponent $b=-2/3$; this value will be rationalised below.

\begin{figure}
\includegraphics[width=0.45\textwidth]{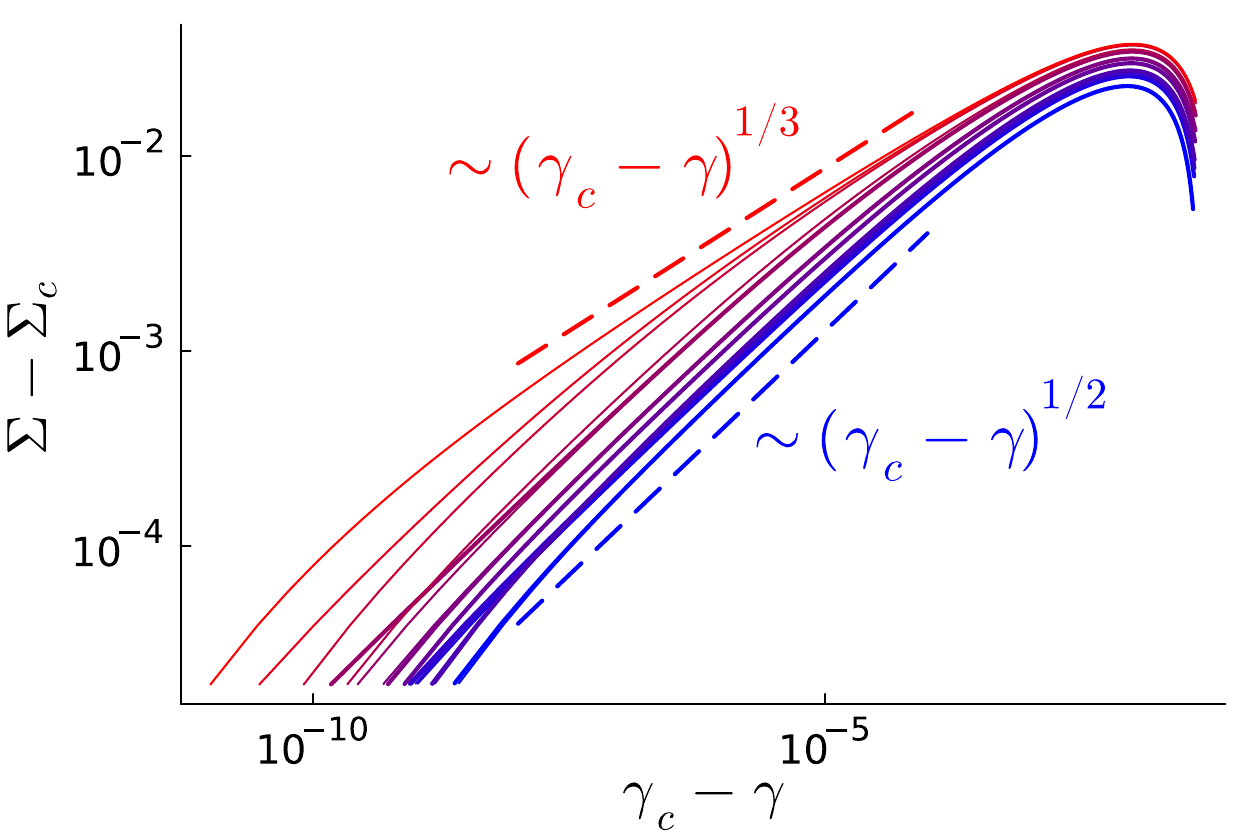}
\caption{\label{fig:stress_exponent}Deviation of the macroscopic stress from its value at the brittle yield point, $\Sigma-\Sigma_c$, shown against $\gamma_c-\gamma$ for the disorder values $R\to R_c^{-}$ considered in Fig.~\ref{fig:effective_exponent}. We see the singular behaviours $\Delta \Sigma \sim (\Delta \gamma)^{1/2}$ and $(\Delta \gamma)^{1/3}$ for $R<R_c$ and $R=R_c$ respectively.} 
\end{figure}

In Fig.~\ref{fig:stress_exponent} we show the corresponding behaviour of the macroscopic stress on approaching the critical disorder from below, $R \to R_c^{-}$. We recall that $\Sigma (\gamma)$ simply follows the integral of $\tY(\gamma)$, so that for $R<R_c$, $\Delta \Sigma \equiv \Sigma-\Sigma_c$ behaves as $\Delta \Sigma\sim (\gamma_c-\gamma)^{1/2}$, i.e.\ displays a square-root singularity on approaching the brittle yield point. At $R=R_c$, one expects from the exponent $b=-2/3$ the critical singularity $\Delta \Sigma \sim (\gamma_c-\gamma)^{1/3}$, which is indeed consistent with the data in Fig.~\ref{fig:stress_exponent}.

\begin{figure}
\includegraphics[width=0.45\textwidth]{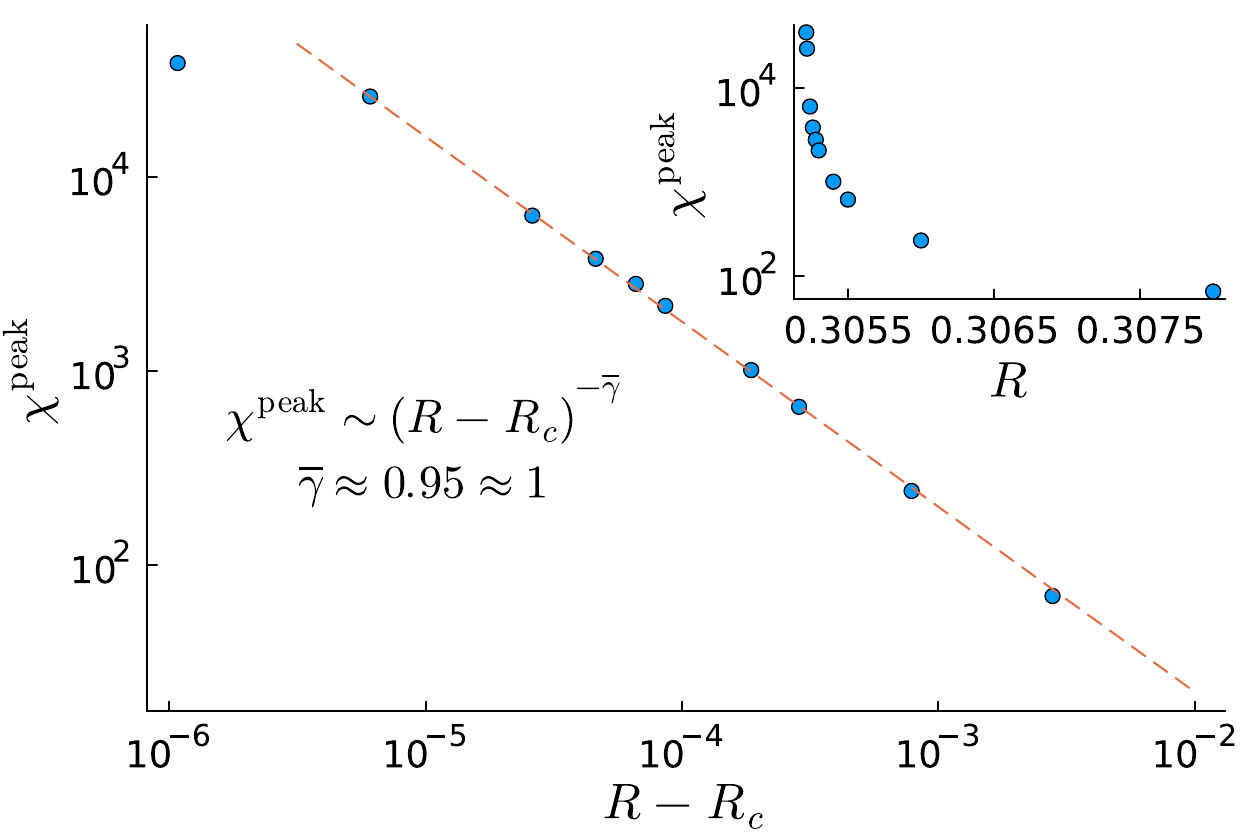}
\caption{\label{fig:over_gamma}Divergence of the peak susceptibility $\chip$, extracted from the numerical solution of the self-consistent dynamics, as $R \to R_c^{+}$, on a log-log scale. The fitted exponent is numerically consistent with $\overline{\gamma}=1$. For the fit we omit the final point, where the value of $\chip$ saturates due to the finite numerical discretisation. Inset: same data on a semi-log scale.} 
\end{figure}

Summarizing thus far, we have found that the exponents $\beta$, $\delta$, describing the sub-critical and critical response of the macroscopic stress correspond to those of simple Landau theory, $\beta=1/2$, $\delta=3$. One may further consider the exponent $\overline{\gamma}$ (we add an overline to distinguish it from the strain variable) describing the divergence of the peak susceptibility as $R \to R_c^{+}$, i.e. $\chip\sim (R-R_c)^{-\overline{\gamma}}$, where we recall that $\chip \sim |Q_0''(1,\gammap)|^{-1}$ is given by the inverse of the boundary curvature of the distribution at the peak. In the simplest scenario, one expects the curvature to vanish linearly in $R-R_c$, in much the same way as the coefficient of the quadratic term vanishes linearly in $T-T_c$ within a Landau picture. The value $\overline{\gamma}=1$ is indeed supported numerically in Fig.~\ref{fig:over_gamma}, where we show the value of $\chip$ extracted from the self-consistent dynamics as $R \to R_c^{+}$.

\begin{figure}
\centering
\includegraphics[width=0.35\textwidth]{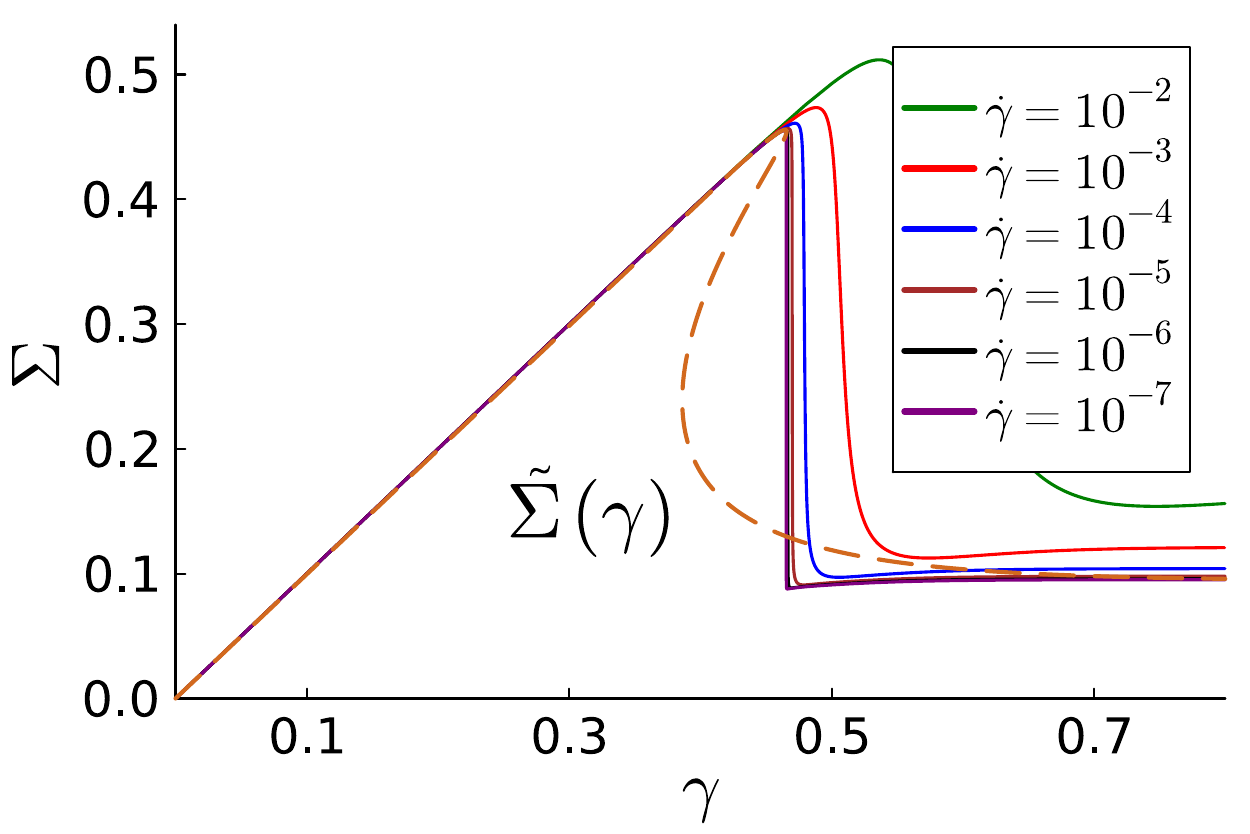}
\caption{\label{fig:sigma_tilde}The stress-strain curves for different shear rates $\dgamma$ shown already in Fig.~\ref{fig:brittle}, along with the alternative solution $\tilde{\Sigma}(\gamma)$ (dashed line) obtained by precluding failure~\cite{popovic_elastoplastic_2018}. This unphysical alternative dynamics leads to a transiently negative shear rate, whereby the system recoils to avoid the $\infty$-avalanche. Importantly, the lower branch of the true quasistatic stress-strain curve is {\it not} recovered after this recoil regime, as one might naively expect by analogy with typical equilibrium phase diagrams.} 
\end{figure}

To understand the Landau values of the exponents characterising the macroscopic observables, we exploit the smoothness of the alternative solution to the dynamics considered in~\cite{popovic_elastoplastic_2018}, the range of validity of which we will also clarify. To arrive at this smooth solution~\cite{popovic_elastoplastic_2018}, one must consider the self-consistent dynamics in terms of plastic strain $\epsilon_p$, where $\md \epsilon_p =\tY \md \gamma$. We give the full details of this transformation in App.~\ref{app:plastic}. One finds an equation of motion
\begin{equation}
    \partial_{\epsilon_p}Q_0=-v (\epsilon_p) \partial_{\sigma}Q_0+\alpha \partial_{\sigma}^2 Q_0 +\delta (\sigma) 
\end{equation}
where the ``velocity'' $v$ is defined as $\lim_{\dgamma \to 0} \tY^{-1}$. Now, as the brittle yield point $\epsilon_{p, c}=\epsilon_p (\gamma_c)$ is approached, one has that $v \to 0$. If one then allows for {\it negative} values of $v$, one may continue the self-consistent dynamics by artificially imposing the quasistatic boundary condition $Q_0(1,\epsilon_p)=0$ even beyond the brittle yield point. This yields a solution $\tilde{\Sigma}(\epsilon_p)$, which when transformed back to total strain, i.e.~$\tilde{\Sigma}(\gamma)=\tilde{\Sigma}\left (\epsilon_p(\gamma)\right)$, results in a transiently negative shear rate, as the system effectively recoils to avoid the $\infty$-avalanche. It is important to emphasise that this regime, and the solution $\tilde{\Sigma}(\gamma)$, are strictly speaking {\it unphysical}, although they may be motivated by introducing a control apparatus with negative stiffness~\cite{popovic_elastoplastic_2018}. Furthermore, one {\it does not} recover the lower branch of the true $\Sigma (\gamma)$ (i.e.\ after the stress drop) once $v$ becomes positive again (see Fig.~\ref{fig:sigma_tilde}), as one would expect e.g.\ from a Maxwell construction.

Nonetheless, assuming smoothness of the underlying $\tilde{\Sigma}(\gamma)$ is enough to give an account of the exponents found above. Considering the inverse function $\gamma(\tilde{\Sigma})$ (i.e.\ the inverse plot of Fig.~\ref{fig:sigma_tilde}), smoothness at the maximum implies 
a parabolic form around $\tilde{\Sigma}_c$ and hence 
the square-root singularity $\Delta \Sigma \sim (\gamma_c-\gamma)^{1/2}$. The curve $\gamma (\tilde{\Sigma})$ can be seen as displaying a local maximum and minimum for $R<R_c$, which merge at $R=R_c$ in a {\it saddle node bifurcation} as argued in~\cite{popovic_elastoplastic_2018}. At $R=R_c$, from symmetry one expects that $\gamma(\tilde{\Sigma})$ is locally cubic, leading to the behaviour $\Delta \Sigma \sim (\gamma_c-\gamma)^{1/3}$. 

\begin{figure}
\centering
\includegraphics[width=0.35\textwidth]{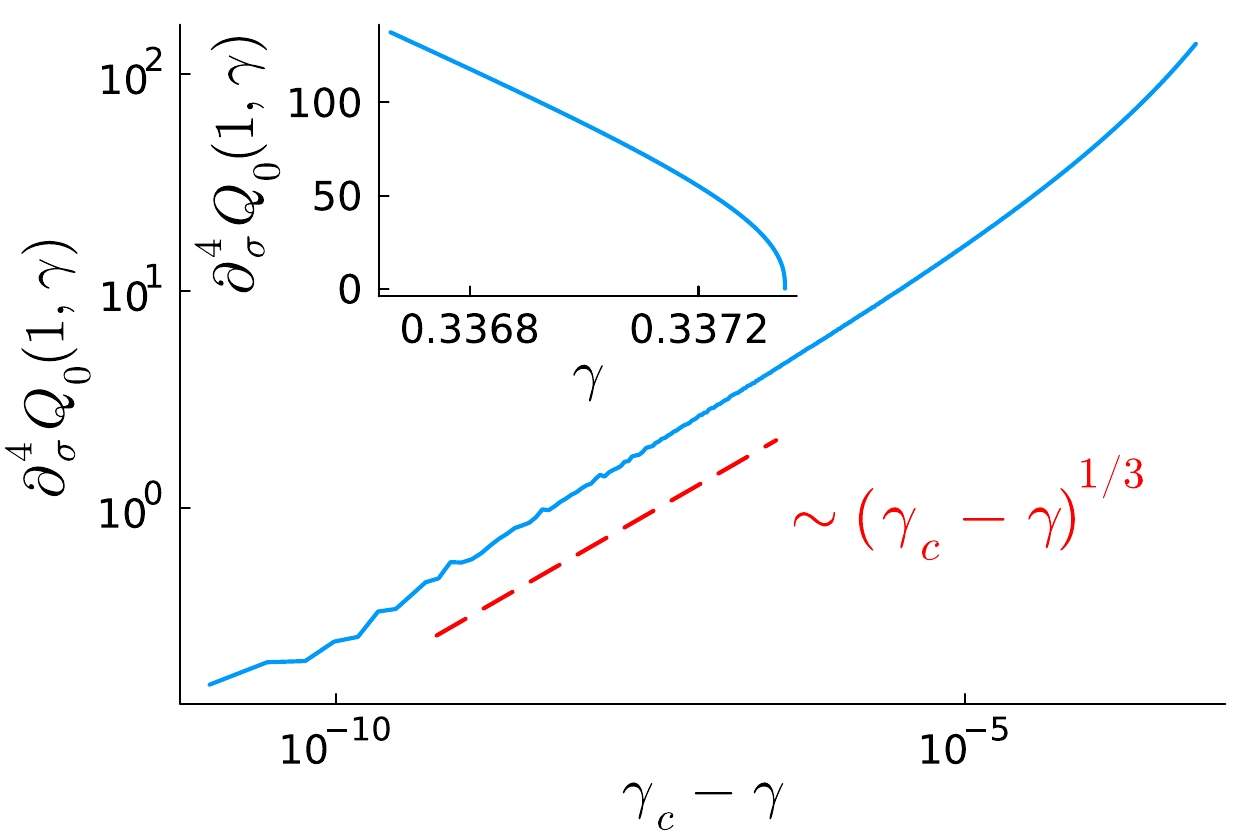}
\caption{\label{fig:fourth_order}Behaviour of the fourth derivative at the yield threshold of the quasistatic solution, $\partial_\sigma^4 Q_0(1,\gamma)$, for $R\approx R_c$, on approaching the singularity at $\gamma_c$, shown on a log-log scale. The data are consistent with the derivative vanishing as $(\gamma_c-\gamma)^{1/3}$, as predicted in the main text. Inset: same data on a linear scale.} 
\end{figure}

This simple assumption of smoothness has non-trivial implications for the threshold ($\sigma=1$) behaviour of the local stress distribution at $R=R_c$. From the above, we recall that the curvature approaching $\gamma_c$ for $R=R_c$ vanishes as $|Q_0''(1,\gamma)|\sim (\gamma_c-\gamma)^{2/3}$, as it corresponds to the inverse of $\tY(\gamma)$. Considering again Eq.~(\ref{eq:curvature_eom}) for the evolution of the curvature at the threshold
\begin{equation}
   \partial_{\gamma}Q_0''(1,\gamma)=-\partial_{\sigma}^3 Q_0(1,\gamma)-\frac{1}{\alpha Q_0''(1,\gamma)}\partial_{\sigma}^4 Q_0(1,\gamma)
\end{equation}
we see that if the third order derivative remains finite (this is checked numerically), a curvature vanishing as $|Q_0''(1,\gamma)|\sim (\gamma_c-\gamma)^{2/3}$ implies that the last term must go as $\sim(\gamma_c-\gamma)^{-1/3}$, and hence the fourth order derivative must vanish as $\partial_\sigma^4 Q_0(1,\gamma)\sim (\gamma_c-\gamma)^{1/3}$. We confirm this behaviour numerically in Fig.~\ref{fig:fourth_order}. At the level of the underlying local stress distribution, therefore, the distinguishing signature of the critical disorder $R_c$ is the vanishing of the fourth order derivative of the stress distribution at the threshold, on top of the vanishing curvature.

These properties of the local stress distribution we have derived in the infinite size limit have a clear physical interpretation in terms of avalanches, through the random walk mapping discussed in Sec.~\ref{subsec:avalanches}. We showed there that the sub-extensive avalanches of the HL model are controlled by $Q_0''(1,\gamma)$, the curvature at the threshold, which must therefore vanish as $\gamma \to \gamma_c^{-}$ both for $R<R_c$ and $R=R_c$. The fourth order derivative does not directly influence the avalanche scalings~\footnote{We note that it is a priori tempting to extend the expansion (\ref{eq:g_of_z}) to higher orders, and consider e.g. a situation where the curvature has vanished and the avalanche cutoff is set by a higher order derivative (e.g. the third). This is not relevant in practice, however, as in reality for finite systems the avalanche scaling is determined by a curvature which can only decay to a finite value vanishing with system size (corresponding to a saturating susceptibility). A ``truly'' vanishing curvature can only be obtained with $N=\infty$ for $\gamma \to \gamma_c^{-}$, but at $\gamma=\gamma_c^{+}$ the random walk mapping is then no longer valid.}, but plays an important role at $R=R_c$, as it changes the way the curvature vanishes. This in turn affects the scaling with system size of the largest sub-extensive avalanches, which as we will see in the following sections scale differently for $R=R_c$ and $R<R_c$. It is important to bear in mind that, once an $\mathcal{O}(N)$ avalanche is triggered (which implies $\mathcal{O}(1)$ changes to the distribution), the random walk mapping in terms of a purely local property (i.e. curvature at threshold) breaks down, and the ensuing $\infty-$avalanche must be analysed from the full dynamics of the model as we did in Sec.~\ref{sec:first_part}.

\begin{table*}[!ht]
    \begin{tabularx}{\linewidth}{sbbb}
   & Mean-field RFIM~\cite{dahmen_hysteresis_1996} \newline ($x$: ``threshold field'' \newline $x=-JM-H$) & ``Ferromagnetic'' EPM~\cite{ozawa_random_2018} \newline ($x$: local stability $\sigma_c-\sigma$ \,; \newline $g(x)$: stress drop distribution) & HL in quasistatic limit \newline ($x$: local stability $\sigma_c-\sigma$)\\
   \hline
   Quenched disorder & Local random field distribution \newline $\rho(f)\sim e^{-f^2/(2R^2)}$ & Initial local stress distribution \newline $P_{\gamma=0}(x) \sim e^{-x/R}-e^{-x/(1-R)}$ \newline with $1/2<R<1$ & Initial local stress distribution \newline $P_{\gamma=0}(\sigma)\sim e^{-\sigma^2/(2R^2)}$ \\
\hline
$R<R_c$ (Brittle)& $\beta=\frac{1}{2}$, \ $\Delta M \sim (H_c-H)^{\beta}$ \newline $\rho(x_c)=1/(2J)$ \newline $\rho'(x_c)\neq 0$ & $\beta=\frac{1}{2}$, \ $\Delta \Sigma \sim (\gamma_c-\gamma)^{\beta}$ \newline $P(x=0,\gamma_c)=P_c$ \newline $P'(x=0,\gamma_c)+P_c g'(0)\neq 0$ & $\beta=\frac{1}{2}$, \ $\Delta \Sigma \sim (\gamma_c-\gamma)^{\beta}$ \newline $Q_0''(x=0,\gamma_c)=0$ \newline $\partial_x^4 Q_0 (x=0,\gamma_c)\neq 0$ \\ \hline
$R=R_c$ (Critical) & $\delta=3$, \ $\Delta M \sim (H_c-H)^{1/\delta}$ \newline $\rho(x_c)=1/(2J)$ \newline $\rho'(x_c)= 0$ & $\delta=3$, \ $\Delta \Sigma \sim (\gamma_c-\gamma)^{1/\delta}$ \newline $P(x=0,\gamma_c)=P_c$ \newline $P'(x=0,\gamma_c)+P_c g'(0)=0$ & $\delta=3$, \ $\Delta \Sigma \sim (\gamma_c-\gamma)^{1/\delta}$ \newline $Q_0''(x=0,\gamma_c)=0$ \newline $\partial_x^4 Q_0 (x=0,\gamma_c)=0$\\ \hline
Avalanches & $P(S)\sim S^{-\tau} e^{-S/S_c}$ \newline $\tau=3/2$  \newline $d_f=0$ \quad ($S_c \sim \chi^2 L^{d_f}$) & $P(S)\sim S^{-\tau} e^{-S/S_c}$ \newline $\tau=3/2$  \newline $d_f=0$ \quad ($S_c \sim \chi^2 L^{d_f}$) & $P(S)\sim S^{-\tau} e^{-(S/S_c)^2+aS/S_c}$ \newline $\tau \approx 1 $ \newline  $d_f=d/2$ \quad  ($S_c \sim \chi L^{d_f}$) \\ \hline
Pseudogap exponent \newline $P(x)\sim x^{\theta}$ & 
\quad \quad \quad \quad \quad ---
& $\theta=0$ & $\theta=1$ \\
   \hline
\end{tabularx}
\caption{\label{T1}Detailed comparison between the mean-field RFIM~\cite{dahmen_hysteresis_1996}, the ``ferromagnetic'' mean-field elastoplastic model with strictly positive interactions~\cite{ozawa_random_2018} and the quasistatic limit of the HL model. In the first two cases, the discontinuity for $R<R_c$ is triggered by the underlying distribution reaching a {\it critical value}, while criticality at $R=R_c$ is associated with a vanishing first derivative. Regarding the ferromagnetic EPM, we show the results for a general stress drop distribution $g(x)$~\cite{ozawa_random_2018} (which can be seen to act as an additional source of disorder~\cite{rossi_emergence_2022}); for  a uniform stress drop (as in HL) where $g'(0)=0$ one would recover the simple RFIM-like condition for criticality $P'(x=0,\gamma_c)=0$. In the quasistatic HL model, on the other hand, although $\beta=1/2$ and $\delta=3$ remain the same, brittle yielding is signalled by a vanishing curvature~\cite{popovic_elastoplastic_2018} and the critical point at $R=R_c$ by a vanishing fourth derivative. This difference in the underlying distributions is at the root of the radically different avalanche behaviour. Note that for the RFIM, we have not defined local ``stabilities'' as in the EPMs, in order to write the conditions for ``brittle'' and ``critical'' behaviour in terms of the fixed $\rho(f)$, as given originally in~\cite{dahmen_hysteresis_1996}: in a field sweep starting from $H=-\infty$ as considered here, a spin $i$ flips when the ``threshold field'' $x$ decreases (due to the combined effect of the external field and the ferromagnetic interaction) below $f_i$.}
\end{table*}


The similarities and differences between the mean-field RFIM, the elastoplastic model with strictly positive interactions, and the HL model in the quasistatic limit, are summarised in Table~\ref{T1}. Although the values of the exponents $\beta$, $\delta$ describing the macroscopic obeservables turn out to be the same, the corresponding behaviour of the underlying distribution is radically different in the HL model. In the presence of sign-varying interactions, which act as a mechanical noise, this behaviour is determined by higher order derivatives of the local stress distribution at the yield threshold, rather than the distribution itself reaching a critical value. This arguably goes to the crux of what is ``missing'' in a standard RFIM-like approach to the problem, where a fractal scaling of avalanches arises solely around the random critical point, whereas the HL model in the quasistatic limit displays system-spanning (but sub-extensive) events at any value of the strain and initial disorder (see further discussion in Sec.~\ref{subsec:criticality}). This situation is in fact reminiscent of a different class of mean-field spin models, which have quenched disorder in the magnetic couplings rather than the local field, i.e.\ spin glasses. In the Sherrington-Kirkpatrick model it was indeed shown~\cite{pazmandi_self-organized_1999} that the system displays marginal stability (referred to there as self-organised criticality) throughout the hysteresis loop, without the need for tuning the initial disorder or the applied external field.


\subsection{\label{subsec:precursors}The brittle yield point as a mean-field spinodal on top of marginality and in the presence of disorder}

In the previous section, we have shown that, in the infinite size limit, within the brittle regime $R<R_c$ the susceptibility displays an inverse square root divergence. We will study now how this determines the fate of individual finite-size samples in the brittle regime, addressing also the important question of whether and how failure can be {\it{forecast}} by observing the avalanche behaviour. Despite some superficial similarities between HL and the RFIM~\footnote{We emphasise that, as explained in Sec.~\ref{sec:mf_theory}, we are comparing here to the behaviour of the {\it mean-field} RFIM. For the RFIM (with short-range ferromagnetic couplings) in any finite dimension, it has been shown~\cite{nandi_spinodals_2016} that spinodal criticality is not of mean-field type due to the depinning and subsequent expansion of rare droplets.}, we will show that the way the $N=\infty$ spinodal limit controls the behaviour of finite-size samples approaching the brittle yield point is qualitatively different in two important ways. Firstly, sign-varying interactions induce a {\it stabilising} effect in finite-size systems by strongly biasing the average yield strain towards higher values. Secondly, criticality in the spinodal region will arise {\it on top} of the marginality of the amorphous solid, reflected in the scaling of avalanches with system size. We note that this second feature is in agreement with an alternative mean-field approach starting from the infinite-dimensional solution for hard spheres~\cite{parisi_theory_2020}, where the yield strain is found to be a spinodal point within a marginally stable phase displaying full replica-symmetry breaking~\cite{rainone_following_2015,urbani_shear_2017} (the avalanche behaviour in this phase, however, is not yet understood~\cite{rainone_following_2016}). 

\begin{figure}
\centering
\begin{subfigure}{0.45\textwidth}
   \includegraphics[width=1\linewidth]{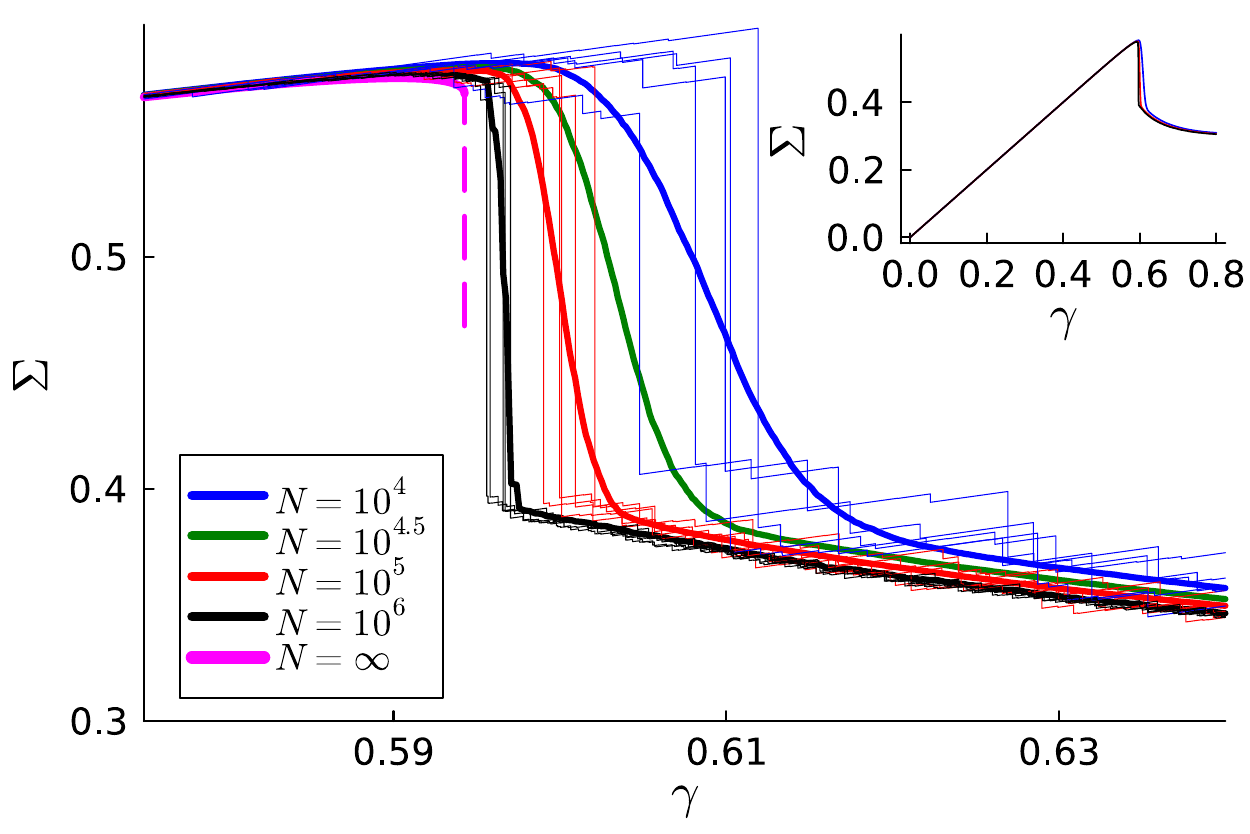}
   \label{fig:noises_1} 
\end{subfigure}

\begin{subfigure}{0.45\textwidth}
   \includegraphics[width=1\linewidth]{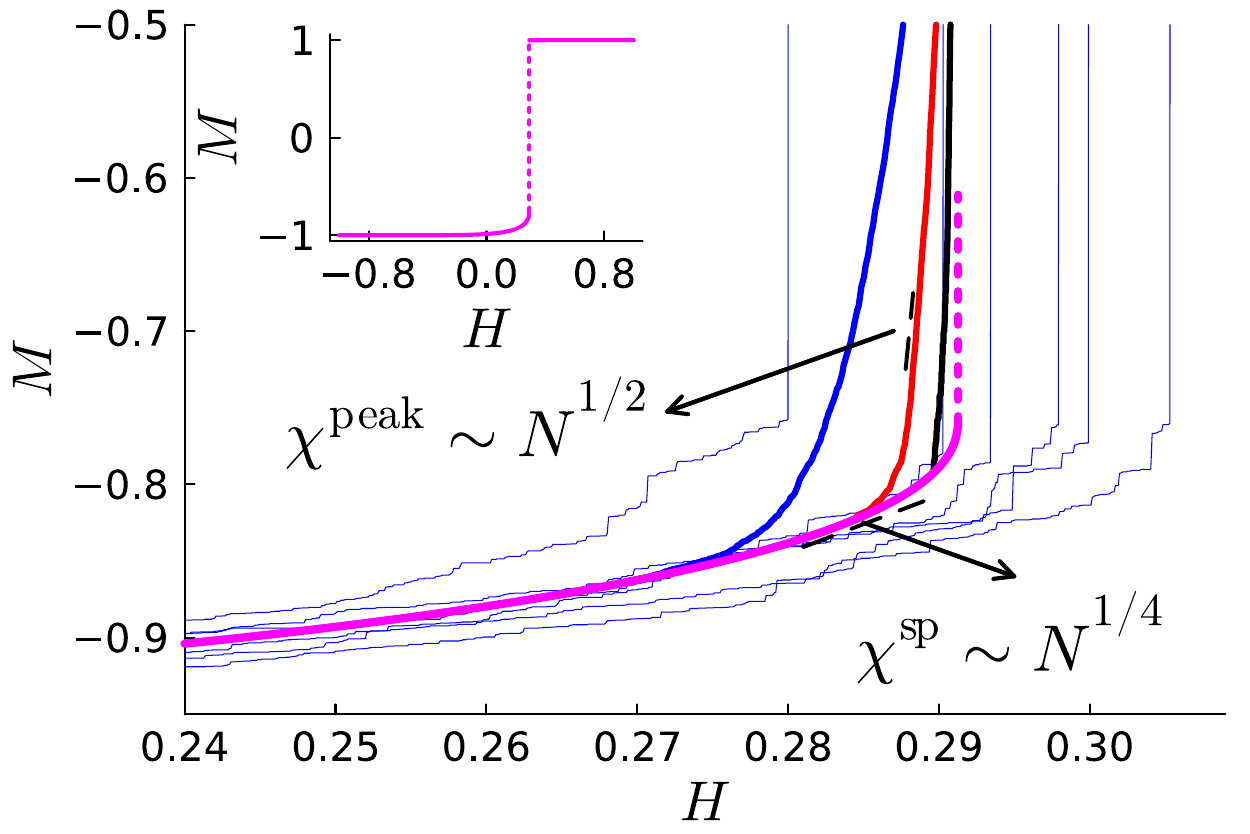}
   \label{fig:noises_2}
\end{subfigure}

\caption{\label{fig:plot_brittle}Top: brittle yielding of finite-sized samples ($R=0.16<R_c(\alpha=0.2)\approx 0.18$): thin lines show macroscopic stress versus strain for single realisations, whereas thick lines indicate ensemble averages. Also shown is the $N=\infty$ solution, which ends in a square root singularity. Inset: corresponding stress-strain curves displayed over the complete strain window up to steady state. Bottom: behaviour of finite-size samples (same color scheme) on approaching the $N=\infty$ (magenta) coercive field $H_c$ in the discontinuous (``brittle'') regime of the RFIM ($R=0.5 R_c$, where $R_c=\sqrt{2/\pi}$ for Gaussian random fields). Thin lines show individual realisations for $N=10^{4}$. Note that the discrete gaps $\Delta H$ between avalanches are expected to scale as $\mathcal{O}(N^{-1})$ in the RFIM~\cite{nampoothiri_gaps_2017}. Arrows indicate (roughly) the two distinct regimes where the divergences $\chi^{\rm peak}$ and $\chi^{\rm sp}$ arise (see text). Inset: magnetisation curve over the whole external field window.}

\end{figure}

At this point, we leave aside an analytical approach and turn again to direct simulations of the HL model in the quasistatic limit. In Fig.~\ref{fig:plot_brittle} (top) we show the macroscopic stress versus strain curves, both for single realisations (thin lines) and ensemble averages (thick lines), for three different system sizes in the brittle regime of the HL model. As in \ref{subsec:avalanches}, we consider again the coupling value $\alpha=0.2$, but now use an initial disorder $R=0.16<R_c$ ($R_c \approx 0.18$ for $\alpha=0.2$). In contrast to the case $R=0.2$ studied in \ref{subsec:avalanches}, which corresponded to a mild overshoot in the $N=\infty$ stress-strain curve (see Fig.~\ref{fig:suscep_N}), with $R=0.16$ there is instead a genuine divergence of the susceptibility with an ensuing $\infty-$avalanche. The smooth ensemble averages for finite $N$ in Fig.~\ref{fig:plot_brittle} (top) are reminiscent of the finite shear rate curves studied in Sec.~\ref{sec:first_part}, with the associated slope (susceptibility) becoming steeper and steeper as $N\to \infty$. We also plot the solution in the infinite size limit (magenta line), which as we know ends in a square-root singularity with an associated spinodal divergence.

To compare the role of the $N=\infty$ spinodal limit in the HL model and the RFIM we show in Fig.~\ref{fig:plot_brittle} (bottom) the behaviour of finite-size systems on approaching the critical coercive field $H_c$ in the RFIM. We recall that, for $N=\infty$ (magenta line), we have the same divergence as in HL, $\chi\sim (H_c-H)^{-1/2}$. We see, however, that the way this determines the fate of finite-size systems near the discontinuity is qualitatively different to Fig.~\ref{fig:plot_brittle} (top), due to the different nature of the interactions. In the RFIM, {\it any} precursor activity preceding the discontinuity has a strictly destabilising effect on the bulk of the distribution, thereby tending to trigger more precursors and bringing the system as a whole closer to the $\infty-$avalanche. This leads naturally to the behaviour observed in Fig.~\ref{fig:plot_brittle} (bottom), where (see e.g.\ the left-most single realisation) the $\infty-$avalanche can be triggered well before the infinite size coercive field $H_c(N=\infty)$. In the HL model (Fig.\ref{fig:plot_brittle} (top)), on the other hand, precursor events may equally well have a stabilising effect, and the possibility of triggering the $\infty-$avalanche before $\gamma_c \equiv \gamma_Y (N=\infty)$ is strongly suppressed, to the extent that, within our numerical samples, we do not observe this for any single realisation. Here, the yield strain $\gamma_Y$ is defined as the strain value at which the $\infty-$avalanche is triggered for a given sample.

This qualitative difference is best visualised by considering the sample-to-sample fluctuations of the yield strain (resp.\ coercive field), which in turn determine the peak divergence of the susceptibility in the brittle (resp.\ discontinuous) regime.  It is straightforward to show~\cite{ozawa_role_2020} that, in both the HL model and the RFIM, for $R<R_c$ the behaviour of $\chi_N$, that is the slope of the average order parameter curve, is well approximated around its peak by
\begin{eqnarray}\label{eq:approx_suscep}
    \chi_{N} (\gamma)\approx \tilde{\chi}_{N} (\gamma)&=&\langle \Delta \Sigma^{\mathrm{max}}\rangle
    P_N(\gamma)\\
    \chi_{N} (H)\approx \tilde{\chi}_{N} (H) &=&\langle \Delta M^{\mathrm{max}}\rangle
    P_N(H) \label{eq:estimate_RFIM}
\end{eqnarray}
where $P_N(\cdot)$ denotes the finite-size distributions of $\gamma_Y$ and its RFIM analogue $H_c$, and the prefactor is the average value of the order parameter jump in the $\infty$-avalanche for a given system size. 

\begin{figure}
\centering

\begin{subfigure}[b]{0.4\textwidth}
   \includegraphics[width=1\linewidth]{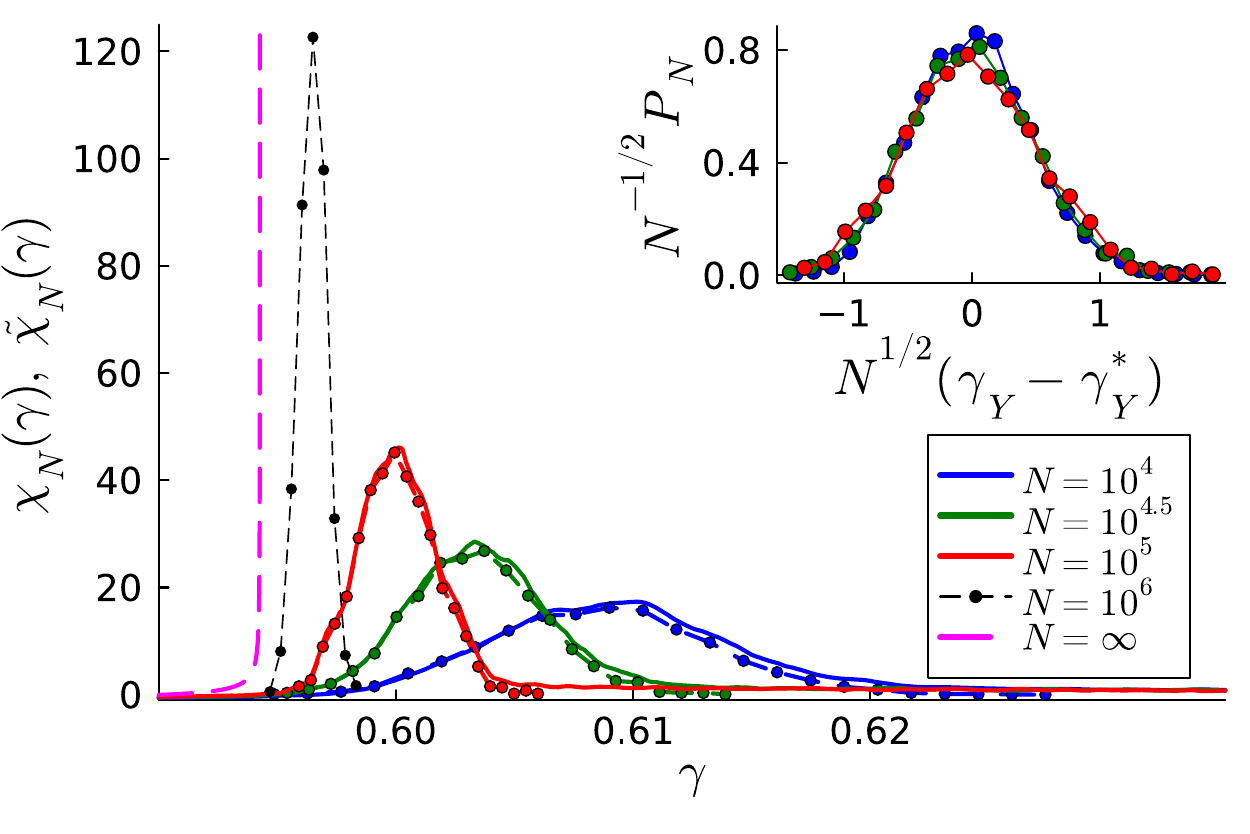}
\end{subfigure}

\begin{subfigure}[b]{0.4\textwidth}
   \includegraphics[width=1\linewidth]{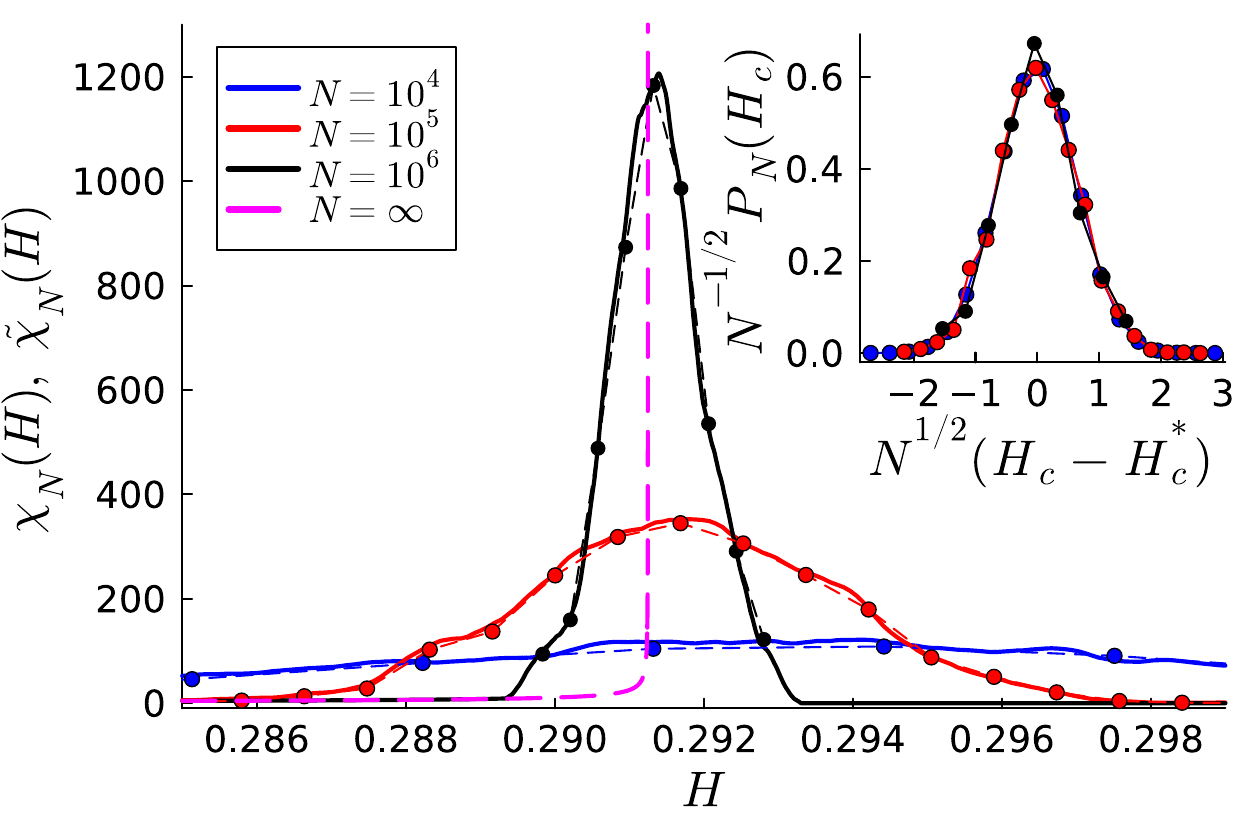}
\end{subfigure}

\caption{\label{fig:susceps_brittle} Finite-$N$ susceptibility curves obtained from the slope of the average order parameter curve (thick lines), along with their estimate (dashed lines with markers) from the finite-size yield strain/coercive field distributions, (\ref{eq:approx_suscep}) and (\ref{eq:estimate_RFIM}), for the HL model (top) and the RFIM (bottom), corresponding to the runs in Fig.~\ref{fig:plot_brittle}. Magenta dashed lines show the behaviour of $\chi$ in the infinite size limit. Insets: histograms of yield strains (top) and coercive fields (bottom), which collapse well according to (\ref{eq:P_N}) with $\fs=1/2$.
}
\end{figure}

In Fig.~\ref{fig:susceps_brittle} we show the susceptibility curves corresponding to Figs.~\ref{fig:plot_brittle}, along with the corresponding estimates (\ref{eq:approx_suscep}) and (\ref{eq:estimate_RFIM}) from the histograms of $\gamma_Y$ and $H_c$, which show good agreement (for the largest system size $N=10^{6}$ in HL, we show only $\tilde{\chi}_N$ due to computational limitations~\footnote{To obtain the $N=10^{6}$ data shown in Fig.~\ref{fig:susceps_brittle} and \ref{fig:conditioned}, we run the dynamics only up to the $\infty-$avalanche for a given sample (which we identify by introducing a threshold on the number of current unstable elements): within the $\infty-$avalanche the computational time scales as $\mathcal{O}(N^2)$, and becomes prohibitive to resolve.}). We see that, in the RFIM case, the finite-size coercive fields are distributed roughly symmetrically around $H_c(N=\infty)$. In contrast, in the HL case we see a strong bias towards larger yield strains for smaller system sizes. In particular, the possibility of yielding before the spinodal limit $\gamma_Y(N=\infty)$ is strongly suppressed. In other words, the different nature of the interactions has a {\it stabilising} effect on the metastable state. This is precisely the trend observed in particle systems: we refer e.g.\ to Fig.~6a in \cite{ozawa_role_2020}. 

\begin{figure}
\centering
\includegraphics[width=0.38\textwidth]{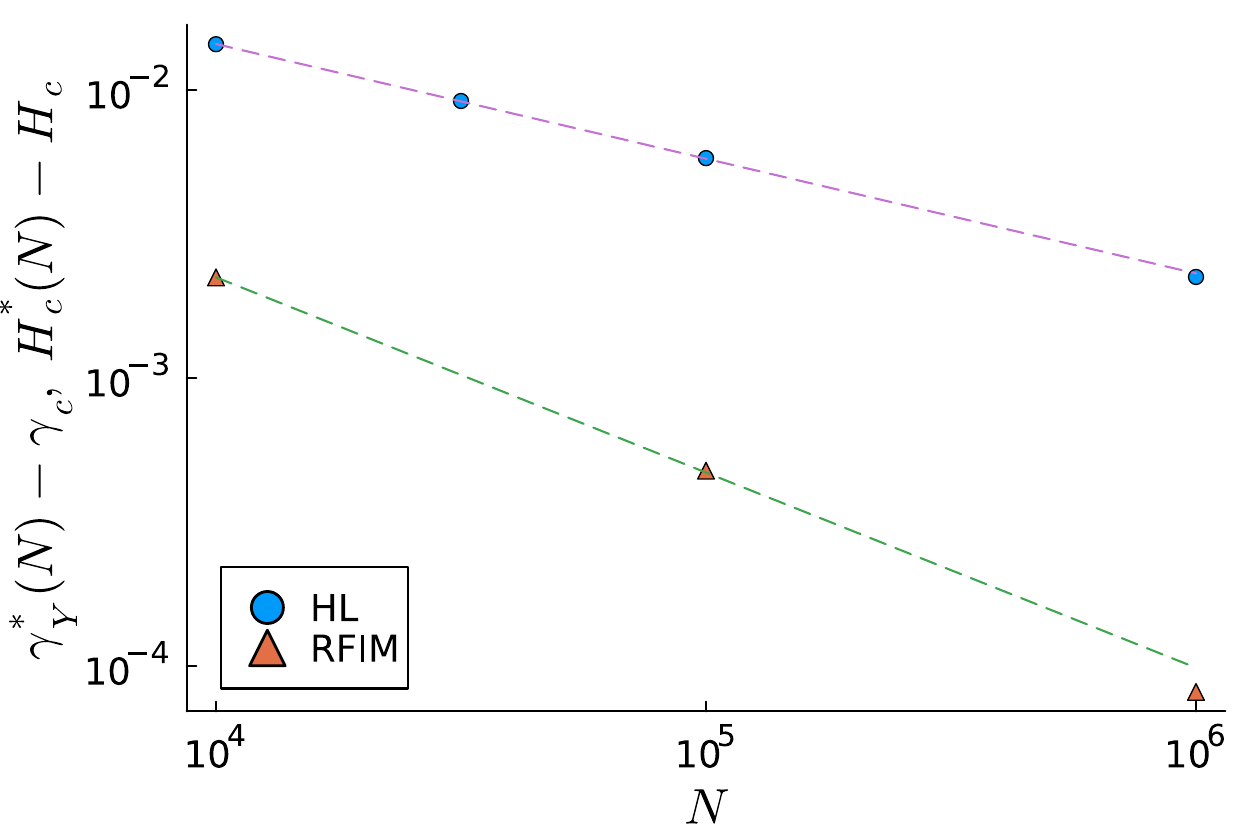}
\caption{\label{fig:moving}Behaviour with system size of $\gamma_Y^{*}(N)-\gamma_c$ in the HL model, compared to $H_c^{*}(N)-H_c$ in the RFIM. Dashed lines indicate power-law fits with exponents $-0.4$ and $-0.67$ respectively.} 
\end{figure}

\begin{figure}
\centering

\begin{subfigure}[b]{0.4\textwidth}
   \includegraphics[width=1\linewidth]{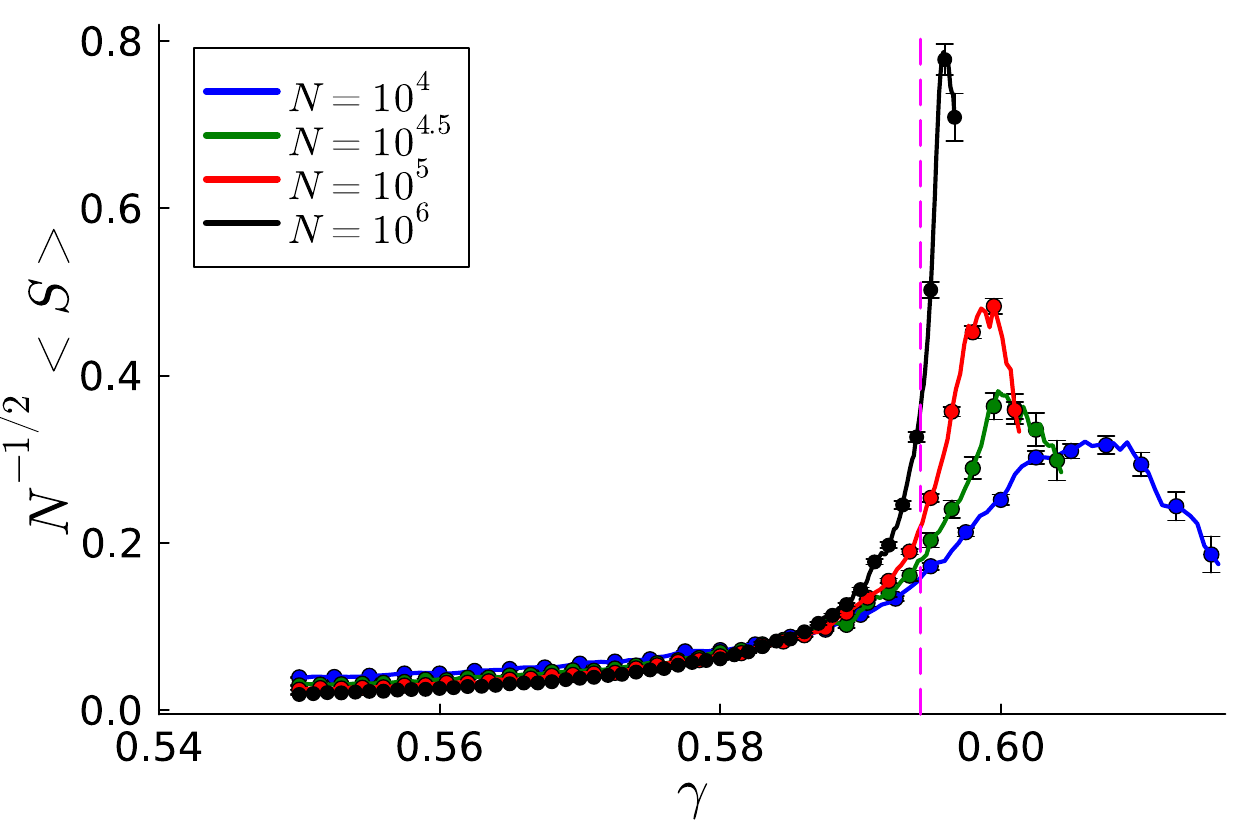}

\end{subfigure}

\begin{subfigure}[b]{0.4\textwidth}
   \includegraphics[width=1\linewidth]{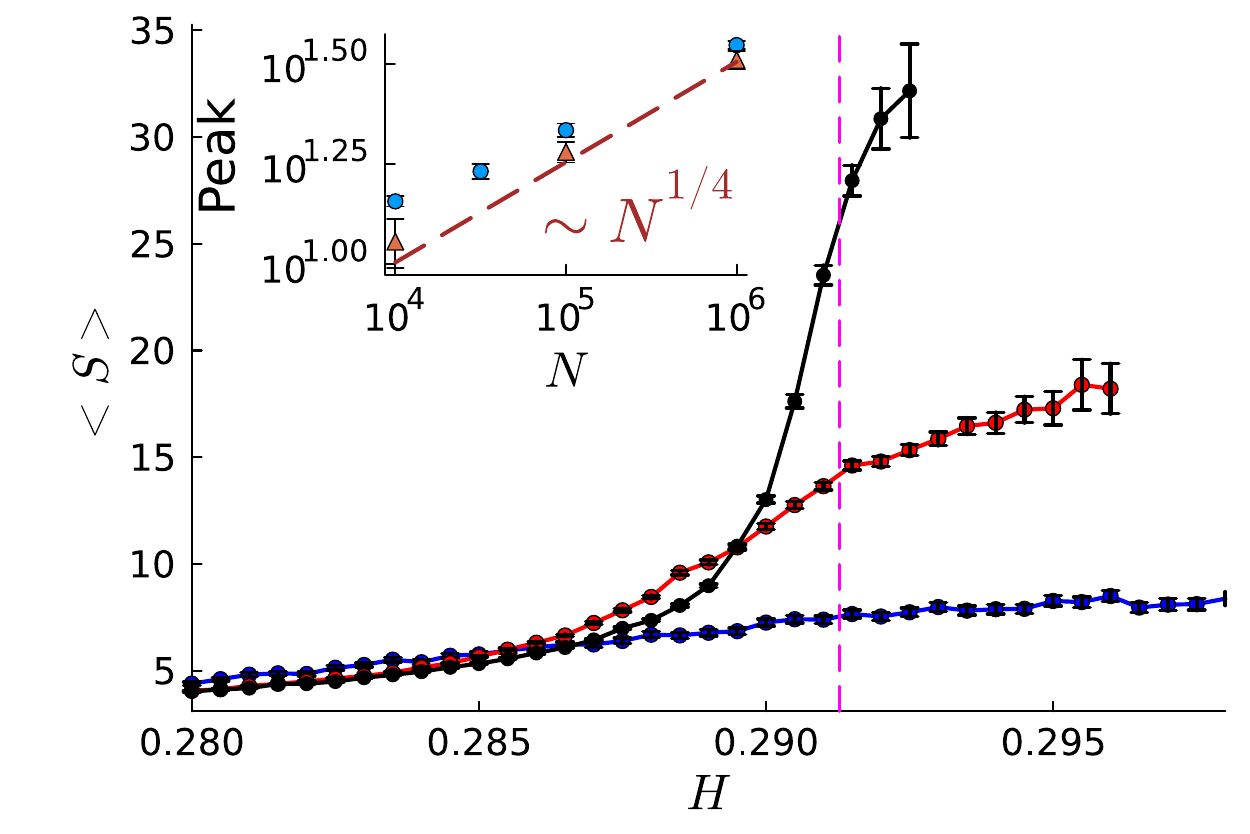}
   
\end{subfigure}

\caption{\label{fig:conditioned}Development of {\it spinodal criticality}. For both HL (top) and the RFIM (bottom), we obtain the average avalanche size within each strain/field window {\it conditioned} on the avalanche occurring before the $\infty-$avalanche for the corresponding run. Dashed magenta lines indicate as reference the values $\gamma_Y (N=\infty)$ and $H_c(N=\infty)$. Inset: peak values of $\langle S \rangle$ in the RFIM (orange triangles) and $N^{-1/2} \langle S \rangle$ in HL (blue circles, scaled by a prefactor for easier comparison), showing agreement with $\chi^{\rm sp}\sim N^{1/4}$.}
\end{figure}

We may quantify the bias in the yield strain for smaller systems by considering the behaviour of $\gamma_Y^{*}(N)-\gamma_c$, where $\gamma_Y^{*}(N)$ is the average yield strain for a given system size, and compare with the equivalent quantity in the RFIM (Fig.~\ref{fig:moving}). We see that the HL data follows a significantly slower power-law with system size in its approach to $\gamma_c$, marking a qualitative difference to the RFIM. Interestingly, the fact that the fitted power-law decay $\sim N^{-0.4}$ is slower than the $N^{-1/2}$ decay of the yield strain fluctuations (see below) suggests that, even for arbitrary large system sizes, the yield strain for any finite system always lies strictly beyond the spinodal limit $\gamma_Y(N=\infty)$.

Turning now to the divergence with system size of the susceptibility, this can be understood from the relation to the yield strain statistics (\ref{eq:approx_suscep}). Indeed, if one assumes that the distribution of yield strains follows a finite-size scaling of the form~\cite{ozawa_role_2020,procaccia_mechanical_2017-1}
\begin{equation}\label{eq:P_N}
        P_N(\gamma_Y)\sim N^{\fs}\mathcal{P} \left((\gamma_Y-\gamma_Y^{*})N^{\fs} \right)
\end{equation}
then the peak slope of the averaged stress-strain will also diverge as $\chip \sim N^{\fs}$. In the HL model, as in the RFIM, the histograms collapse well following the form (\ref{eq:P_N}), with $\fs=1/2$~\footnote{For HL, we leave out the $N=10^{6}$ data due to insufficient statistics to accurately measure the peak of the histogram: see footnote on computational limitations above.}. This is the value found for the yield strain distribution in $3d$ particle simulations, consistent with the divergence $\chip \sim N^{1/2}$ measured there~\cite{ozawa_role_2020}. Interestingly, in $d=2$ the authors of Ref.~\cite{ozawa_role_2020} measured stronger fluctuations with $\fs =1/3$.

Finally, we turn now to the second qualitative difference to the RFIM given above, namely the arisal of spinodal criticality on top of marginality, as well as the important practical question of the role of precursors approaching the brittle yield point. Although the finite-size effects are qualitatively different in the two models, it is clear that in both cases, by continuity, as $N\to \infty$ the average magnetisation or stress curves should follow more and more of the $N=\infty$ solution. We see this clearly in Fig.~\ref{fig:plot_brittle} (bottom) for the RFIM: indeed, given that deviations for finite $N$ are due to the triggering of $\infty-$avalanches, and these are triggered at values of the external field that display $\mathcal{O}(N^{-1/2})$ fluctuations around $H_c$, one expects the finite $N$ curves to follow the limiting solution up to a field value $H_{\rm sp}(N)$ obeying $H_c-H_{\rm sp} (N)\sim \mathcal{O}(N^{-1/2})$. At this point, the susceptibility is of order $\chi^{\rm sp}\sim N^{1/4}$. Note that this refers to the divergence of the susceptibility within the metastable state before the triggering of an $\infty-$avalanche, and is hence distinct from the divergence in the peak region $\chip \sim N^{1/2}$ discussed above. The distinct regimes where the divergences arise are (roughly) indicated by the arrows in Fig.~\ref{fig:plot_brittle} (bottom). It is interesting to compare the scaling of $\chi^{\rm sp}$ in the RFIM to the scaling at the mean-field Ising spinodal in an external field for a clean system (i.e.\ without disorder), where criticality is destroyed instead by {\it thermal} fluctuations out of the metastable state. These fluctuations scale~\cite{colonna-romano_anomalous_2014} as $\mathcal{O}(N^{-2/3})$, implying a divergence with system size $\chi^{\mathrm{sp}}\sim N^{1/3}$ for the magnetic susceptibility at the spinodal.

In the HL model, the finite $N$ average stress-strain curves appear to be following an analogous trend to the finite $\dgamma$ curves approaching the $\dgamma=0^{+}$ spinodal (see Fig.~\ref{fig:spinodal}, top): smaller $N$, like larger $\dgamma$, makes the average curve display a larger and smoother overshoot. This is precisely 
what is 
found in particle simulations (see e.g.\ Fig.~1b in \cite{ozawa_role_2020}). For the finite $\dgamma$ case (see discussion of Fig.~\ref{fig:spinodal} in Sec.~\ref{sec:breakdown}), we saw that the finite shear rate curves ``come off'' the quasistatic limiting solution at $\gamma_c-\gamma_{\rm sp}\sim \mathcal{O}(\dgamma^{1/2})$. Given that $\chi\sim (\gamma_c-\gamma)^{-1/2}$, then at this point $\chi^{\rm sp}=\mathcal{O}(\dgamma^{-1/4})$. In the finite $N$ case, 
it is plausible to conjecture that 
the relevant scale is given again by 
$\mathcal{O}(N^{-1/2})$ 
so that $\gamma_c-\gamma_{\rm sp}\sim \mathcal{O}(N^{-1/2})$ and $\chi^{\rm sp}\sim N^{1/4}$ as in the RFIM.

To test these predictions, we may study the growth of the average avalanche size as one approaches the brittle yield point. That is, we restrict the avalanche statistics only to events that occur {\it before} the $\infty-$avalanche for a given run. This is similar in spirit to what is done to study the divergence of the susceptibility at the thermal spinodal in a mean-field system without disorder~\cite{colonna-romano_anomalous_2014}, where one must restrict the Monte Carlo sampling to configurations within the metastable state. In Fig.~\ref{fig:conditioned} we show the average avalanche size obtained in this manner for both the RFIM and the HL model. In both cases one observes the development of spinodal criticality as the system size is increased. Importantly, in the HL case this occurs {\it on top} of the already system-spanning avalanche behaviour; to study the effect of the spinodal, one must consider the rescaled avalanche size $N^{-1/2}\langle S \rangle $. In both cases, one expects the peak values of $\langle S \rangle$ (RFIM) or $N^{-1/2}\langle S \rangle$ (HL) to scale as the peak $\chi^{\rm sp}\sim N^{1/4}$ of the susceptibility at the spinodal. We indeed find (inset) good agreement in the case of the RFIM; for HL, the data are also consistent with this asymptotic scaling though with larger 
pre-asymptotic corrections.

We conclude with some final remarks regarding the role of precursors, and the possibility of predicting failure. We see from Fig.~\ref{fig:conditioned} (top) that the (slow) spinodal divergence $\chi^{\rm sp}\sim N^{1/4}$ can be understood as an increasing ratio between the size of the avalanches in the spinodal regime, and the baseline size provided by the final section of the quasi-elastic branch (take, e.g. $\gamma=0.56$ in Fig.~\ref{fig:conditioned}). For a system size $N=10^{6}$, this ratio is already around eight. One must however bear in mind that the baseline size of avalanches is very small. Expressing the average stress release per avalanche $\langle S \rangle /N$ (in adimensional units) as a fraction of the total elastic stress carried by the material (roughly $\Sigma \approx0.5$), for the case $N=10^{6}$ this fraction goes from roughly $0.02 \%$ to $0.16 \%$. Therefore, despite the factor of $8$, one must note that for well-annealed samples even the largest precursors in the spinodal regime may remain undetectable from a practical point of view.



\subsection{\label{subsec:criticality}The critical disorder $R_c$: criticality on top of marginality}

In Sec.~\ref{subsec:avalanches}, where we characterised the avalanche behaviour of the HL model, we noted the key feature that avalanches sizes diverge both for $N\to \infty$ (growing system size), an effect we referred to as marginality, and for $\chi \to \infty$, i.e. growing transient susceptibility, which occurs for any sample in the brittle regime $R<R_c$. We will consider now samples prepared at the critical disorder $R=R_c$, which we recall separates ductile from brittle yielding. As in the RFIM, the critical disorder will play the role of a random critical point, but importantly, as was the case for the spinodal studied in Sec.~\ref{subsec:precursors}, criticality will have to emerge {\it on top} of the underlying marginality of the amorphous solid.

\begin{figure}
\centering
\includegraphics[width=0.45\textwidth]{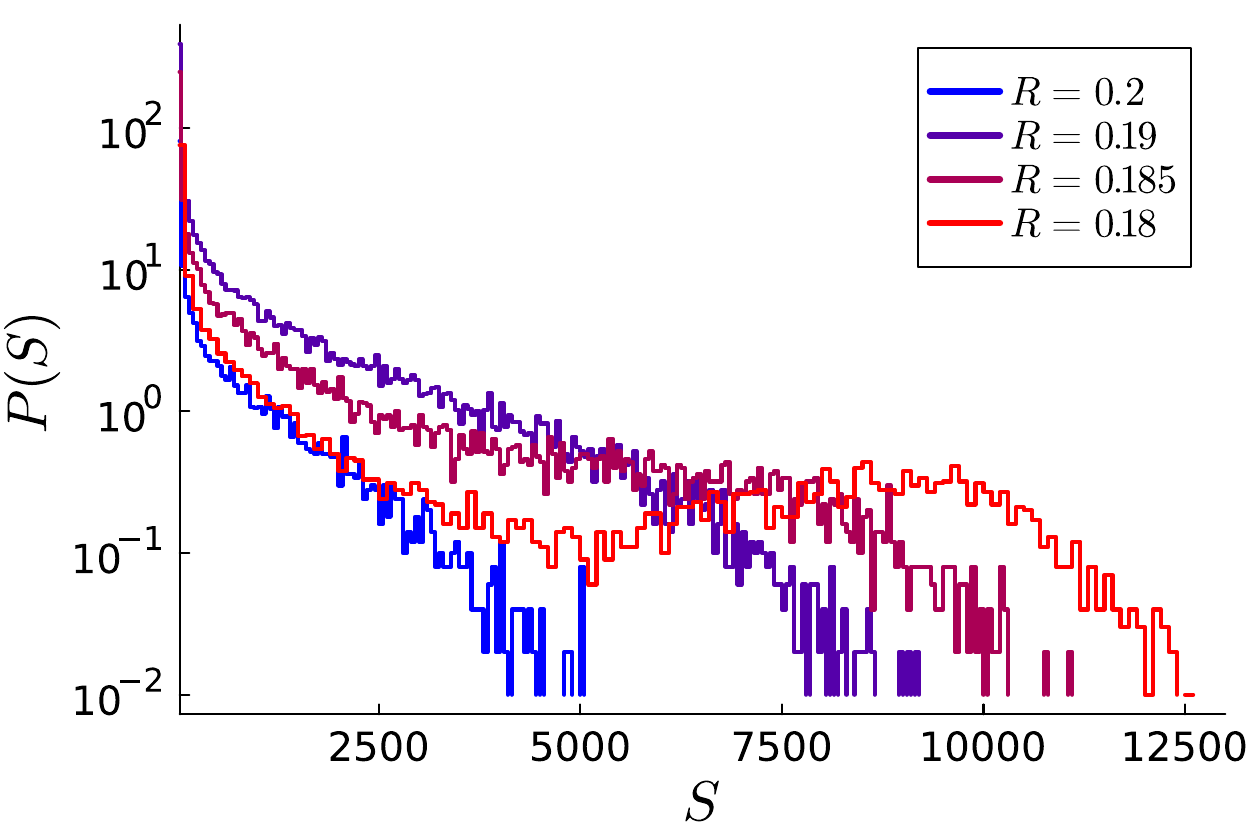}
\caption{\label{fig:approach_Rc}Avalanche distributions extracted around the peak of the susceptibility, for $R$ values approaching the critical disorder from above ($N=10^{5}$, $\alpha=0.2$). The distribution for $R=0.2$ corresponds to the ``Peak $N=10^{5}$'' distribution analysed in Fig.~\ref{fig:ava_non_collapsed}, now shown on a semi-log scale. For the lowest value $R=0.18$, the distribution becomes bimodal, with a ``bump'' indicating the appearance of $\infty-$avalanches.} 
\end{figure}

In Fig.~\ref{fig:approach_Rc}, we firstly show the distribution of avalanches extracted around the peak of the susceptibility for different $R$ values, starting with the case $R=0.2>R_c$ (displaying a mild overshoot) studied in Sec.~\ref{subsec:avalanches}. Decreasing $R$ from this value towards $R_c$, we see that the distribution indeed becomes extremely broad, until eventually it develops a ``bump'' and becomes bimodal, reflecting the appearance of macroscopic $\infty-$avalanches. 

In the RFIM case, this behaviour is typically analysed by considering the avalanche distribution integrated over an entire hysteresis loop for a given disorder. Here, we are considering an integration solely over the ``peak region'', given that the number of avalanches in the steady state of plastic flow diverges. For a full analysis in the $3$d RFIM, we refer e.g. to~\cite{perez-reche_finite-size_2003} (compare e.g.\ Fig.~1 there). It is important to note that, unlike for $R<R_c$, where it is simple to remove the single $\infty-$avalanche and study only the precursor behaviour, extracting the critical avalanches at $R=R_c$ is much more subtle. Indeed, in the RFIM, the number of macroscopic avalanches at the critical disorder is expected to diverge in the infinite size limit~\cite{perez-reche_finite-size_2003}, so that singling them out from the full statistics is a non-trivial task. We will return to the question of critical scaling below. 

\begin{figure}
\centering
\includegraphics[width=0.45\textwidth]{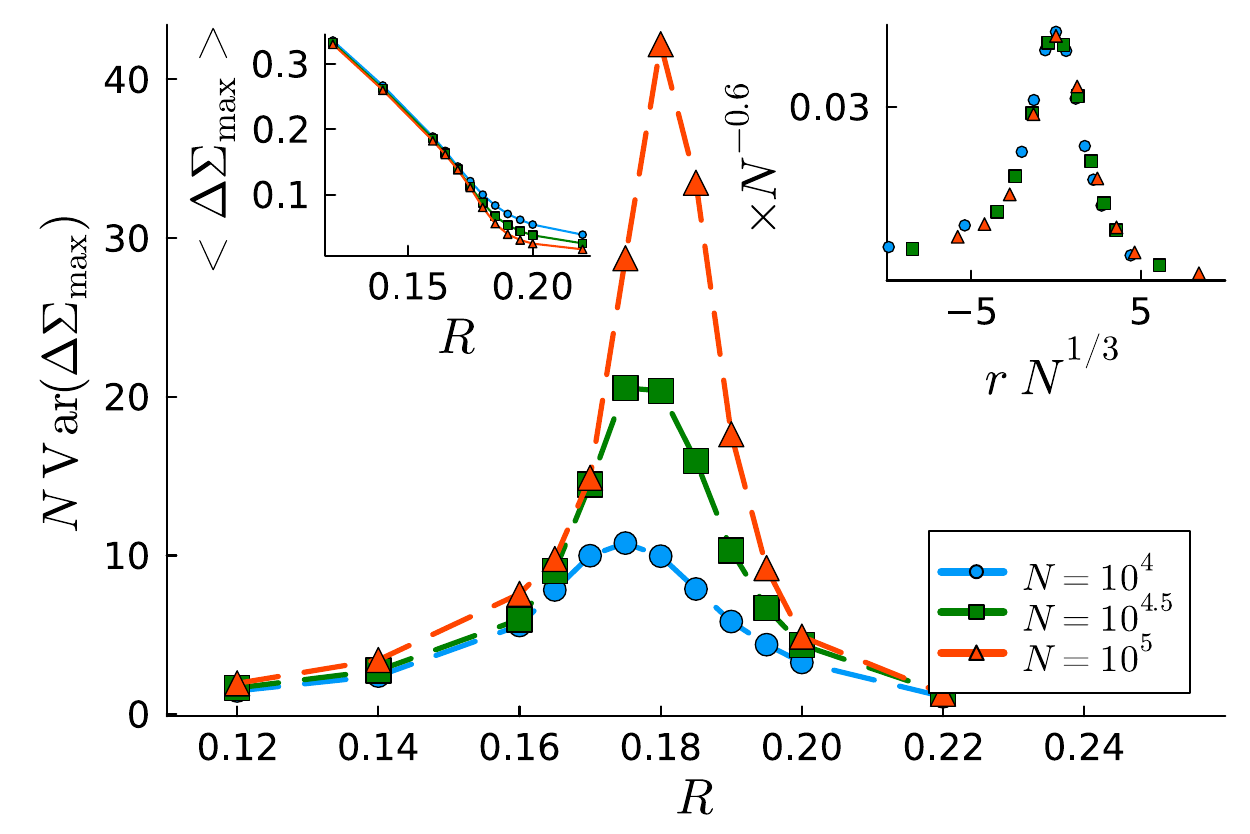}
\caption{\label{fig:plot_maxs}Fluctuations of the maximum stress drop (\ref{eq:var_sigma}) for a range of $R$ values spanning the critical disorder. One observes a clear peak around a critical value $R_c(N)$, with the height of the peak growing with system size (compare e.g.\ Fig.~3B in \cite{ozawa_random_2018}). Inset (left): $\langle \Delta \Sigma^{\rm max}\rangle $ for the same $R$ values; in the ductile regime, $\langle \Delta \Sigma^{\rm max}\rangle \sim N^{-1/2}$, while in the brittle regime $\langle \Delta \Sigma^{\rm max}\rangle=\mathcal{O}(1)$. Inset (right): finite-size scaling collapse of the data in the main figure (see text).} 
\end{figure}

\begin{figure}
\centering
\includegraphics[width=0.45\textwidth]{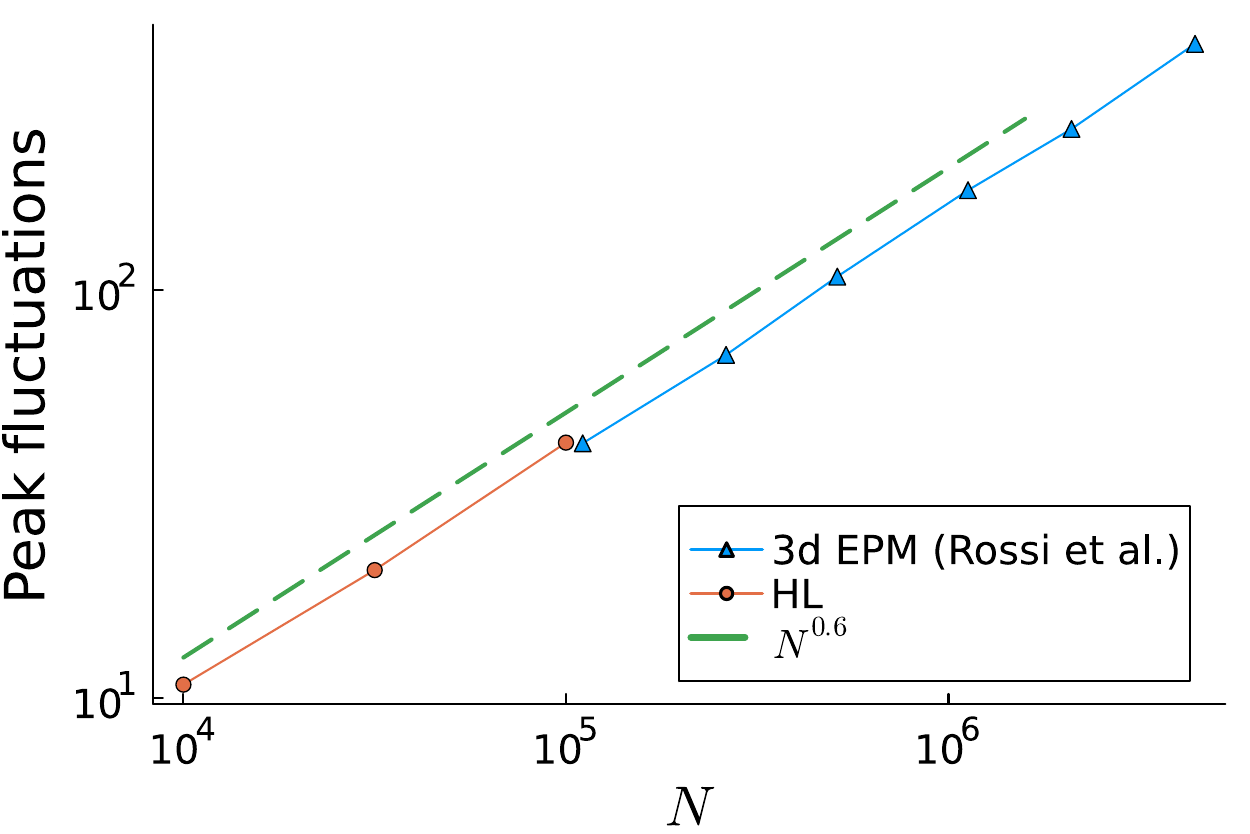}
\caption{\label{fig:peak_HL_EPM}Peak values (across $R$) of the maximum stress drop fluctuations, extracted from Fig.~\ref{fig:plot_maxs} for the HL model and from Fig.~S1d of\ Ref.~\cite{rossi_finite-disorder_2022} for a $d=3$ elastoplastic model, for different system sizes. We find in both cases a divergence consistent with $\sim N^{0.6}$. } 
\end{figure}

\begin{figure}
\centering
\includegraphics[width=0.45\textwidth]{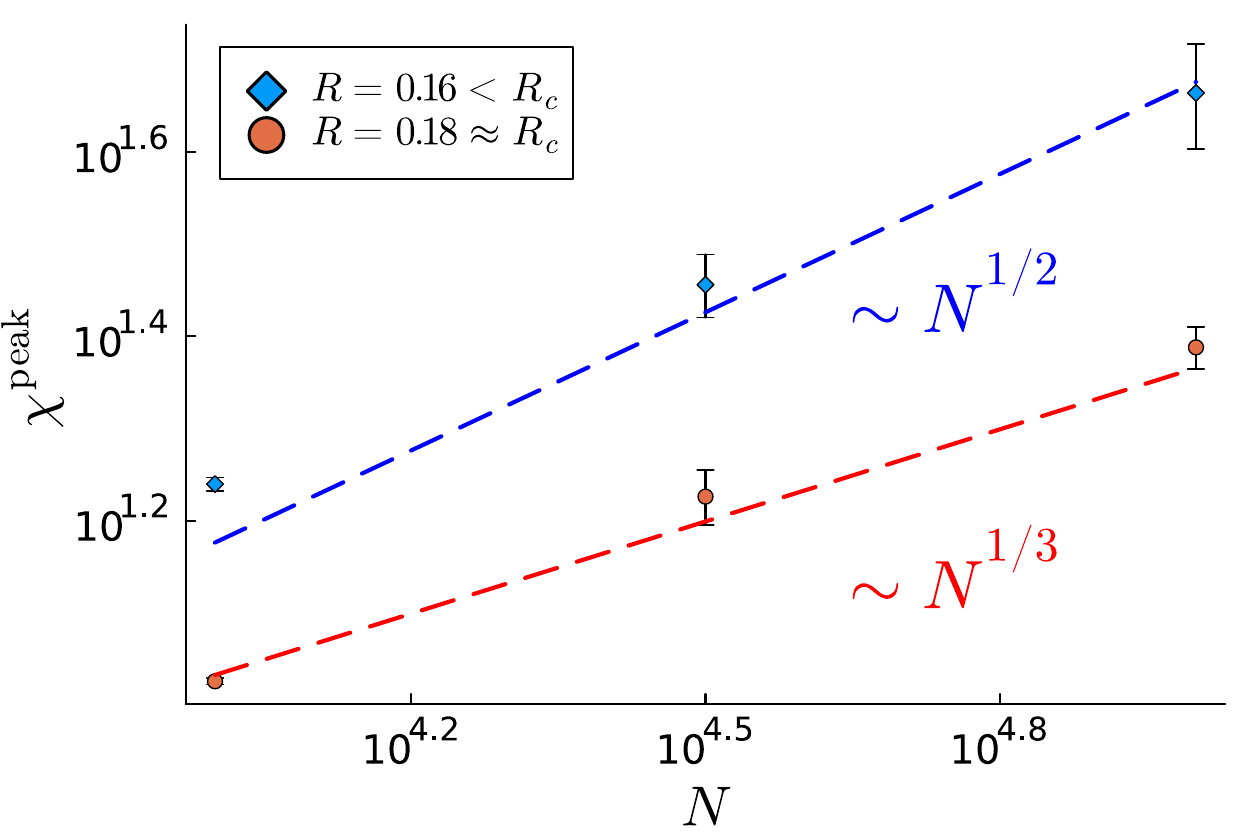}
\caption{\label{fig:chips}Peak value of the susceptibility (\ref{eq:chi_con}), extracted for $R=0.16<R_c$ (see Fig.~\ref{fig:susceps_brittle}) and for $R=0.18\approx R_c$. Within the limited system size range, the divergences are consistent with those of the mean-field RFIM.
} 
\end{figure}

Before doing so, we study the critical fluctuations of the largest stress drop, following previous studies~\cite{ozawa_random_2018,rossi_finite-disorder_2022} aiming at a characterisation of the random critical point in amorphous solids. For a single realisation, one may define the quantity $\Delta \Sigma ^{\rm max}$ as the largest recorded macroscopic stress drop throughout the entire loading protocol. One can then study the variance of sample-to-sample fluctuations
\begin{equation}\label{eq:var_sigma}
    N \ {\rm{Var}}[\Delta \Sigma ^{\rm max}]=N \left(\langle {\Delta \Sigma ^{\rm max}}^2\rangle -{\langle \Delta \Sigma ^{\rm max} \rangle}^2\right)
\end{equation}
where brackets indicate averages over realisations of both the initial condition and the dynamics. The quantity $\Delta \Sigma^{\rm max}$ has been considered in particle simulations~\cite{ozawa_random_2018} as an ``order parameter'' distinguishing ductile and brittle yielding, while its fluctuations (\ref{eq:var_sigma}) have been used to identify a critical point separating the two regimes. As in these simulations one expects also in the HL model that (\ref{eq:var_sigma}) will be $\mathcal{O}(1)$ well within the ductile or brittle regimes, where $\Delta \Sigma^{\rm max}=S^{\rm max}/N$ should show Gaussian fluctuations around a mean value; on the other hand, the variance should display a strong peak around the critical disorder $R_c$, where the distribution of $\Delta \Sigma^{\rm max}$ becomes abnormally broad. We confirm this numerically in Fig.~\ref{fig:plot_maxs}. Although due to computational limitations our system size range is restricted to a decade, we attempt a finite-size scaling collapse in the standard manner, defining the rescaled disorder $r=(R-R_c(N))/R$ as in~\cite{rossi_finite-disorder_2022}. Within the modest system size range, we find a good collapse with $\sim N^{-1/3}$ and $\sim N^{0.6}$ scalings on the horizontal and vertical axes respectively. The first of these is consistent with the scaling of the peak susceptibility at criticality $\chip (N)\sim N^{1/3}$ discussed further below. Indeed, we saw in Sec.~\ref{subsec:TD_limit} that $\chi \sim (R-R_c)^{-\overline{\gamma}}$ with the Landau value $\overline{\gamma}=1$, hence $r\sim N^{-1/3}$ if the susceptibility for a finite size saturates at $\chip (N)\sim N^{1/3}$. Note that the natural size variable for the scaling collapse here is the number of elements $N$ (or system volume), as is the case for fully-connected models~\cite{botet_large-size_1983,colonna-romano_anomalous_2014} or for spatial models above their upper critical dimension~\cite{fytas_finite-size_2023}. 

We may compare the above scalings to those obtained at the critical disorder of a $3$d lattice elastoplastic model~\cite{rossi_finite-disorder_2022}. Although in the main article the authors consider an alternative order parameter that scales differently with system size (related to the fraction of sites that have yielded within a plane), we may turn to the fluctuations of the largest stress drop shown in the SM (Fig.~S1d). From there, we extract the peak (across $r$) value of the fluctuations for different linear system sizes $L$ and plot them as a function of $N=L^{3}$ along with the corresponding mean-field data in Fig.~\ref{fig:peak_HL_EPM}. Although the mean-field (HL) data are limited to a narrower range, we find interestingly that the form of the divergence is very close to the one found in $d=3$. Further analytical work on the sample-to-sample fluctuations is needed to understand the exponent $\approx 0.6$ and the apparent similarity to the behaviour in $3$ dimensions~\footnote{As for the exponent controlling the location of the finite-size critical disorder, where we predict a scaling $R_c(N=\infty)-R_c(N)\sim N^{-1/3}$, in \cite{rossi_far--equilibrium_2023} a scaling $L^{-1/\nu}$ with $\nu\approx 1.6$ is reported, which in $d=3$ translates into a somewhat slower decay with exponent $\approx -0.2$.}. It would also be interesting to study the equivalent finite-size scaling (i.e.\ Fig.~\ref{fig:plot_maxs}) in the mean-field RFIM (where one would instead consider the fluctuations of the largest magnetisation jump; we are not aware of this having been analysed in the literature), in order to understand whether (and in what way) the peak scaling is affected by the underlying marginality. Due to the rapidly increasing computational time ($\mathcal{O}(N^2)$), we cannot at this point pin down 
the value of the exponent in HL numerically by accessing larger system sizes in Fig.~\ref{fig:plot_maxs}.

From the analysis of the avalanche behaviour performed in Sec.~\ref{subsec:avalanches}, we already expect that, as elsewhere, the critical avalanche distribution as well as the critical avalanche scaling (i.e.\ the largest avalanches at criticality) will be radically different in the HL model. Nevertheless, we find an intriguing similarity between the two models concerning the scaling with system size of the peak susceptibility $\chip$, shown in Fig.~\ref{fig:chips} for the HL model. The data are broadly consistent with a divergence $\chip\sim N^{1/3}$ at criticality. This is the exponent expected for the mean-field RFIM (see also App.~\ref{app:RFIM} for numerics confirming this scaling in the RFIM). A simple way of obtaining the value $1/3$ is by considering the finite-size scaling at the upper critical dimension $d_{u}=6$: with the Landau exponents $\overline{\gamma}=1$ and $\nu =1/2$, one has $\chip\sim L^{\overline{\gamma}/\nu}\sim L^{2}
\sim N^{1/3}$ (this can also be stated as replacing $L$ in the standard expressions by $L_{\rm eff}=L^{d/d_{u}}=N^{1/d_{u}}$; this approach to scaling above the upper critical dimension has recently been confirmed explicitly for the RFIM~\cite{fytas_finite-size_2023}~\footnote{We note here the relation to the corresponding scalings in the fully connected non-disordered Ising model, $\chip\sim N^{1/2}$ and $N^{1/3}$ at the critical point and at the spinodal, respectively. These can be obtained~\cite{colonna-romano_anomalous_2014} from the mean-field relation $\chi\sim L^2$ and the upper critical dimensions $d_u=4$,$6$ at the critical point and the spinodal, respectively. In the RFIM, from the correspondence to the clean system for $d-2$ dimensions~\cite{dahmen_hysteresis_1996}, one expects $d_u=6$,$8$ at the critical point and spinodal respectively, and hence the scalings $N^{1/3}$ and $N^{1/4}$ , consistent with those given in the text.}). This is also consistent with the fractal dimension of the critical avalanches in the RFIM (which takes the mean-field value $d_f=4$ in $d_u=6$, see Eq.~(21) in Ref.~\cite{dahmen_hysteresis_1996}), implying that $S_{c}\sim N^{2/3}$ and hence $\chi^{\rm peak}\sim N^{1/3}$ from $S_c\sim \chi^2$.

The fact that the divergence at criticality for HL is also consistent with $\chip\sim N^{1/3}$ points towards a simple explanation. In the previous subsection we saw that, on approaching the spinodal, the divergence of the susceptibility is cut off for finite system sizes on the scale $\mathcal{O}(N^{-1/2})$ of the yield strain/coercive field deviations, yielding $\chi^{\rm sp}\sim N^{1/4}$. At criticality, where $\chi\sim (\gamma_c-\gamma)^{-2/3}$, assuming the same $\mathcal{O}(N^{-1/2})$ deviations, this yields $\chip \sim N^{1/3}$. Overall, a plausible explanation for the common $\chip\sim N^{1/3}$ divergence is therefore the joint consequence of Landau exponents for the infinite size limit (reflecting analyticity), and the normal fluctuations $\mathcal{O}(N^{-1/2})$ of the yield strain/coercive field in both models. We finally note that from the scaling $\chi^{\rm peak}\sim N^{1/3}$ one expects the critical avalanches in the HL model to scale as $S_c\sim \chi^{\rm peak} N^{1/2} \sim N^{5/6}\gg N^{2/3}$, i.e. much larger than the corresponding avalanches in the RFIM.

\section{\label{sec:quantitative}Quantitative predictions: testing against simulation/experiment}

We conclude with some final remarks concerning the physical assumptions of the theory, as well as the testability and broader implications of its predictions. Turning to the first point, the physical picture underpinning the HL dynamics (as outlined in the introduction), in terms of localised rearrangements interacting through an elastic propagator, is widely regarded~\cite{nicolas_deformation_2018,barrat_heterogeneities_2011,argon_plastic_1979,desmond_measurement_2015,maloney_amorphous_2006,tanguy_plastic_2006,puosi_time-dependent_2014} as a faithful description of disordered solids that can ``flow''. Being a coarse-grained, mesoscopic approach, it can surmount the stark differences with respect to microscopic properties spanning a wide range of systems, e.g.\ from metallic glasses to foams or colloids~\cite{nicolas_deformation_2018}. Such a description of course does not apply to {\it all} amorphous solids, but is certainly appropriate for the particle simulations of model glass formers~\cite{ozawa_random_2018,singh_brittle_2020,ozawa_role_2020,ozawa_rare_2022} to which we have mainly compared our results. 

As regards testability of the theory, we summarise the main quantitative predictions and how they may be compared to particle simulations (and eventually experiments). Starting with the behaviour at finite shear rates, we have already noted there that the qualitative behaviour of the stress-strain curves is in excellent agreement with particle simulations~\cite{singh_brittle_2020}. As to the quantitative predictions we make, such as the scalings of the peak susceptibility or the shifting of the brittle yield strain with shear rate, these can be tested against more detailed measurements along the lines of~\cite{singh_brittle_2020}. To assess our predictions regarding the universal form of the yield rate tails, one could in principle also consider ensemble-averaged measurements, during the brittle yielding process, of quantities that are known to correlate well with the rate of plastic rearrangements (along the lines of the mean-squared velocity shown in~\cite{ozawa_rare_2022}). We must note, however, that the temporal propagation and growth of the $\infty-$avalanche is, at least in $d=2$, 
unlikely to be well-described by mean-field. In finite dimensions the $\infty-$avalanche is localised within a shear band, and in $d=2$, the physics is of plastic rearrangements propagating purely along a line~\cite{dasgupta_microscopic_2012,pollard_yielding_2022}: a mean-field approach which considers the global level of mechanical noise on a given region due to the entire rest of the system must therefore break down. Notwithstanding this limitation regarding the dynamical details of the $\infty-$avalanche itself, mean-field may describe well the behaviour at finite $\dgamma$ we have analysed (and hence e.g. the shifting of the brittle yield strain or the scaling of the peak susceptibility), where the systems ``flows around'' the $\infty-$avalanche and yields over a large number of overlapping avalanches instead of a single macroscopic one. We stress that this overlapping is ultimately the physical content of the boundary layers described by the HL model for a finite driving rate $\dgamma$. The finite tail probability of elements with $|\sigma|>\sigma_c$ at any given time corresponds to a pool of unstable regions where a plastic rearrangement can be triggered; given that these elements are uncorrelated, they may effectively be thought of as being located at random locations throughout the system and lead to avalanches of plasticity that must overlap with each other, as studied in particle simulations~\cite{oyama_instantaneous_2021}.

Turning to the avalanche behaviour, our thorough analysis of the HL model should provide the foundation for a full understanding of avalanches in the model with power-law noise ($\mu=1$)~\cite{lin_mean-field_2016}. In accounting via the noise distribution for the spatial decay $r^{-d}$ of the propagator, this model is expected to give improved predictions in the quasistatic limit. In fact, given that the avalanche exponent $\tau$ is known not to display significant variation between $d=2$ and $d=3$~\cite{nicolas_deformation_2018}, it is reasonable to assume that it may be well explained by mean-field theory (which as always should hold best for large $d$). In Sec.~\ref{subsec:extension_PL}, we show numerically that the values measured with power-law noise ($\tau \approx 1.35$, $1.39$) are indeed in broad agreement with most measurements available in the literature of lattice simulations, as well as experiments on metallic glasses~\cite{sun_plasticity_2010} (see further below). The above remark on the dimension dependence of $\tau$ does not apply to the fractal dimension $d_f$, which is clearly strongly affected by the geometry of the propagator, forming e.g.\ line-like structures $d_f\approx 1$ in $d=2$ (note that HL predicts the correct value of $d_f=d/2=1$ in $d=2$, but this should be regarded as a fortunate coincidence). In $d=3$, we may take as reference the value $d_f \approx 1.3$ measured in~\cite{liu_driving_2016} both for lattice and particle simulations (note that this value will depend somewhat on the choice of cutoff function), implying roughly $S_c\sim N^{0.43}$, and compare this with values from the power-law noise model (see Sec.~\ref{subsec:extension_PL}), where we measure $S_c\sim N^{0.41}$. Taking these new results on avalanches (not studied in~\cite{lin_mean-field_2016}) together with the original results there~\cite{lin_mean-field_2016} regarding the pseudogap exponent $\theta$, this all supports the general view that, although $d=2$ is clearly ``special'', mean-field can provide fairly good predictions already for $d=3$.

As to the behaviour approaching the brittle yield point, as discussed in Sec.~\ref{subsec:precursors}, we have already noted there that we recover the qualitative finite-size trend (i.e.\ a strong stabilising bias on the yield strain for smaller system size; this comparison could of course be made more quantitative), as well as the $\kappa=1/2$ (Eq.~\ref{eq:P_N}) fluctuations of the yield strain found in $d=3$ particle simulations~\cite{ozawa_role_2020}. A key point to test in particle simulations is whether the mean-field spinodal criticality we predict is ``saved'' in $d=3$ by the long-range nature of the elastic interactions. This would require measurements along the lines of Fig.~\ref{fig:conditioned}, where one would have to replace the scaling $N^{1/2}\langle S \rangle$ by the appropriate power $N^{1-\tilde{\alpha}}\langle S \rangle$ in order to collapse the pre-spinodal avalanche sizes. Note that this is an inherently statistical question, concerning ensemble averages and finite-size scaling; it is clear that, even in mean field, there are individual samples, particularly when the system size is small (see Fig.~\ref{fig:plot_brittle}), that yield ``abruptly'' without any appreciable precursors. 

\begin{figure}
\centering

\begin{subfigure}[b]{0.4\textwidth}
   \includegraphics[width=1\linewidth]{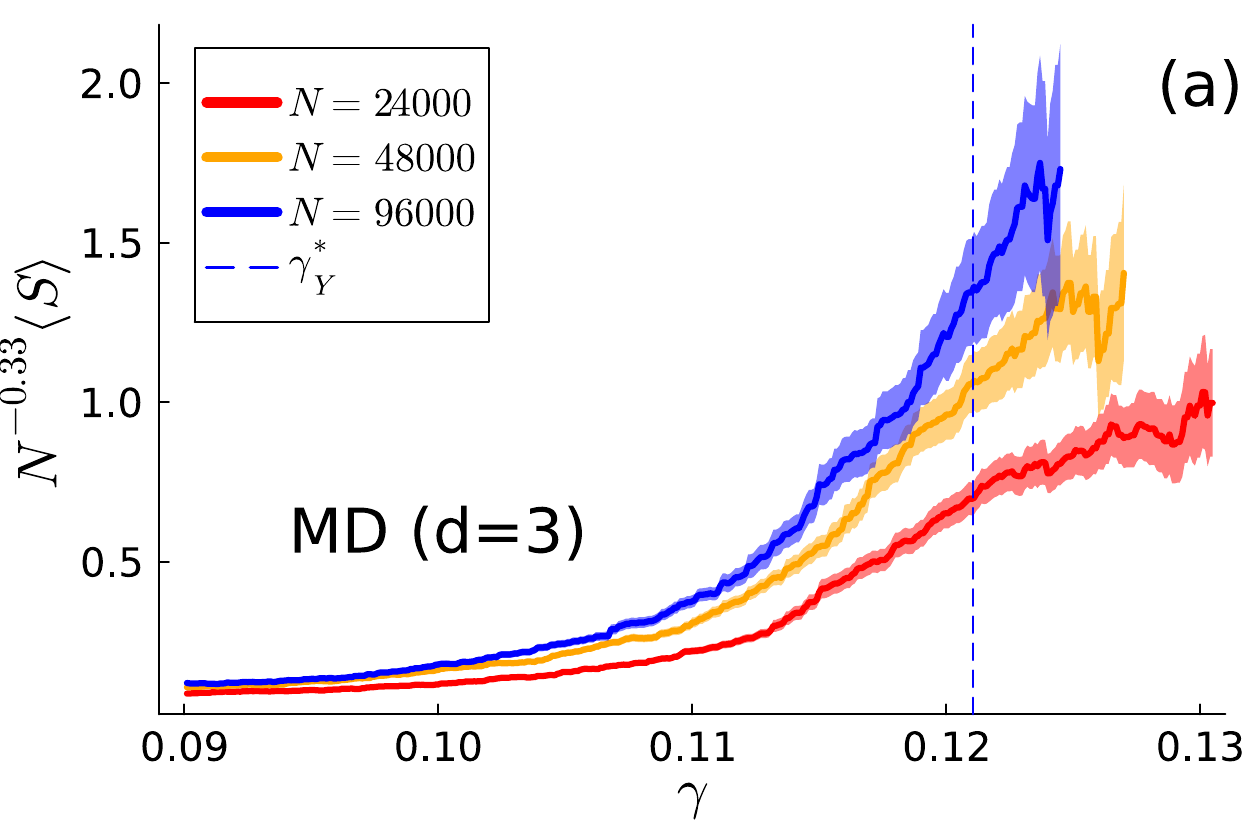}

\end{subfigure}

\begin{subfigure}[b]{0.4\textwidth}
   \includegraphics[width=1\linewidth]{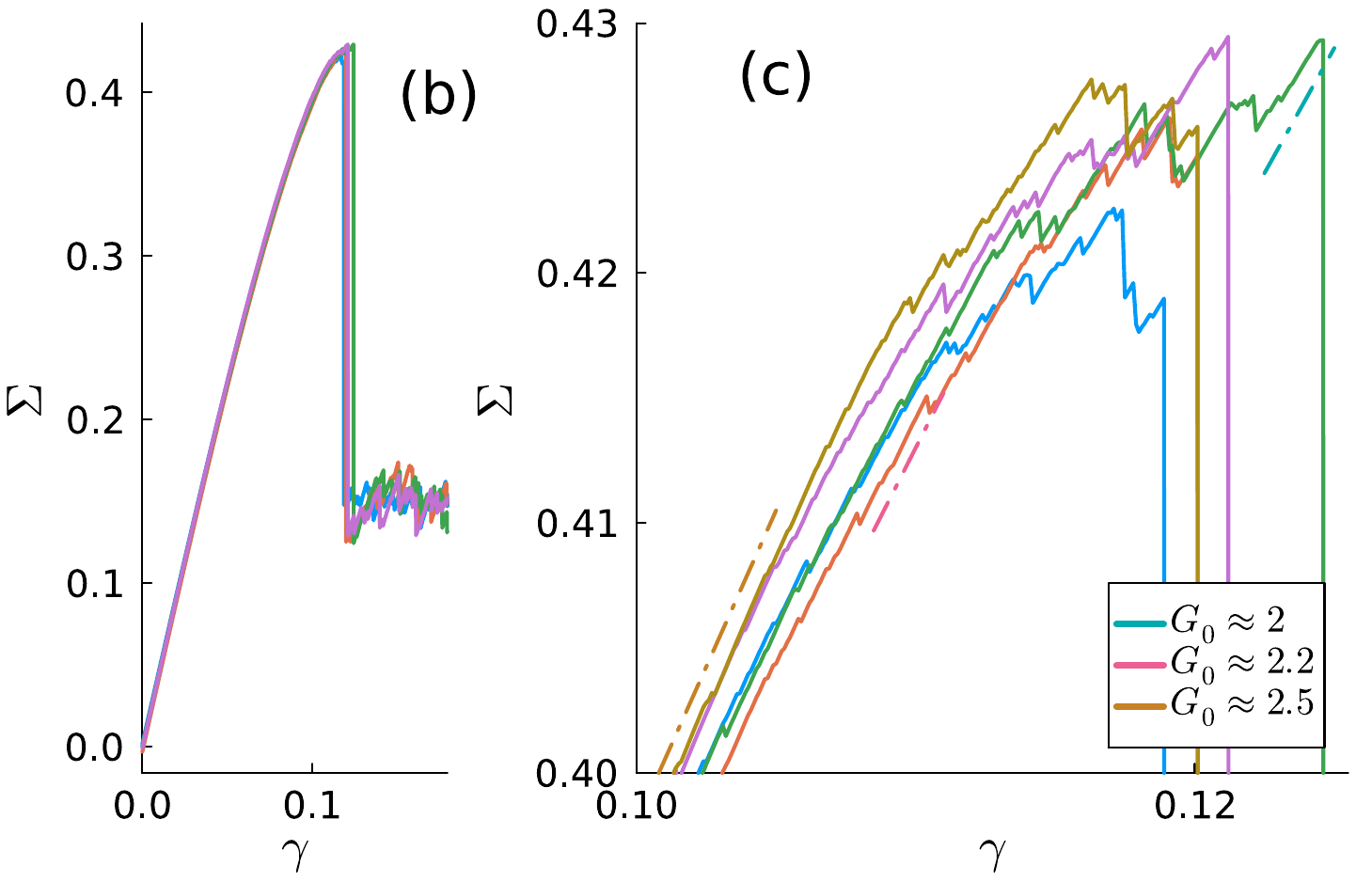}
   
\end{subfigure}

\caption{\label{fig:MD_data} Approaching the brittle yield point in quasistatic MD (molecular dynamics) simulations of an atomic glass former~\cite{ozawa_random_2018}. (a) Average avalanche size (from average stress drop, see text) within strain windows of width $\Delta \gamma =9 \cdot 10^{-3}$ conditioned (as in Fig.~\ref{fig:conditioned}) on the avalanche occurring before the $\infty-$avalanche for a given run. Vertical dashed lines indicate as reference the average yield strain for the largest system size $N=96000$. (b) Stress-strain curves over the entire window for 5 single realisations with $N=96000$. (c) Zoom-in for these 5 realisations approaching their yield point. As detailed in the text, the local shear modulus $\Go$ is obtained from an ensemble average over positive slope segments at each $\gamma$. To illustrate the softening, we indicate (roughly) the modulus (in MD units) of three elastic segments approaching the yield point.}
\end{figure}

In Fig.~\ref{fig:MD_data}, we make an important first step in this direction. For the atomic glass model studied in~\cite{ozawa_random_2018}, we extract stress drops for the most brittle sample (corresponding there to $T_{\rm ini}=0.062$) following the same method employed there. For each strain interval $\Delta \gamma$, $\Delta \sigma_i=\sigma_i-\left(\sigma_{i-1}+\Go \Delta \gamma \right)$ is evaluated, and stress drops (corresponding to negative $\Delta \sigma_i$) due to avalanches are then extracted by setting a threshold given by $\Delta \sigma= -c/N$ (with $c=0.1$~\cite{ozawa_random_2018}). Here, $\Go$ is the local shear modulus, which we extract from a strain-dependent ensemble average of the slope of loading segments (i.e.\ those with positive slope) for each system size. Within the strain window of interest approaching the yield point, the average modulus shows a softening of around $20\%$, as illustrated for individual realisations corresponding to the largest system $N=96000$ in Fig.~\ref{fig:MD_data}c. The average avalanche size (from $\langle S \rangle=N \langle \Delta \sigma \rangle$) per strain interval, conditioned on occurring before the $\infty-$avalanche for a given run (as in Fig.~\ref{fig:conditioned}) is shown in Fig.~\ref{fig:MD_data}a. We recover the two main features discussed in mean-field: the pre-spinodal avalanches grow with system size reflecting marginal stability; once rescaled to remove this effect, the curves then display a divergence with system size around the yield point. A larger system size range and improved statistics will be necessary to characterise in detail the marginal stability properties in the pre-spinodal regime, as well as the divergence in system size of $\chi^{\rm sp}$ after rescaling. From the theoretical side, we note that the HL model in its simplest form (\ref{eq:hl_first}) presents a unique fixed $\Go$ and is hence unable to capture the softening of the modulus approaching the yield point; this may be possible in the disordered variant of the model~\cite{agoritsas_relevance_2015}, which accounts for a distribution $\rho(E)$ of yield energies (see also Sec.~\ref{subsec:hetero} below), by introducing a distribution of $\Go$ values correlated with the yield energy $E$. It would also be interesting to study the approach to the spinodal in the model with $\mu=1$ power-law noise~\cite{lin_mean-field_2016}, which as discussed above may reproduce avalanche scalings closer to those measured in $d=3$. 

More generally, although the model we have considered implicitly assumes that loading conditions and sample geometry are such that the opening of a crack is avoided~\cite{barlow_ductile_2020,ozawa_random_2018}, our results regarding the approach to brittle yielding are also relevant in the context of material fracture~\cite{alava_statistical_2006,zapperi_first-order_1997}. There, mean-field treatments of models such as the fiber bundle model assume a uniform load redistribution and typically lead to the same avalanche criticality as the RFIM~\cite{zapperi_avalanches_1999}. However, in e.g.\ model amorphous solids under athermal quasistatic expansion~\cite{dattani_universal_2022}, (Eshelby-like) quadrupolar plastic events were observed prior to cavitation and fracture, suggesting that long-range sign-varying interactions are also relevant in this context. Recent simulation work on the precursors to fracture of silica under tensile stress, on the other hand, were well described by predictions of the fiber bundle model~\cite{shekh_alshabab_criticality_2024}. This raises an existing question in the context of material fracture~\cite{zapperi_first-order_1997}, worth exploring further, regarding the role of material properties and loading conditions in determining the relevance of sign-varying long-range interactions and the associated development or not of marginal stability (or self-organised criticality) in the approach to fracture.

Finally, turning to the random critical point (Sec.~\ref{subsec:criticality}), we have measured (and partially explained) in mean-field the fluctuations of the largest stress drop, which is the only property of the critical point measured thus far in the literature~\cite{ozawa_random_2018,rossi_finite-disorder_2022}. Although $d=3$ particle simulations show a distinct peak growing with system size~\cite{ozawa_random_2018}, a finite-size scaling analysis has not yet been performed, so that we have instead compared our results to $d=3$ lattice simulations~\cite{rossi_finite-disorder_2022}, finding a good agreement within the limited system size range. Eventually, one could go beyond the peak fluctuations and  compare the full shape of the distribution of the largest stress drop at the critical level of annealing, as measured recently in atomistic metallic glass simulations~\cite{su_atomic_2022}. A theoretical characterisation of this (presumably universal) non-Gaussian distribution is beyond the scope of this work, and is left as a task for future studies.

\section{\label{sec:discussion}Discussion: universality class and ``completeness'' of the theory}
As outlined in the introduction, the main appeal of the RFIM paradigm is that the random critical point can be tackled using renormalisation group (RG) scaling arguments. The beauty of the RG approach lies of course in the concept of a universality class, arising from the existence of a fixed point in the multidimensional space of model parameters. Scale invariance in particular gives access to a full understanding of the integrated avalanche distribution~\cite{dahmen_hysteresis_1996,perez-reche_finite-size_2003}. It is clear that, for the HL model, the fundamental difference to the RFIM universality class lies in the fractality of avalanches (which we recall holds for any value of $R$, not just $R=R_c$ as in the RFIM), leading to a different form of scale invariance; the sign-varying interactions in addition lead to a different value of $\tau$. 

Throughout this paper, we have 
restricted 
our comparisons to the standard ferromagnetic RFIM. At the time of submitting, a pre-print~\cite{rossi_far--equilibrium_2023} has appeared showing simulation results on a variation of the RFIM with interactions of the Eshelby form on a lattice in both $d=2$ and $3$. This 
reproduces quite well (at least in $d=3$) the critical fluctuations of the largest stress drop found previously in lattice EPMs~\cite{rossi_finite-disorder_2022} (to which we have also compared our results in Sec.~\ref{subsec:criticality}), as well as the characteristic band-like spatial structure of the $\infty-$avalanche, and further underscores 
the fact that Eshelby-like interactions lead to a distinct universality class with respect to the standard RFIM. We note, however, that at present this formulation does not allow for analytical progress, nor does it appear to display~\cite{rossi_far--equilibrium_2023} any significant traits of the marginal stability found in amorphous solids: the authors attribute this to the difference between spin flip dynamics $s_i=\pm 1$ and yielding of the real-valued local stress. In future work, it would nonetheless be interesting to compare the scale invariance properties of the avalanche distribution to the models defined here in terms of an effective mean-field mechanical noise.
Note that the nature of the noise (via the noise exponent $\mu$) will also affect the scale invariance properties and the universality class, given that the avalanches are expected to scale differently with system size and display a different exponent $\tau$ (see preliminary results in Sec.~\ref{sec:outlook}).


We may finally discuss the ``completeness'' of the theory, and the ultimate relevance of the standard RFIM paradigm to the problem of yielding in amorphous solids (within the elastoplastic framework). Regarding the first point, strictly speaking we have gained insight into the {\it full} problem from two special limits, $\dgamma \to 0$ with $N=\infty$, and $N \to \infty$ with $\dgamma=0^{+}$. Of course, in reality one will deal with a situation where both $N$ and $\dgamma$ are finite; if thermal fluctuations are present, there will also be a non-vanishing $T$. Starting from a description in terms of avalanches of plasticity, it is interesting to think of these three parameters as perturbations away from the idealised limit of athermal, quasistatic shear in an infinite system, i.e.\ $N=\infty$, $\dgamma=0^{+}$, $T=0$. Away from this limit, each of the parameters cuts off the maximal extent of the avalanche of plasticity caused by a single rearrangement; a qualitative phase diagram for this (in the steady state) has recently been proposed in Ref.~\cite{korchinski_dynamic_2022}. Indeed, at finite $\dgamma$ or $T$ a new avalanche can be triggered before the previous one can finish, making avalanches overlap (for the case of thermal avalanches, we refer to the recent work~\cite{tahaei_scaling_2023}); a finite system size $N$, on the other hand, induces a cutoff on the largest avalanche size that can be reached. A ``full'' theory of yielding in amorphous solids would have to account for the three-dimensional axes ($N$, $\dgamma$, $T$) and all associated crossovers and scaling functions as a function of the quenched disorder $R$; in this sense, 
we are only beginning to scrape the complexity of the full problem. 
Regarding the standard RFIM paradigm, strictly speaking, the only thing that has survived in our mean-field elastoplastic analysis is the Landau-like behaviour in the idealised limit ($N=\infty$, $\dgamma=0^{+}$, $T=0$), as well as the standard finite-size fluctuations of the yield strain. Returning to our previous point, it is then clear that the full complexity of the problem demands an alternative mean-field approach such as the one presented here.

\section{\label{sec:outlook}Outlook: extensions of the theory}

\subsection{\label{subsec:extension_PL}Beyond the Gaussian approximation: extension to power-law mechanical noise}
Turning to the outlook, arguably the most interesting direction would be to extend our analysis to a mean-field elastoplastic description with power-law mechanical noise, which in the $N=\infty$ limit is defined by the master equation incorporating L\'{e}vy noise (see~\cite{parley_aging_2020} for details).  
Regarding the $\infty$-avalanche in the brittle regime, basing ourselves on the analysis of the aging problem~\cite{parley_aging_2020} one may conjecture 
how some of our scaling predictions would change with the noise exponent $\mu$. One expects that the tail $f_{-}(z)\sim \left(-z\right)^{-2}$ will be replaced by a $\mu$-dependent power law $f_{-}(z)\sim \left(-z\right)^{-\mu/(\mu-1)}$, and a stretched exponential $f_{-}(z)\sim \exp{\left(-B\sqrt{-z}\right)}$ for $\mu=1$ (with $B$ an initial-condition dependent constant). 
Overall, therefore, one expects the onset of the infinite avalanche as $\dgamma\to 0$ to be {\it{sharper}} for $\mu <2$. Turning to the divergences of the peak susceptibility, for $R<R_c$ one expects $\chip \sim \dgamma^{-1}$ independently of the noise exponent, as this is just a consequence of the yielding of a finite fraction of the system within the $\infty$-avalanche. At $R=R_c$, on the other hand, one would conjecture, following the HL model, that the form of the divergence might be related to the critical scalings under shear at $A=A_c$. These scalings are at this point an entirely open question~\cite{lin_microscopic_2018}.


Turning to the quasistatic $\dgamma=0^{+}$ behaviour in the infinite size limit, a priori one expects again the smoothness argument to hold as in the HL model, which would lead to Landau exponent values. For the case $\mu=1$, however, it would be interesting to study the non-trivial behaviour of the pseudogap exponent $\theta$ during the transient. Indeed, for $\mu=1$, one does not expect~\cite{lin_mean-field_2016} a fixed strain-independent value of $\theta$ as in the HL model; instead 
the degree of marginal stability is tuned by the dynamics. In particular, $\theta$ would presumably display a discontinuous jump across the brittle yield point, as observed in particle simulations~\cite{ozawa_random_2018}. Also interesting would be to determine the conditions satisfied by the local stress distribution, in the infinite size limit, at $\gamma=\gamma_c$, both for $R<R_c$ (the spinodal limit) and $R=R_c$ (the critical point): rather than by the curvature and the fourth derivative at the threshold as in HL (see Table~\ref{T1}), one expects the behaviour to be determined by non-local properties involving the L\'{e}vy propagator.

\begin{figure}
\centering
\includegraphics[width=0.4\textwidth]{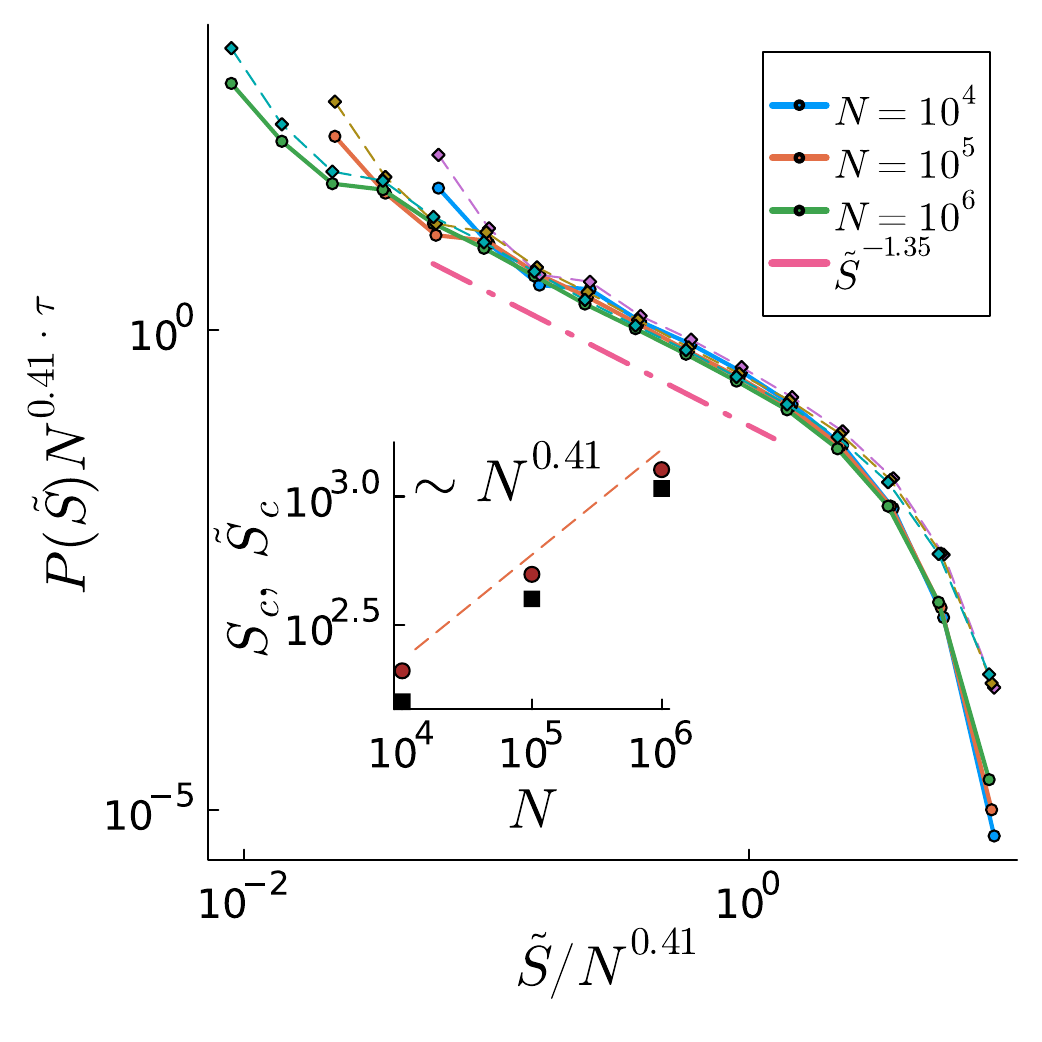}
\caption{\label{fig:avalanches_mu_1}Avalanche distribution in steady state for $\mu=1$, for coupling $A=0.47$, with the upper cutoff as originally defined in~\cite{lin_mean-field_2016}, for three different system sizes. The distributions are very well fitted by the form (\ref{eq:p_of_S}), from which we extract the exponent $\tau$ as $1.347\pm 0.037$ (for $N=10^6$, by fitting to Eq.~\ref{eq:p_of_S} in the range $S\in [10,2600]$), and the scaling of the cutoff size $\tilde{S}_c\sim N^{0.41}$, which we use to collapse the distributions. Note that we show both the statistics of the number of rearrangements $S$ (dashed lines) and the correspondingly scaled stress drop $\tilde{S}=N\Delta \Sigma$ (full lines). These show quantitative differences in the tail due to a non-negligible number of yield events at $-\sigma_c$, but do not change the 
qualitative behaviour.
Inset: scaling of the fitted cutoffs $S_c$ (circles) and $\tilde{S}_c$ (squares); both diverge roughly as $\sim N^{0.41}$.} 
\end{figure}

Finally, an important open question is the mean-field avalanche behaviour for $\mu <2$, in particular for the improved mean-field model with $\mu=1$. A priori, one possible scenario would be for the $\tau \approx 1$ behaviour of the HL model to be a universal feature common to the whole family of mean-field models with $\mu <2$. We leave a full characterisation of the $\mu=1$ avalanche behaviour, which would involve relating the statistics to the finite-size shape of $P(x)$ as done here for HL, for future work. At this point we can nonetheless straightforwardly adapt the numerical scheme developed in Sec.~\ref{sec:second_part}, replacing the HL Gaussian noise kicks by the power-law kick distribution (\ref{eq:power_law_noise}) with $\mu=1$, and measure the avalanche statistics at least in steady state. We consider two values of the coupling $A$, $A=0.32$ (shown in App.~\ref{app:avalanches_mu_1}) and a slightly larger value $A=0.47$. From these steady state results, shown for $A=0.47$ in Fig.~\ref{fig:avalanches_mu_1}, we can discard the scenario described above: we find again a distribution of the form (\ref{eq:p_of_S}), but the fitted avalanche exponent takes a distinctly higher value  $\tau \approx 1.35$ (for $A=0.32$, we measure instead $\tau \approx 1.39$, see App.~\ref{app:avalanches_mu_1}). The scaling of the cutoff size is also clearly distinct from HL, with a weaker divergence $S_c\sim N^{0.41}$ (inset of Fig.~\ref{fig:avalanches_mu_1}).

While an attempt at a thorough theoretical understanding of these mean-field predictions is beyond our scope here, it is plausible to speculate that the higher value of $\tau$ is related to the different scaling of far-field stress kicks discussed in Sec.~\ref{sec:mf_theory}. Indeed, we saw from the first-passage mapping for HL (Sec.~\ref{subsec:avalanches}) that what pushes down the value of $\tau$ is the finite-size plateau in $P(x)$, representing an over-abundance of barely stable sites, which is in turn related to the $\mathcal{O}(N^{-1/2})$ scaling of the smallest stress kicks. The HL model strongly overestimates the typical size of these far-field kicks (see Fig.~\ref{fig:noises}), which for $\mu=1$ instead scale as $\mathcal{O}(N^{-1})$. One therefore expects for $\mu=1$ a much weaker finite-size plateau in $P(x)$, as we confirm in App.~\ref{app:avalanches_mu_1}, which may in turn be 
the origin of the higher value of $\tau$. 

As to the measured values $\tau \approx 1.39$ and $1.35$, we firstly compare these to those found in lattice elastoplastic models, both for $d=2$ and $d=3$. Values reported in the literature (note the following list is far from exhaustive) are mostly in the same ballpark: in strain-controlled simulations of three different $2d$ implementations, the authors of~\cite{ferrero_criticality_2019} found $\tau \approx 1.33$ irrespective of the dynamical rules; also in the $2d$ strain-controlled model studied in~\cite{budrikis_avalanche_2013}, the avalanche distribution at so-called criticality (corresponding to the steady state under quasistatic shear considered here), was found to follow the form (\ref{eq:p_of_S}) with a Gaussian tail, and values $\tau \approx 1.34, \ 1.36$ were measured for the avalanche exponent depending on the short-range details of the propagator used; in~\cite{liu_driving_2016}, the slightly lower values $\tau \approx 1.28$ and $\tau \approx 1.25$ were measured in $2d$ and $3d$ respectively; finally, in the recent work~\cite{richard_mechanical_2023}, avalanches were studied and compared in both $2d$ elastoplastic models and particle simulations, measuring $\tau \approx 1.3$ for both cases. Turning to experimental measurements of avalanche statistics in amorphous materials, these are arguably still somewhat trailing behind predictions from elastoplastic modelling approaches~\cite{nicolas_deformation_2018}. At the large end of the particle size spectrum, granular packings of photoelastic disks studied in~\cite{bares_local_2017} provide a particularly fruitful experimental setup. Statistics of global energy drops under pure shear were recorded in a range spanning over three decades, with an exponent $\tau \approx 1.24$. At the other end of the particle size spectrum, stress drop distributions in ductile metallic glasses (under compression) were found to follow exponents in the range $1.37\ldots1.49$~\cite{sun_plasticity_2010}.

Regarding the universality of avalanches inherent to plastic yielding of amorphous solids, the work of~\cite{budrikis_universal_2017} on tensorial models suggested that there {\it are} universal features independent of dimensionality and loading (in particular, an exponent $\tau \approx 1.28$ is measured there for the avalanche distribution). Given that such results are clearly distinct from the ``classical'' prediction $\tau=3/2$ from mean-field depinning or the RFIM, an important task for future work is to gain a full understanding of the avalanche statistics for the family of mean-field elastoplastic models with mechanical noise given by (\ref{eq:power_law_noise}); our analysis of the HL model is a first step in this direction.

As a final remark, we note that our present discussion of power-law noise has been limited to noise distributions of the form (\ref{eq:power_law_noise}), with a lower cutoff of $\mathcal{O}(N^{-1/\mu})$ derived under the assumption of stress ``kicks'' from single isolated rearrangements. Alternative scalings arise in the coarse-grained noise approach of~\cite{ferrero_properties_2021}, where the stress ``kick'' considered 
arises instead as the aggregate effect of avalanches. A refined effective mechanical noise approach could also aim to account for the screening effects~\cite{lemaitre_anomalous_2021,hentschel_eshelby_2023} that have recently been elucidated in amorphous solids. We expect that the thorough analysis we have offered for simple Gaussian noise (i.e.\ the paradigmatic HL model) will provide a solid basis for such refinements.

\subsection{\label{subsec:hetero}The question of the initial condition and the origin of brittleness}

A natural direction in which to extend the model is the inclusion of disorder in the form of a non-trivial  distribution of local yield thresholds $\rho(\sigma_c)$~\cite{agoritsas_relevance_2015,parley_mean-field_2022}; in the original HL model studied here, this ``renewal'' distribution corresponds to a delta function. Such a distribution can also be re-cast in terms of elastic yield energies $E=\sigma_c^2/2$, (with a modulus $\Go$ set to unity) as in \cite{parley_mean-field_2022}. In this augmented description, commonly referred to as ``disordered'' HL model~\cite{agoritsas_relevance_2015}, the current state of each block is characterised by two variables, its local stress $\sigma_i$ and current local yield threshold $\sigma_{c,i}$, and hence, in the $N \to \infty$ limit, by a joint distribution $P(\sigma,\sigma_c,\gamma)$. Despite this additional variable, we expect all the qualitative scaling results found here to remain valid. The finite $\dgamma$ results in Part 1, on the one hand, depend only on the boundary layer scalings, which follow the same form in the augmented model~\cite{agoritsas_relevance_2015}. The avalanche behaviour for finite $N$ studied in Part 2, on the other hand, depends on the distribution of stabilities $x_i=\sigma_{c,i}-\sigma_i$, which we expect to follow the same linear form for small $x$, cut off by a finite-size dependent plateau.

Although qualitative changes are not expected, the inclusion of a distribution of yield thresholds is important in two ways. Firstly, in deriving the model from the KEP~\cite{bocquet_kinetic_2009} (kinetic elastoplastic) framework, one finds that the contribution to the mechanical noise due to a yield event should scale with the square of the yield threshold $\sigma_c^2$ that has been overcome. In other words, defining a (dimensionless) coupling $\hat{\alpha}$, the noise due to an event at site $j$ acting on any other site should be sampled from a Gaussian with standard deviation $\sigma_{c,j} \sqrt{2 \hat{\alpha}/N}$. This means that, instead of considering a narrowing of the distribution of the local stress, an initial distribution of constant shape can become more brittle simply by shifting towards higher yield thresholds, as this leads to a large effective value of the coupling controlling the transient behaviour. Once in steady state, it is then known~\cite{agoritsas_relevance_2015} that the same state is reached as for a uniform yield stress-independent coupling of strength $\alpha=\hat{\alpha}\langle {\sigma_c^2 \rangle}_{\rho}$.

The second important aspect is that the inclusion of $\rho(\sigma_c)$ allows one to relate directly to physical properties of the annealing-dependent inherent states that are subjected to shear. Indeed, empirical measurements in model glasses~\cite{barbot_rejuvenation_2020}, using the frozen matrix method~\cite{adhikari_soft_2023,rottler_thawed_2023}, show that the typical magnitude of yield stresses extracted from the renewal distribution $\rho(\sigma_c)$ correspond to those found in inherent states obtained by quenching a supercooled liquid around the mode-coupling temperature $T_{\rm MCT}$~\cite{barbot_rejuvenation_2020}. Samples prepared by quenches from parent temperatures $T_{\rm ini}<T_{\rm MCT}$ display rapidly increasing values of the average yield stress (see 
Fig.~7 in~\cite{barbot_rejuvenation_2020}). This was proposed as an explanation for the fact that the ductile-brittle transition is also found to be in the vicinity of $T_{\rm MCT}$, and is a consequence of the {\it emerging contrast} between yield stresses in the as-quenched state and in the renewal distribution $\rho(\sigma_c)$.

\begin{figure}
\centering

\begin{subfigure}[b]{0.4\textwidth}
   \includegraphics[width=1\linewidth]{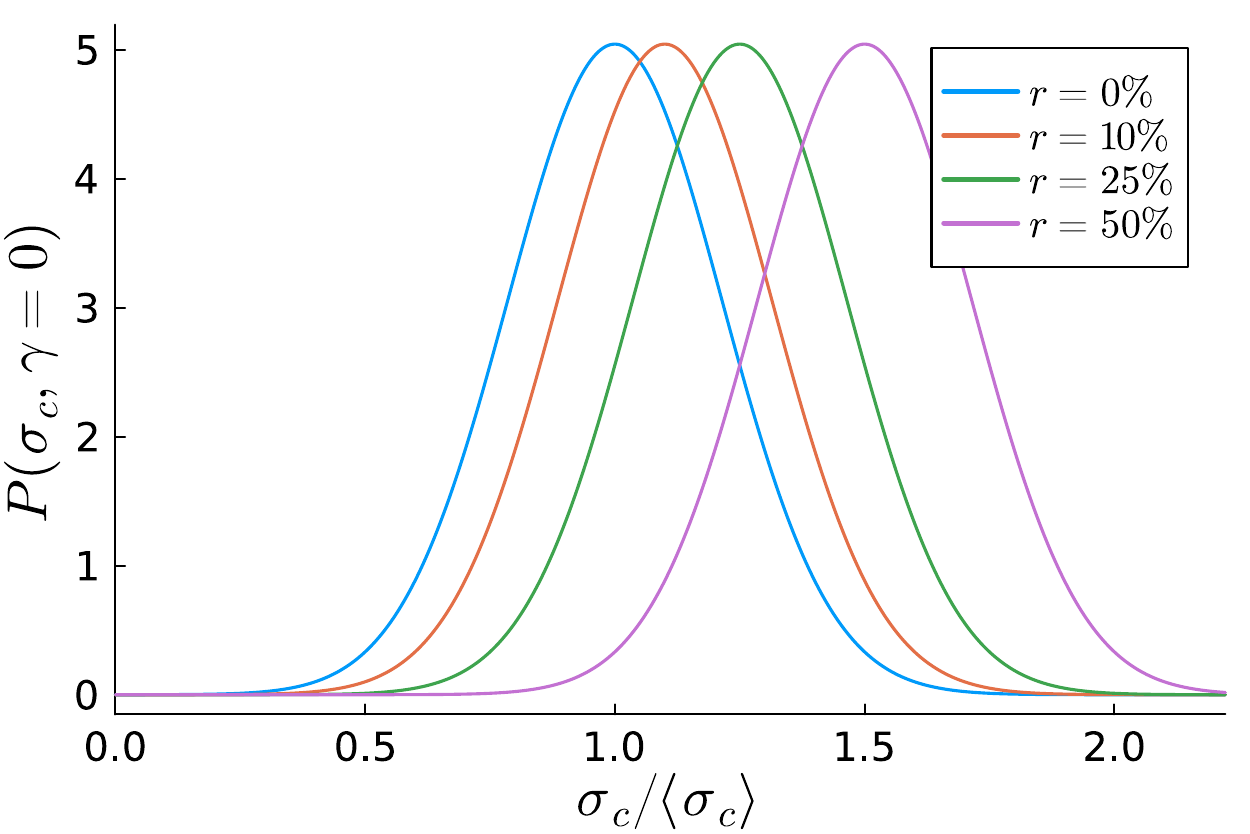}
   \label{fig:ics_shifted}
\end{subfigure}

\begin{subfigure}[b]{0.4\textwidth}
   \includegraphics[width=1\linewidth]{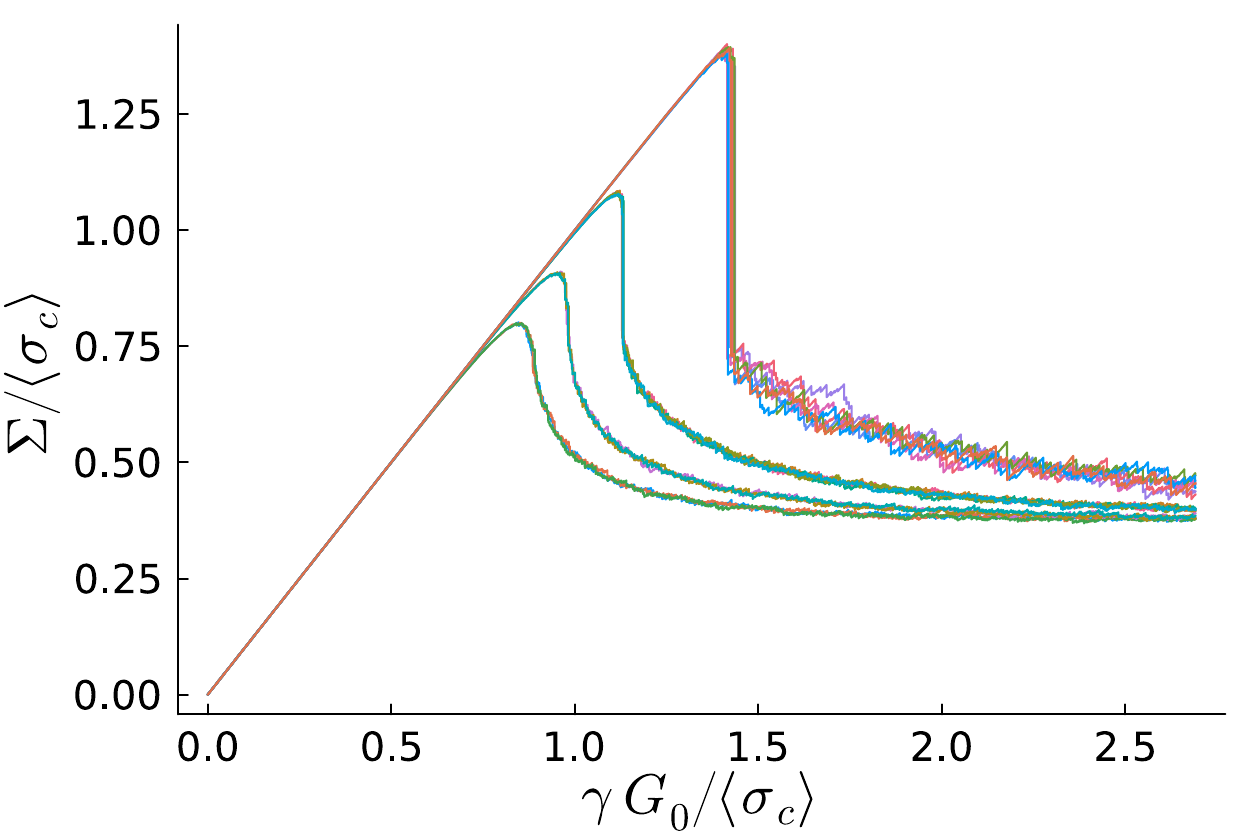}
   \label{fig:brittle_shifted} 
\end{subfigure}

\caption{\label{fig:emerging_contrast}Ductile-brittle transition induced by stronger yield stresses in the initial state. $\hat{\alpha}=0.2$ (see text), and $\rho(E)$ (with $E=\sigma_c^2/2$) of the same Gaussian form as in~\cite{parley_mean-field_2022}. Top: preparation of the initial condition $P(\sigma, \sigma_c, \gamma=0)=P_r (\sigma_c)\delta(\sigma)$, where $r$ denotes the relative increase of the average yield stress with respect to the average value in the distribution $P_{r=0}(\sigma_c)$ (which we approximate here by a Gaussian) found at the threshold under cyclic shear~\cite{parley_mean-field_2022}. Bottom: quasistatic stress-strain curves for increasing $r$ (ductile to brittle). Shown for each $r$ are $5$ individual samples, with $N=10^{5}$. Averages $\langle \cdot \rangle$ are with respect to $\rho(\sigma_c)$.}
\end{figure}

The mechanism described above for the disordered HL model, of increased brittleness due to stronger yield thresholds in the initial state, is hence arguably more in line with this physical scenario. In Fig.~\ref{fig:emerging_contrast} we test this mechanism numerically; we consider $\hat{\alpha}=0.2$, and a renewal distribution $\rho(\sigma_c)$ of the same shape and parameters as in~\cite{parley_mean-field_2022}. We consider initial conditions of the form $P(\sigma,\sigma_c,\gamma=0)=P_r(\sigma_c) \delta (\sigma)$ (the second factor, assumed here for simplicity, implies that the distribution of residual stresses $x_i=\sigma_{c,i}-\sigma_i$ has the same form $P_r(x)$ as the first factor). Here, $r=0$ corresponds as reference to the yield stress distribution found in the threshold state under cyclic shear for $\hat{\alpha}=0.2$ and the given $\rho(\sigma_c)$. An analytical theory for this state was derived in~\cite{parley_mean-field_2022}, and it can be interpreted as the limit of mechanical annealing due to cyclic shear. The distribution $P_{r}(\sigma_c)$ is then simply obtained by increasing the average yield stress by a factor $r$ (see Fig.~\ref{fig:emerging_contrast}, top). As seen in Fig.~\ref{fig:emerging_contrast} (bottom), a ductile-to-brittle transition is indeed observed.

A full characterisation of the disordered HL model, which would involve relating the renewal distribution $\rho(\sigma_c)$, the threshold distribution under cyclic shear, and the ductile-brittle transition under uniform shear within an empirically justified (see below) thermal annealing protocol, for a given coupling $\hat{\alpha}$, is left for future studies~\footnote{We clarify the relation to observations in~\cite{parley_mean-field_2022} regarding the yielding of well-annealed samples under oscillatory shear. These samples were also obtained by essentially shifting the initial yield stress distribution via a factor $\beta$ and remained ductile under uniform shear, while here we argue that brittle yielding can also occur. The difference arises, firstly and most importantly, because the effect of a growing mechanical noise with yield threshold (expressed through the so-called closure relation~\cite{agoritsas_relevance_2015}) on shear startup was not considered in~\cite{parley_mean-field_2022}. Secondly, a significantly smaller ($\alpha=0.043 {\langle \sigma_c^2 \rangle}_{\rho}$) coupling constant was chosen, reducing the possibility of observing brittle behaviour.}. Ultimately, this mean-field mesoscopic theory of the mechanical behaviour of amorphous solids should take as input an empirical characterisation of the yield stress distributions, which in turn should be explained by theories of the glass transition, accounting for the origin (via the complex energy landscape) of the local yield barriers in the initial amorphous state itself. 
This issue is now all the more important given that the local yield stresses have been shown to correlate well not only with rearrangements under athermal shear, but also in the relaxation under finite temperature~\cite{lerbinger_relevance_2022}, so that elastoplastic models have been suggested to capture some of the key features of dynamical heterogeneities in supercooled liquids~\cite{chacko_elastoplasticity_2021,ozawa_elasticity_2023,tahaei_scaling_2023}.

\begin{acknowledgements}
    The authors thank Gilles Tarjus, Misaki Ozawa, Kirsten Martens, Matthieu Wyart and Marko Popovi\'c for comments and helpful discussions. We also thank Misaki Ozawa for providing the MD data in Fig.\ref{fig:MD_data}, Saverio Rossi for sharing the EPM data in Fig.~\ref{fig:peak_HL_EPM}, and Marko Popovi\'c for sharing the numerical routine for implementing the self-consistent dynamics.
    Simulations were run on the GoeGrid cluster at the University of Göttingen, which is supported by the Deutsche Forschungsgemeinschaft (project IDs 436382789; 493420525).
\end{acknowledgements}

\appendix

\section{\label{app:details}Boundary layer method}
We give here further details on the analytical boundary layer expansion setup introduced in Sec.~\ref{sec:background}. The leading order equations of motion for $Q_0$ (\ref{eq:Q0_eom}) and $Q_1$ are given by 
\begin{equation}\label{eq:Q0_app}
    \partial_{\gamma}Q_0=-\partial_{\sigma}Q_0+c_0(\gamma)\partial_{\sigma}^2 Q_0 +\frac{c_0(\gamma)}{\alpha}\delta(\sigma) 
\end{equation}
\begin{equation}\label{eq:Q1_app}
    \partial_{\gamma}Q_1=-\partial_{\sigma}Q_1 + c_0(\gamma)\partial_{\sigma}^2 Q_1+c_1(\gamma)\partial_{\sigma}^2 Q_0 +\frac{c_1(\gamma)}{\alpha}\delta (\sigma)
\end{equation}

\subsection{Steady state solution}
\begin{figure}
\centering
\begin{subfigure}[b]{0.3\textwidth}
   \includegraphics[width=1\linewidth]{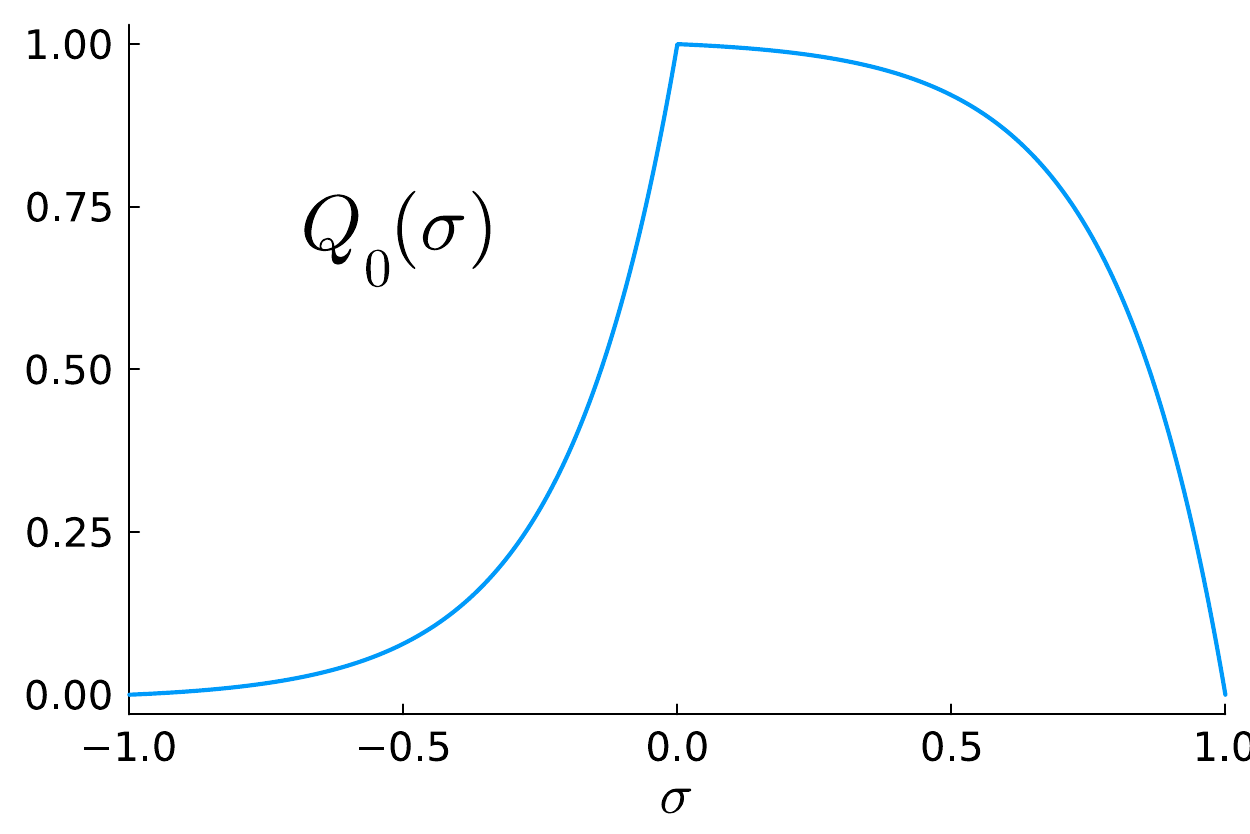}
   \label{fig:Q0_app}
\end{subfigure}

\begin{subfigure}[b]{0.3\textwidth}
   \includegraphics[width=1\linewidth]{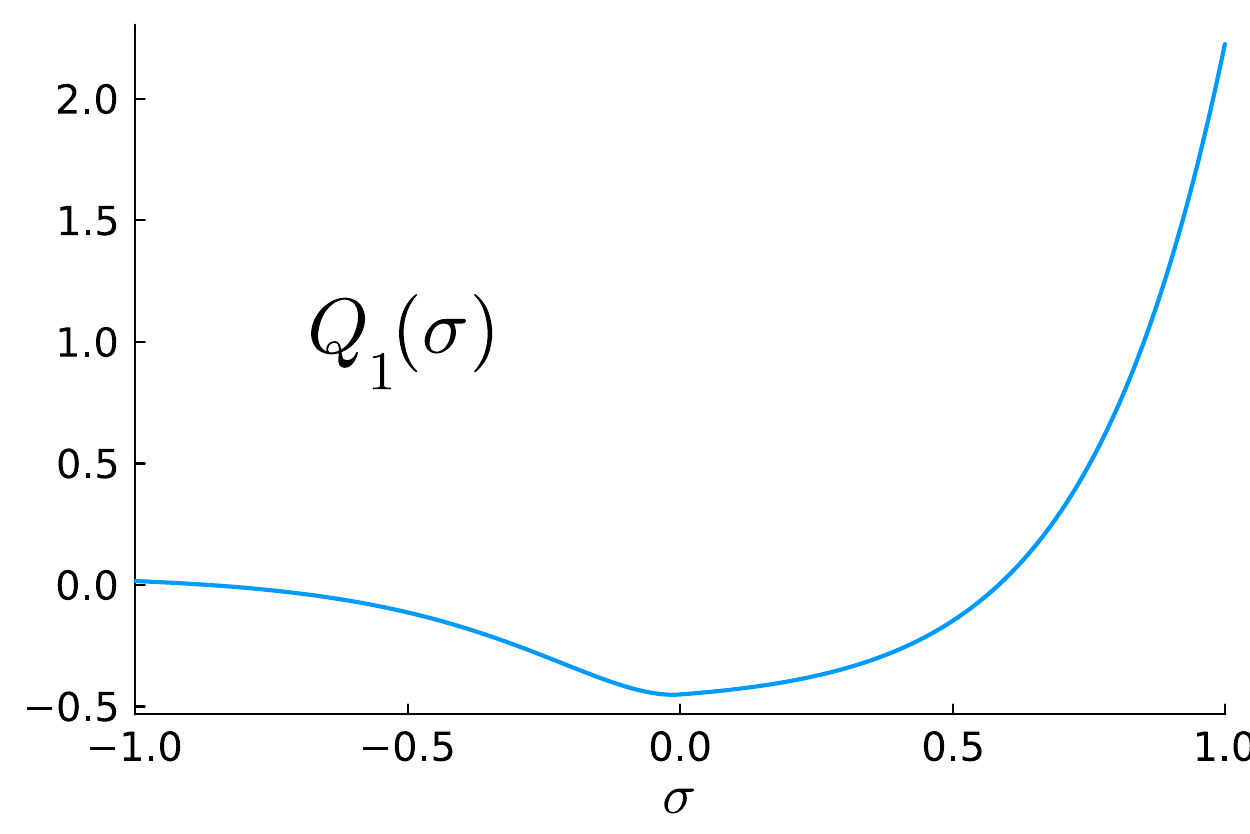}
   \label{fig:Q1_app} 
\end{subfigure}
\caption{\label{fig:Q0_Q1}Steady state solutions of $Q_0$ and $Q_1$ for $\alpha=0.2$. $Q_1$, which gives the leading order correction to the local stress distribution as the shear rate is increased above the quasistatic limit, has its main contribution around the yield threshold $\sigma=1$ as one would expect intuitively.}

\end{figure}

We discuss firstly the analytical solution of the pair of equations (\ref{eq:Q0_app}) and (\ref{eq:Q1_app}) in steady state, where the left hand side is set to zero. The resulting equations must be solved together with the boundary conditions for $Q_0$ and $Q_1$, which requires solving in parallel for the exterior functions, alongside the conditions $\int \mathrm{d}\sigma\,  Q_0=1$ and $\int \mathrm{d}\sigma\, Q_1=0$.  This is useful to gain intuition for the roles and general forms of $Q_0$ and $Q_1$, the steady state forms of which are shown for $\alpha=0.2$ in Fig.~\ref{fig:Q0_Q1}. Although we do not give the full functional forms, which may be found in~\cite{puosi_probing_2015}, we note for convenience the equations determining $c_0$ and $c_1$ in steady state:
$c_0$ is given by the transcendental equation
\begin{equation}
    c_0\tanh{\left(\frac{1}{2c_0}\right)}=\alpha
\end{equation}
while $c_1$ can be determined from $c_0$ by
\begin{equation}
    c_1=-c_0^{3/2}\frac{\sinh{\left(\frac{1}{c_0}\right)}+\frac{1}{c_0}}{\sinh{\left(\frac{1}{c_0}\right)}-\frac{1}{c_0}}
\end{equation}

\subsection{Deriving the self-consistent dynamics}
We show here how to derive the leading order self-consistency equations, which allow one to evolve the dynamics by only considering the interior functions $\{Q\}$. This amounts effectively to integrating out the boundary layers, which we recall is important in order to study the quasistatic limit where these become challenging to resolve numerically.

To obtain the 0th order self-consistency equation (\ref{eq:self_consistent}), we simply consider 
\begin{equation}
    \frac{\partial \Sigma_0}{\partial \gamma}=\frac{\partial}{\partial \gamma}\int \md \sigma \, \sigma \, Q_0(\sigma,\gamma)=\int \md \sigma \, \sigma \, \partial_{\gamma}Q_0(\sigma,\gamma)
\end{equation}
and insert the equation of motion (\ref{eq:Q0_eom}) for $Q_0$. Integrating by parts and rearranging one indeed finds (\ref{eq:self_consistent}) determining $c_0(\gamma)$ solely from properties of $Q_0$.

This method can be extended to higher orders, e.g. to close the equation of motion for $Q_1$ (\ref{eq:Q1_app}) by inferring $c_1$ from $\Sigma_1$, $Q_0$ and $Q_1$. We give the resulting expression obtained from considering the dynamics of $\Sigma_1$, although, as noted in the main text, the self-consistency equation becomes rather cumbersome and we have not found a numerically stable way of integrating it. 
One finds
\begin{multline}
    c_1(\gamma)=\left(|Q_0'(1,\gamma)|-Q_0'(-1,\gamma)\right)^{-1}\\ \bigg(-\left(Q_1(1,\gamma)+Q_1(-1,\gamma)\right)+c_0(\gamma)\left(Q_1'(1,\gamma)+Q_1'(-1,\gamma)\right)\\-c_0(\gamma)\left(Q_1(1,\gamma)-Q_1(-1,\gamma)\right)-\frac{\partial \Sigma_1}{\partial \gamma}\bigg)
\end{multline}

\subsection{Details on breakdown}
\begin{figure}
\centering
\includegraphics[width=0.38\textwidth]{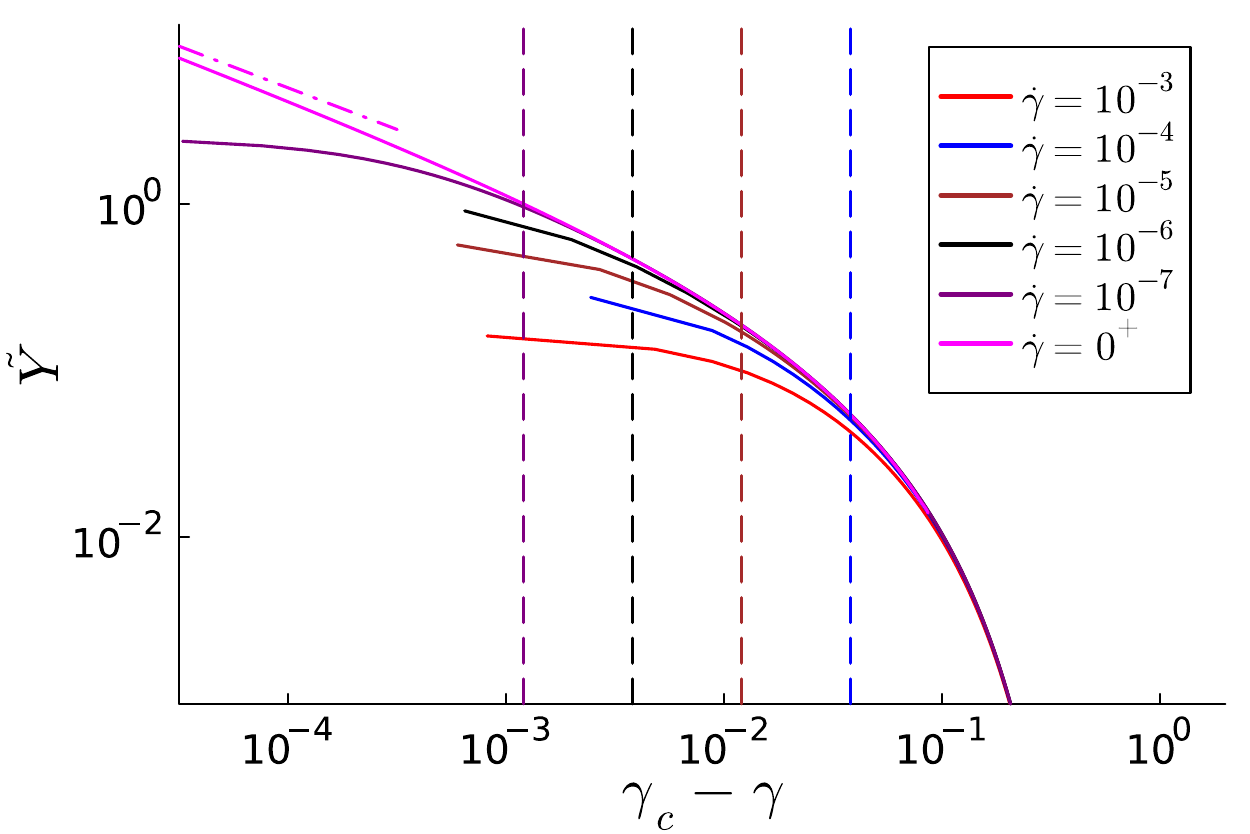}
\caption{\label{fig:distance_spinodal}Rescaled yield rate $\tY$ against $\gamma_c-\gamma$ for different values of the shear rate $\dgamma$. Colored vertical dashed lines indicate $B \dgamma^{1/2}$ for each corresponding shear rate, with a prefactor $B$ found numerically. Dash-dotted line shows the spinodal divergence of the $\dgamma=0^{+}$ solution, obtained from the self-consistent dynamics.} 
\end{figure}

We provide here an additional figure and some further details concerning the breakdown of the boundary layer expansion on approaching the brittle yield point, discussed in Sec.~\ref{sec:breakdown}. In Fig.~\ref{fig:distance_spinodal}, we show the same data for the rescaled yield rate as in Fig.~\ref{fig:spinodal}, but plotted now against $\gamma_c-\gamma$, where we recall that $\gamma_c$ is the brittle yield point in the quasistatic limit. This supports the claim in the main text that the boundary layer expansion breaks down for $\gamma_c-\gamma_{\rm sp}=\mathcal{O}(\dgamma^{1/2})$.

On the other hand, one expects from (\ref{eq:expa_yield_rate}) that the breakdown occurs when the second term of the expansion, $\dgamma^{1/2} c_1/\alpha $, becomes of the same order as the leading term $c_0/\alpha$, implying that, if the breakdown occurs for $\gamma_c-\gamma_{\rm sp}\sim \dgamma^{1/2}$, $c_1\sim \partial_{\gamma}c_0$, and therefore, following the expansion (\ref{eq:expa_yield_rate}),
\begin{equation}
    \tY (\gamma)\sim A_0 (\gamma_c-\gamma)^{-1/2}-A_1 (\gamma_c-\gamma)^{-3/2}\dgamma^{1/2}+\mathcal{O}(\dgamma^{3/2})
\end{equation}
This should be valid on approaching $\gamma_c$; we have introduced two constants $A_0,A_1>0$ and have used the fact that $c_1$ is in general a negative correction.

\section{\label{app:universality}Universality of the $\infty$-avalanche tails}
\begin{figure}
\centering
\includegraphics[width=0.38\textwidth]{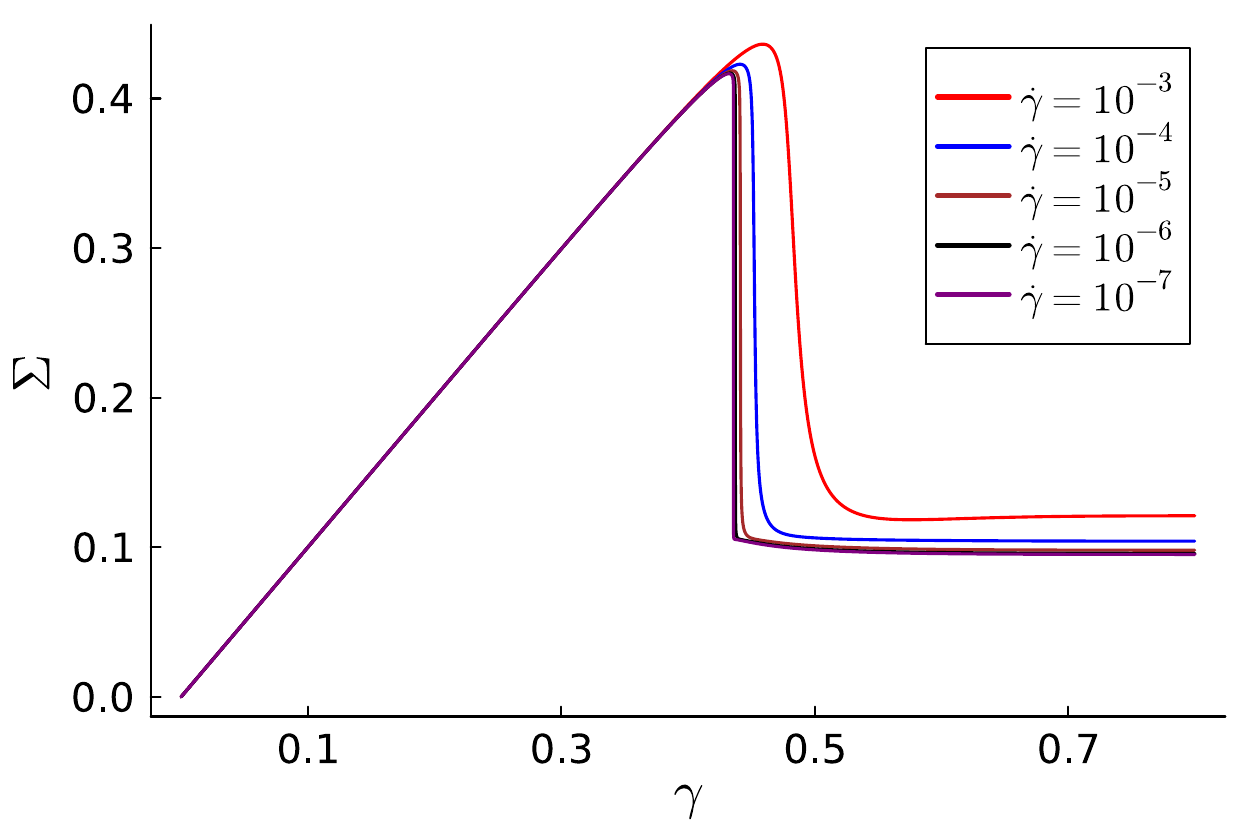}
\caption{\label{fig:pl_stress}Stress versus strain curves for different shear rates $\dgamma$, for the power-law initial distribution defined by (\ref{eq:power_law_ic}).} 
\end{figure}

\begin{figure}
\centering
\includegraphics[width=0.38\textwidth]{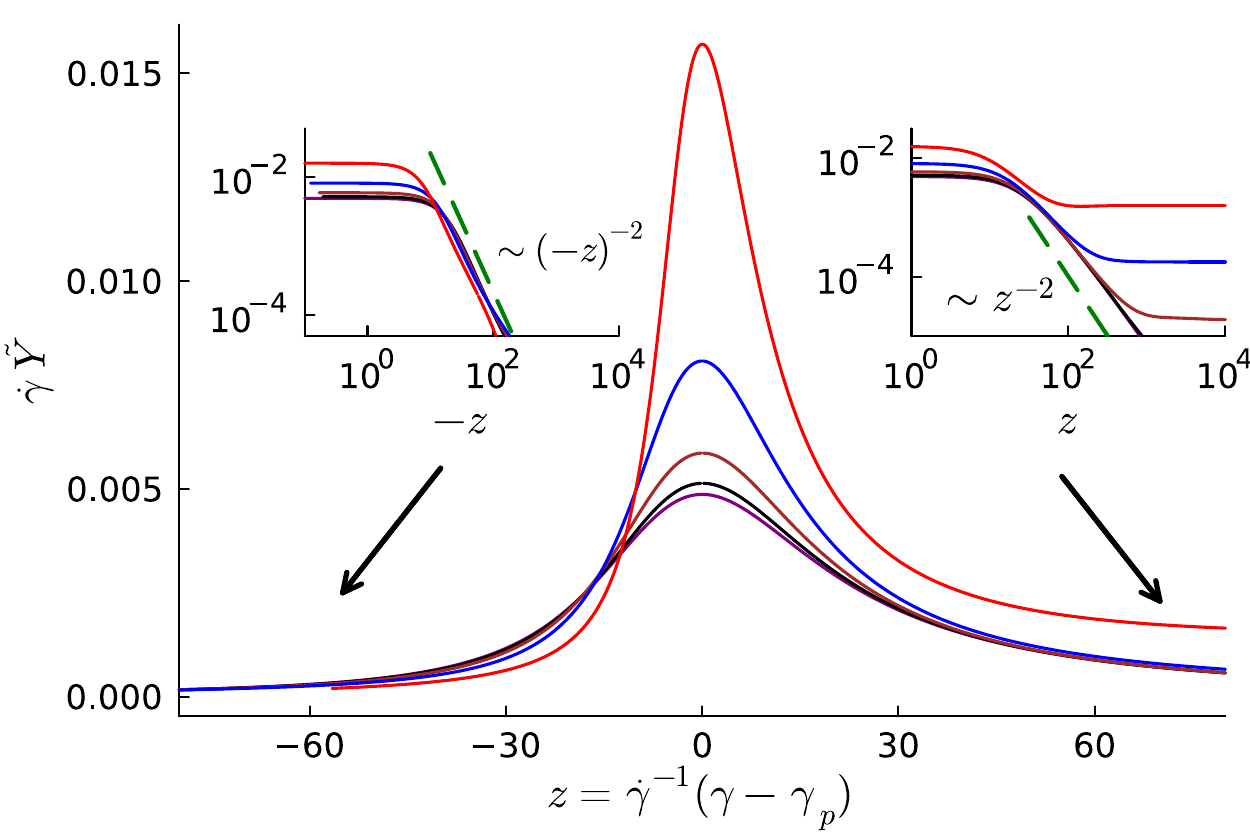}
\caption{\label{fig:pl_yield_rate}Collapse of the $\tY (\gamma)$ curves for different shear rates, analogous to Fig.~\ref{fig:yield_rate} but for
the power-law initial condition. Insets: tails of the $\infty$-avalanche, which as before follow $(\pm z)^{-2}$.} 
\end{figure}

To support our claim that the tails of the $\infty$-avalanche are universally described by the athermal aging exponent, independently of the form of the initial condition, we show here numerics analogous to those in Sec.~\ref{sec:first_part} but for a different family of initial distributions. In particular, we consider a power-law form
\begin{equation}\label{eq:power_law_ic}
    P(\sigma,\gamma=0)\propto (1-\sigma^2) \ (1-|\sigma|)^{\delta} \quad \mathrm{for} \quad |\sigma|<1
\end{equation}
with $\delta=4.1$ chosen to give roughly the same standard deviation $R\approx 0.2$ as the Gaussian distribution considered in Sec.~\ref{sec:first_part}. In Fig.~\ref{fig:pl_stress} we show the stress versus strain curves starting from this initial condition for different shear rates, with the coupling fixed to $\alpha=0.45$ as in Sec.~\ref{sec:first_part}. Paralleling Fig.~\ref{fig:yield_rate}, we then show in Fig.~\ref{fig:pl_yield_rate} the rescaled yield rate $\tY$ curves for different shear rates, collapsed by scaling both axis appropriately. In the insets we show that the tails are also described by the universal power-laws $(\pm z)^{-2}$.

\section{\label{app:collapse_stress}Scaling collapse of macroscopic stress}

\begin{figure}
\centering
\includegraphics[width=0.38\textwidth]{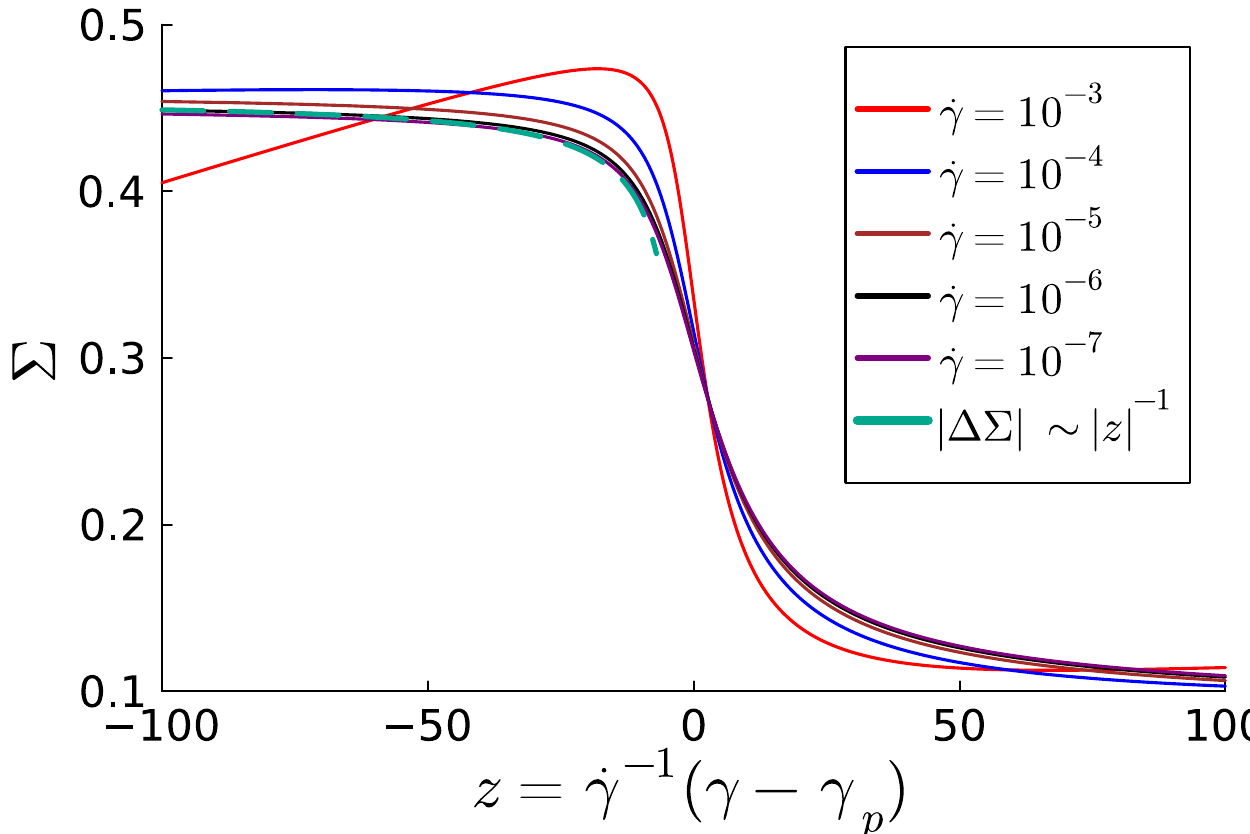}
\caption{\label{fig:stress_collapse}Macroscopic stress-strain curves for the same runs as shown in Fig.~\ref{fig:brittle} of the main text, collapsed by rescaling the strain axis following (\ref{eq:collapse_stress}). Dashed line indicates the predicted universal onset of the stress drop, which is a consequence of the universal tail of the rescaled yield rate discussed in the main text.} 
\end{figure}

We discuss in this appendix the scaling collapse for the macroscopic stress within the $\infty$-avalanche, mirroring the discussion of the yield rate scaling in the previous section. From on the one hand the scaling form of the rescaled yield rate (\ref{eq:ansatz}) with $b=1$ and, on the other, the evolution equation for the macroscopic stress (\ref{eq:stress_eom}), we expect $\Sigma (\gamma)$ to behave within the peak region as
\begin{equation}\label{eq:collapse_stress}
    \Sigma (\gamma)=g \left(z,\dgamma \right), \quad z = \dgamma^{-1}(\gamma-\gamma_p)
\end{equation}
If one defines $\Delta \Sigma$ as the deviation of the stress from its peak value (i.e.\ $\Sigma (\gamma_c^{-})$), one expects, as noted in the main text, the stress deviation before the onset of the stress drop to inherit the universal negative tail of the rescaled yield rate $\tY \sim \dgamma^{-1}f_{-} (z)$ with $f_{-}(z)\sim \left(-z\right)^{-2}$. This implies that the stress deviation at the onset of the stress drop should universally follow
\begin{equation}
    |\Delta \Sigma|\sim \left(-z\right)^{-1} \quad \mathrm{for} \quad |z|\gg 1, \ z<0 
\end{equation}
In Fig.~\ref{fig:stress_collapse} we show the scaling collapse using (\ref{eq:collapse_stress}) of the macroscopic stress curves $\Sigma(\gamma)$ for different shear rates, for the same runs as in Fig.~\ref{fig:brittle} of the main text. The curves do indeed collapse onto a master curve for $\dgamma \ll 1$, with the predicted tail behaviour.

\section{\label{app:scaling_Rc}Small shear rate behaviour at $R=R_c$}

We provide in this Appendix further details on the small shear rate expansion at the critical disorder $R=R_c$, as well as additional numerical evidence for the associated scalings.

We first set up the general scaling ansatz in the framework of the matched asymptotic expansion~\cite{olivier_glass_2011}. With the assumption of rational exponent values, this involves the introduction of unknown integers $s$ and $l$. To study the transient regime, we introduce a further integer $n$, with $\dgamma^{n/s}$ dictating the strain scale around the peak where the critical expansion holds. One can then write down an ansatz for the probability distribution in the exterior ($|\sigma|>1$) and the interior ($|\sigma|<1$) respectively, in the following manner
\begin{equation}\label{eq:interior_Rc}
    P(\sigma,\gamma)=\sum_{k=1}^{\infty}\dgamma^{k/s}\fext_k^{\pm}\left(\dgamma^{-l/s}(\pm \sigma-1),\dgamma^{-n/s}(\gamma-\gamma_p)\right) 
\end{equation}
\begin{equation}\label{eq:exterior_Rc}
    P(\sigma,\gamma)=\sum_{k=0}^{\infty}\dgamma^{k/s}Q_k \left(\sigma,\dgamma^{-n/s}(\gamma-\gamma_p)\right)
\end{equation}

We will not attempt here to deduce a priori the values of $s$, $l$ and $n$ in the critical regime. Rather, we will restrict ourselves to showing that the values $s=5$, $l=2$ which apply for the critical rheology of the model (see discussion in main text) lead to a consistent dynamics, and support this via numerical evidence. In the following, we shall denote the rescaled stress and strain variables, $\dgamma^{-l/s}(\pm \sigma-1)$ and $\dgamma^{-n/s}(\gamma-\gamma_p)$, by $z_1^{\pm}$ and $z_2$ respectively. 

The ansatz (\ref{eq:exterior_Rc}) for the exterior region implies the following expansion for the yield rate
\begin{multline}
    Y(z_2)=d_1(z_2)\dgamma^{\frac{1+l}{s}}+d_2(z_2)\dgamma^{\frac{2+l}{s}}+\dots \\ =d_1(z_2)\dgamma^{\frac{3}{5}}+d_2(z_2) \dgamma^{\frac{4}{5}}+\dots
\end{multline}
where the coefficients $d_1$, $d_2$ are given by the total integrals of $\fext_1^{\pm}$ and $\fext_2^{\pm}$ respectively. Now, the key aspect of the scaling at criticality is that the first coefficient $d_1$ vanishes. This implies vanishing $\fext_1^{\pm}$, which in turn implies (through the boundary conditions)
\begin{equation}\label{eq:bc_Q1}
    \fext_1^{\pm}(0,z_2)=Q_1(\pm 1,z_2)=0
\end{equation}
The vanishing of $Q_1$ at the boundary therefore ensures that the boundary layer (represented by $\fext_2^{\pm}$) can have width and height given by $Y^{1/2}$ (in this case scaling as $\dgamma^{2/5}$, given that $Y\sim \dgamma^{4/5}$ to leading order), which is always expected due to the diffusive fluctuations of the model (see discussion in Sec.~\ref{sec:background}). In addition to (\ref{eq:bc_Q1}), one has also the two following boundary conditions
\begin{equation}
    Q_0(\pm 1,z_2)=0
\end{equation}
\begin{equation}\label{eq:bc_T1}
    \frac{\partial}{\partial \sigma}Q_0(\pm 1,z_2)=\frac{\partial}{\partial z_1}\fext_1^{\pm}(z_1=0,z_2)
\end{equation}
We may now insert the forms (\ref{eq:interior_Rc}) and (\ref{eq:exterior_Rc}) into the master equation~(\ref{eq:hl_adim}) and derive the equations of motion order by order. We fix $n=2$, given that one expects the natural strain scale to be provided by the width $\sim \dgamma^{2/5}$ of the boundary layer. 

In the interior region, we firstly have at $\mathcal{O}(\dgamma^{3/5})$
\begin{equation}
    \partial_{z_2}Q_0(\sigma,z_2)=0
\end{equation}
This equation is important, because (as stated in the main text) it ensures that the quasistatic solution remains fixed throughout the scaling region around the peak, which we recall shrinks to zero (in terms of strain) in the $\dgamma\to 0$ limit.

At the next order $\mathcal{O}(\dgamma^{4/5})$
\begin{equation}\label{eq:eom_Q1_Rc}
    \partial_{z_2}Q_1(\sigma,z_2)=\alpha d_2 \partial_{\sigma}^2 Q_0(\sigma,z_2)+d_2 \delta (\sigma)
\end{equation}
The right hand side is non-zero except at $\sigma=1$, and is balanced by the evolution of $Q_1$: it would only be zero for a $Q_0$ consisting of two line segments (which is incompatible with normalization of $Q_0$ for $\alpha<1/2$). At the boundary $\sigma=1$, where we know $Q_0$ has fixed curvature of zero (see Sec.~\ref{subsec:TD_limit}), this ensures that $Q_1(1,z_2)$ remains fixed to zero throughout the peak. Note that the convective term $\sim \partial_{\sigma}Q_0$ is absent from (\ref{eq:eom_Q1_Rc}). Being of higher order it appears instead in the $\mathcal{O}(\dgamma)$ equation
\begin{multline}
    \partial_{z_2} Q_2(\sigma,z_2)=-\partial_{\sigma}Q_0(\sigma,z_2) +\alpha d_2(z_2)\partial_{\sigma}^2 Q_1(\sigma,z_2)\\ +\alpha d_3(z_2)\partial_{\sigma}^2 Q_0(\sigma,z_2)+d_3(z_2)\delta(\sigma)
\end{multline}

In the exterior region, we have to leading order the following equation for $\fext_2^{\pm}$
\begin{equation}
    \alpha d_2(z_2)\frac{\partial^2}{\partial z_1^2}\fext_2^{\pm}(z_1,z_2)-\fext_2^{\pm}(z_1,z_2)=0
\end{equation}
This implies a decaying exponential profile for $\fext_2^{\pm}$. We recall finally that its slope at $z_1=0$ is set by the boundary condition (\ref{eq:bc_T1}), implying in particular for the positive boundary
\begin{equation}\label{eq:slope_T1}
    \partial_{z_1}\fext_2^{\pm}(z_1=0,z_2)=\frac{\partial}{\partial \sigma} Q_0 (1,z_2)\approx -\frac{1}{\alpha}
\end{equation}
where the last property holds generally for quasistatic loading (\ref{eq:bc_dQ0}) around the yield point, where one can safely neglect yielding at the opposite threshold $\sigma=-1$.

\begin{figure}
\centering
\includegraphics[width=0.38\textwidth]{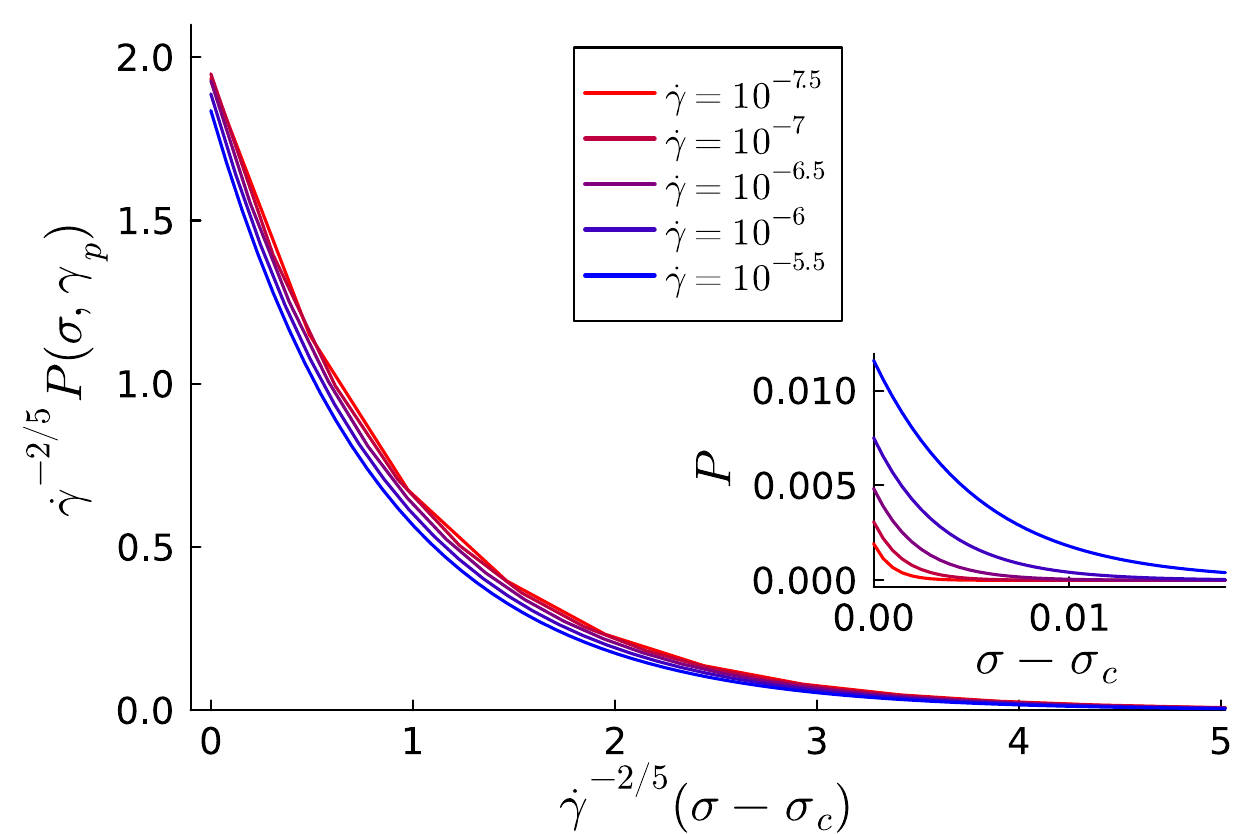}
\caption{\label{fig:BL_Rc}Local stress distribution above the positive threshold $\sigma_c$, extracted at the peak of the yield rate $\gamma_p(\dgamma)$, for a subset of the shear rates shown in Fig.~\ref{fig:Rc_scaling}. We collapse these according to the critical boundary layer scalings. Inset: same data before rescaling.}
\end{figure}

\begin{figure}
\centering
\includegraphics[width=0.38\textwidth]{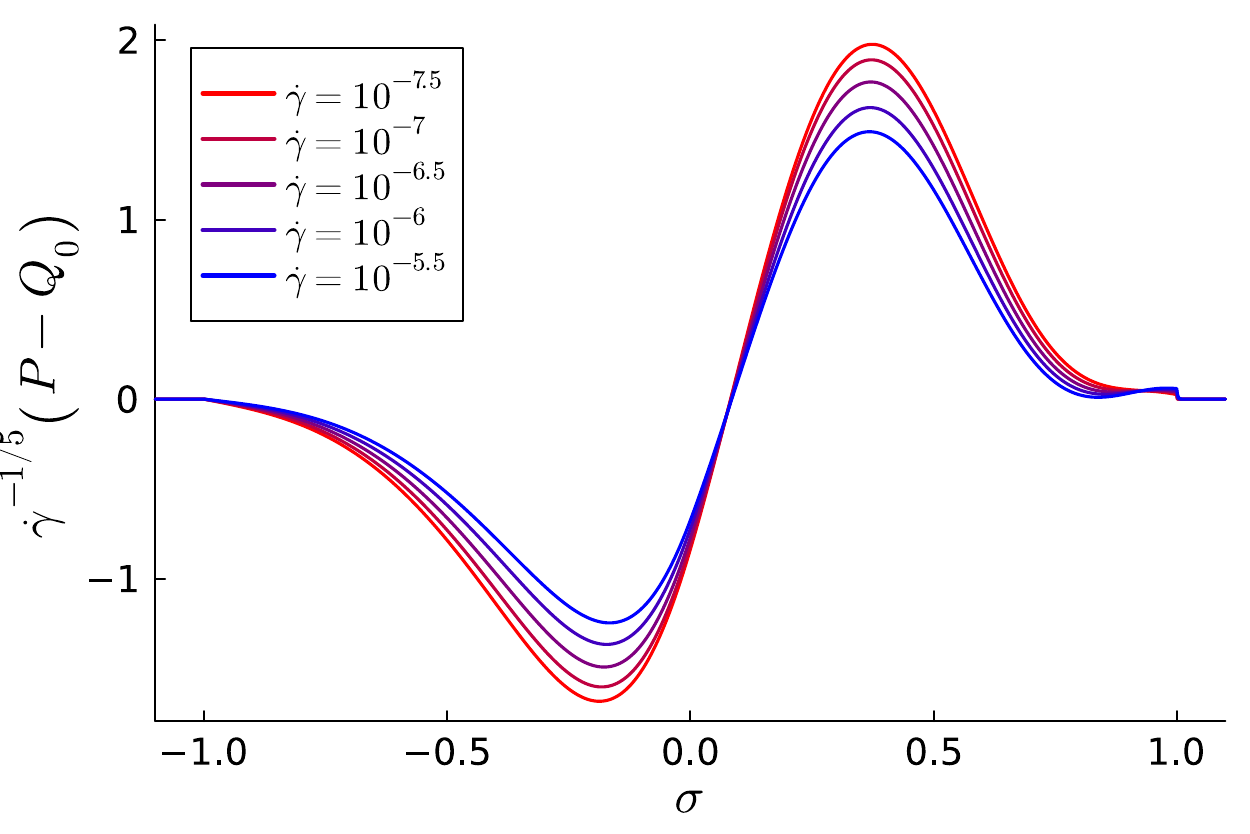}
\caption{\label{fig:Q1s_Rc}Numerical estimate of the leading order interior function, obtained as detailed in the text. For the smallest shear rates this collapses towards a $Q_1(\sigma,\gamma_c)$ that vanishes at the relevant boundary $\sigma=1$, as predicted in the critical scaling regime (contrast Fig.~\ref{fig:Q0_Q1} bottom, corresponding to the steady state). } 
\end{figure}

In Figs.~\ref{fig:BL_Rc} and \ref{fig:Q1s_Rc} we provide numerical support for the main implications of the scalings we have just described. In Fig.~\ref{fig:BL_Rc}, we show the distribution extracted at the peak of the yield rate $\gamma_p(\dgamma)$ for a range of shear rates, and collapse the exterior boundary layer according to the above scalings. 
For Fig.~\ref{fig:Q1s_Rc}, we firstly determine the $0$th order distribution $Q_0(\sigma,\gamma)$ from the self-consistent dynamics (magenta line in inset of Fig.~\ref{fig:Rc_scaling}), again at its peak $\gamma_c \equiv \gamma_p(\dgamma=0^{+})$. We then subtract this from the full distribution $P(\sigma,\gamma_c)$ obtained for finite shear rates at the same strain value. Rescaled by $\dgamma^{-1/5}$, this then gives access to $Q_1(\sigma,\gamma)$ in the $\dgamma \ll 1$ limit. Importantly, we confirm that (within numerical precision) this leading order interior correction does go to zero at the boundaries. This contrasts with the standard scaling expansion, e.g.\ for the steady state shown in Fig.~\ref{fig:Q0_Q1}. There the value of $Q_1$ at the boundary $\sigma=1$ is always finite, implying that the leading order exterior function is of the same order.  

\section{\label{app:RFIM}Finite-size scaling at $R=R_c$ in the RFIM}
\begin{figure}
\centering
\includegraphics[width=0.45\textwidth]{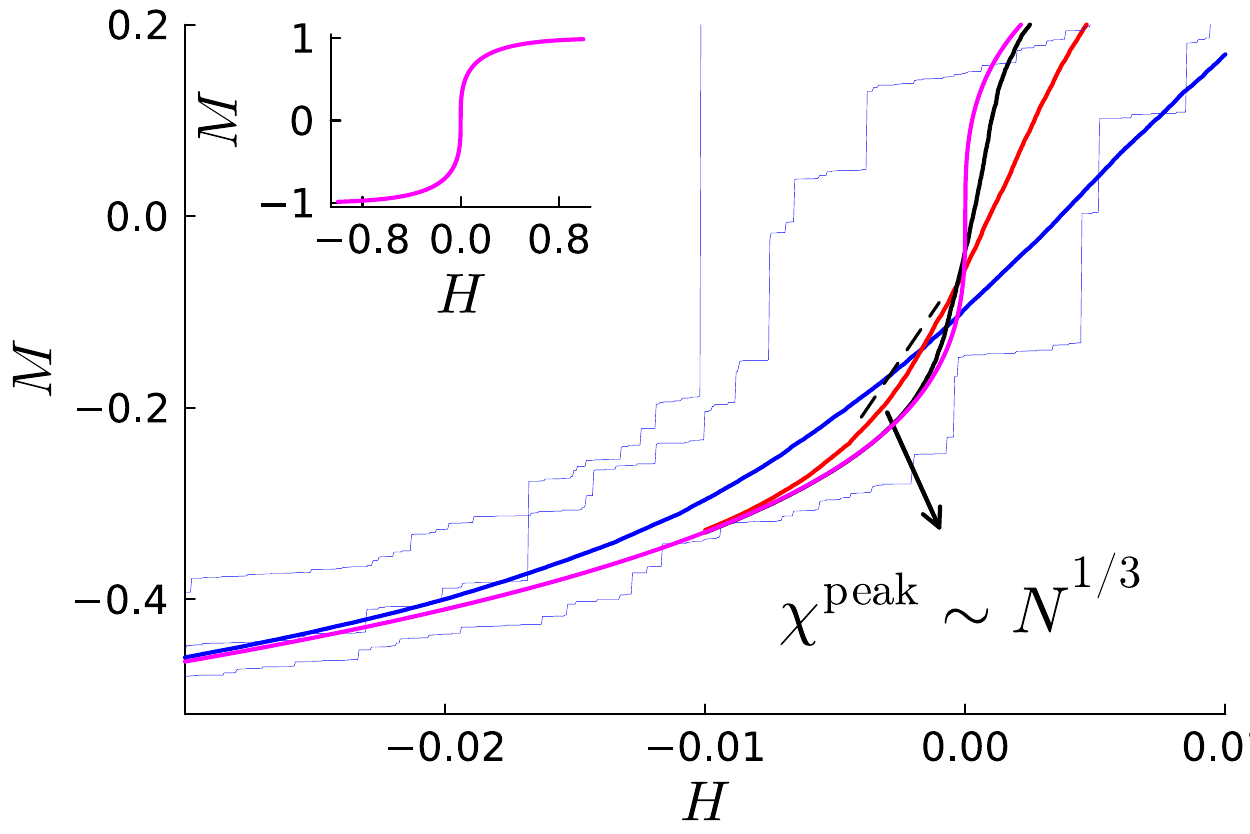}
\caption{\label{fig:Rc_RFIM}Behaviour of finite-size samples (same color scheme as in Fig.~\ref{fig:plot_brittle}) approaching the $N=\infty$ (magenta) coercive field $H_c=0$ at the critical disorder $R_c$ of the RFIM. Thin lines show individual realisations for $N=10^{4}$. In Fig.~\ref{fig:susceps_Rc_RFIM}, we show that the fluctuations of the coercive field are again $\mathcal{O}(N^{-1/2})$, while $\chip\sim N^{1/3}$ from $\chi \sim (H_c-H)^{-2/3}$; the arrow indicates roughly the regime where the divergence of $\chip$ arises. Inset: magnetisation curve over the whole external field window.} 
\end{figure}

We show here two additional figures regarding the finite-size scaling at the critical disorder in the RFIM. The first (Fig.~\ref{fig:Rc_RFIM}) is the equivalent of Fig.~\ref{fig:plot_brittle} (bottom) shown in the main text for $R<R_c$. We see how the finite-$N$ ensemble averages follow the $N=\infty$ limiting behaviour up to a scale set by the coercive field fluctuations $\mathcal{O}(N^{-1/2})$. Assuming that the susceptibility $\chi \sim (H_c-H)^{-2/3}$ saturates for finite sizes on this scale, one has $\chi^{\mathrm{peak}}\sim N^{1/3}$ as discussed in the main text.  

\begin{figure}
\centering
\includegraphics[width=0.38\textwidth]{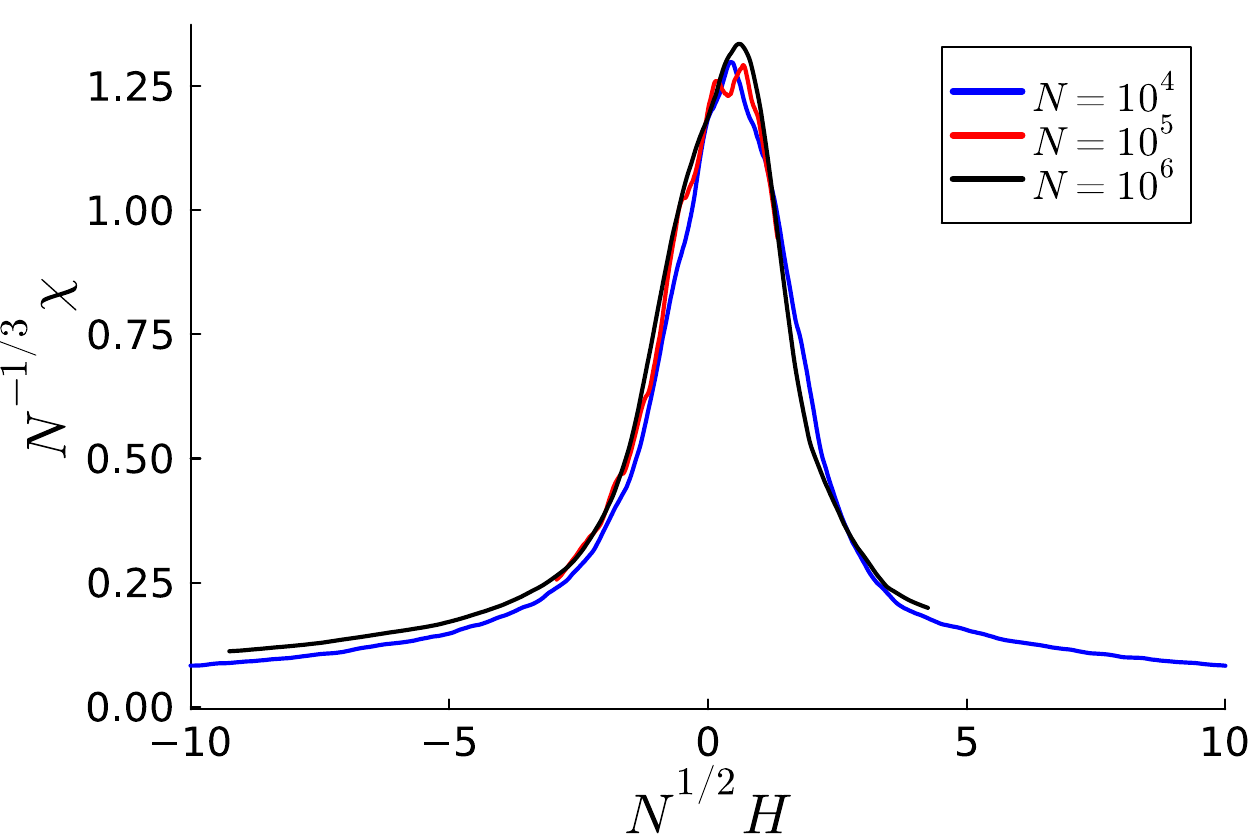}
\caption{\label{fig:susceps_Rc_RFIM}Behaviour of the susceptibility (slope of average magnetisation curve) at the critical disorder $R_c$ of the RFIM (corresponding to the runs in Fig.~\ref{fig:Rc_RFIM}). We confirm the $\sim N^{1/3}$ scaling of the peak susceptibility and the $\mathcal{O}(N^{-1/2})$ scaling controlling the width of the coercive field fluctuations. Note that the field axis is not recentered, indicating that the fluctuations are roughly symmetric around $H_c(N=\infty)=0$.} 
\end{figure}

In Fig.~\ref{fig:susceps_Rc_RFIM}, we show the corresponding susceptibility curves calculated as the slope of the average magnetization vs field. 
The 
collapse supports the scalings $\sim N^{-1/2}$ for the fluctuations and $\sim N^{1/3}$ for the peak height.

\section{\label{app:disconnected}Additional data for susceptibilities}
We provide in this appendix additional figures concerning the scaling with system size of the connected and disconnected susceptibilities. The disconnected susceptibility quantifies the sample-to-sample fluctuations of the order parameter, and is defined as
\begin{equation}
    \chi_{\mathrm{dis}}(\gamma)=N \left(\langle \Sigma ^2(\gamma)\rangle -{\langle \Sigma (\gamma) \rangle}^2 \right)
\end{equation}
where the average is again over both realisations of the initial disorder and of the dynamics. This is in contrast to the standard ``connected'' susceptibility considered in the main text, 
defined as the slope of the order parameter averaged over realisations. In the mean-field RFIM 
the disconnected susceptibility defined above scales as the square of the (connected) susceptibility~\cite{ozawa_random_2018}. In the discontinuous regime, where the order parameter has a finite jump, one expects the peak value of $\chid$ to be $\mathcal{O}(N)$,
and hence $\chip \sim N^{1/2}$ as we found in Sec.~\ref{subsec:precursors}. 
In the mean-field elastoplastic model with ``ferromagnetic'' interactions~\cite{ozawa_random_2018} it has been shown~\cite{rossi_emergence_2022} that an {\it{effective}} random field strength can be defined by exploiting a formal analogy within a path integral formalism. This again implies the relation $\chi_{\mathrm{dis}}^{\mathrm{peak}}\propto \left(\chip\right)^{2}$ as in the mean-field RFIM, even though the ferromagnetic elastoplastic model does not  
a priori present any random field coupled to the local stabilities. It remains an open question whether the HL model also allows for such an analogy; as pointed out in Ref.~\cite{rossi_emergence_2022} the calculations in this case are expected to be much more involved, starting from the fact that there is no explicit analytical expression for the average order parameter $\langle \Sigma (\gamma)\rangle $. 

\begin{figure}
\centering
\includegraphics[width=0.38\textwidth]{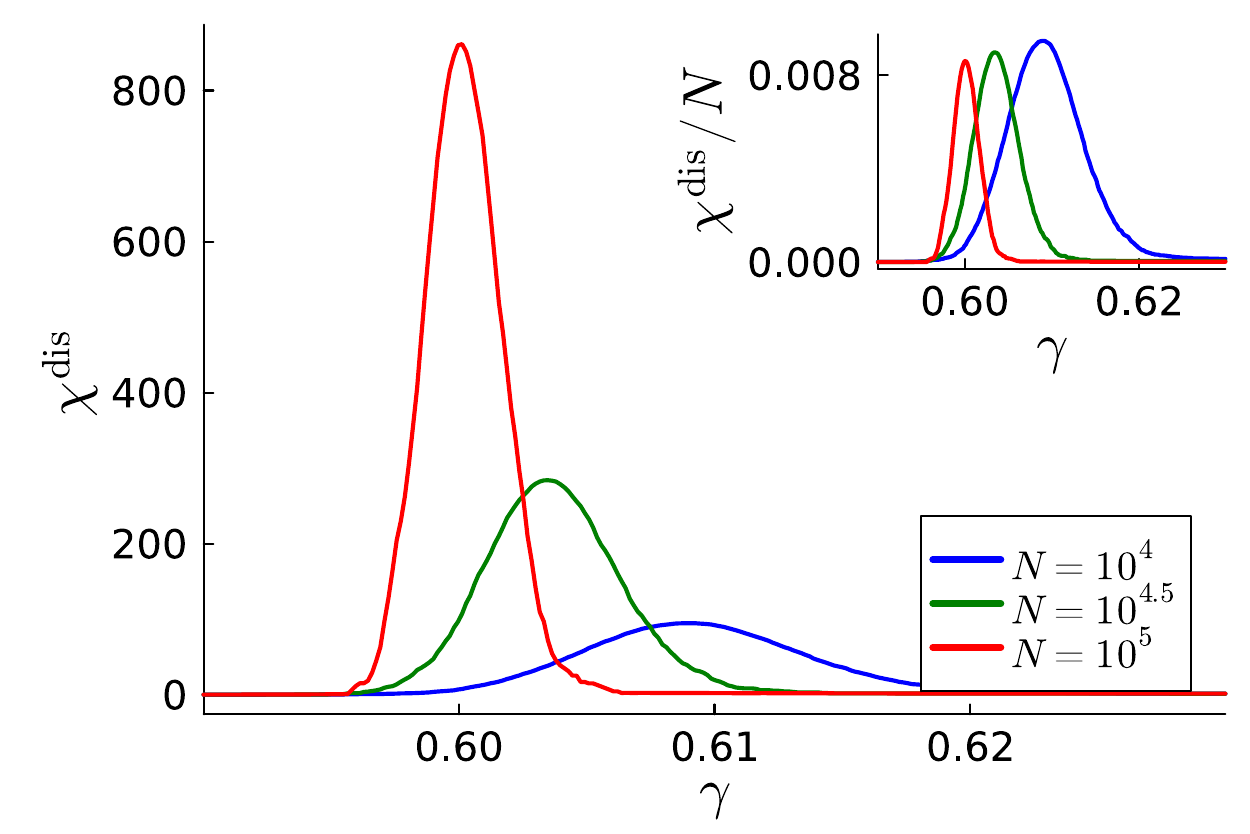}
\caption{\label{fig:chi_dis_R_016} Disconnected susceptibility (see definition in text) as a function of strain for $R=0.16$ (brittle) and three different system sizes. Inset: same data rescaled by $N$, consistent with $\chi_{\mathrm{dis}}^{\mathrm{peak}}\sim N$.} 
\end{figure}

In Fig.~\ref{fig:chi_dis_R_016}, we show the disconnected susceptibility as a function of strain for $R=0.16$ (brittle regime), for three different system sizes. We see (inset) that the peak value $\chi_{\mathrm{dis}}^{\mathrm{peak}}\equiv \max_{\gamma} \chi_{\mathrm{dis}}(\gamma)$ indeed scales as $\chi_{\mathrm{dis}}^{\mathrm{peak}}\sim N$. Accordingly, plotting the peak values against the corresponding peak values of the (connected) susceptibility, we find (Fig.~\ref{fig:square}) a good agreement with the law $\chi_{\mathrm{dis}}^{\mathrm{peak}}\propto \left(\chip\right)^{2}$, which applies in the mean-field RFIM~\cite{ozawa_random_2018}.

\begin{figure}
\centering
\includegraphics[width=0.38\textwidth]{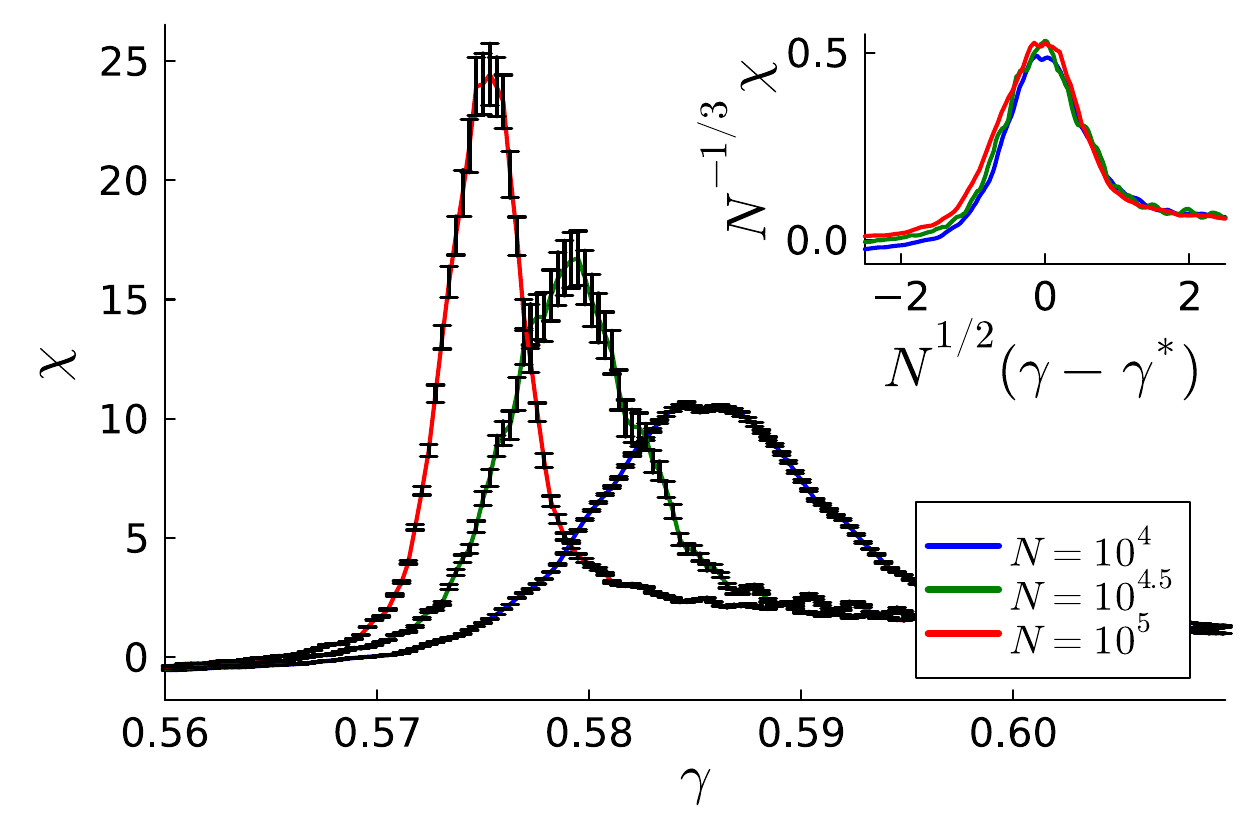}
\caption{\label{fig:chis_Rc}Susceptibility (slope of average stress-strain curve) for $R=0.18\approx R_c$ and three different system sizes, from which we extract the peak values shown in Fig.~\ref{fig:chips} of the main text. Inset: same data rescaled by $N^{1/3}$, consistent with $\chip\sim N^{1/3}$.} 
\end{figure}

\begin{figure}
\centering
\includegraphics[width=0.38\textwidth]{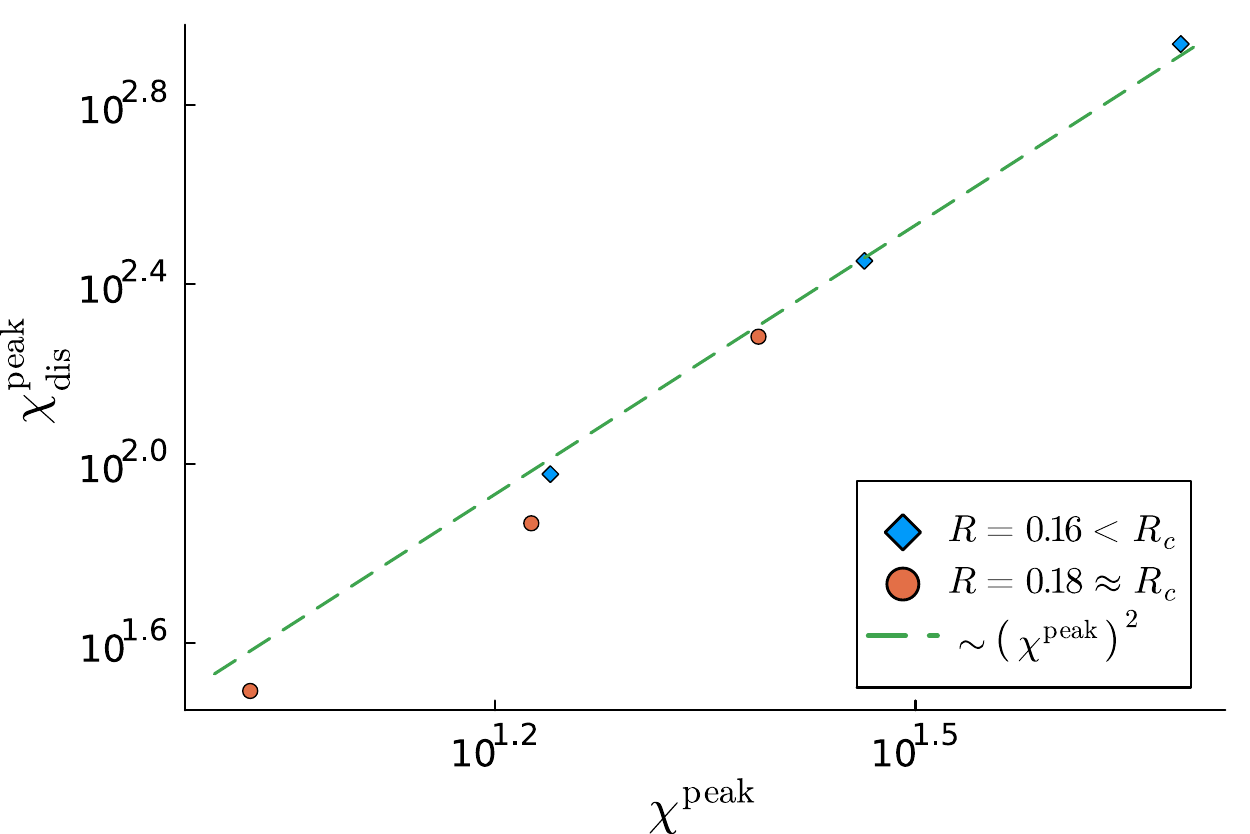}
\caption{\label{fig:square}Parametric plot of the peak disconnected susceptibilities against the peak susceptibilities for $R=0.16$ (brittle) and $R=0.18\approx R_c$. For $R=0.16$ we find good agreement with $\chi_{\mathrm{dis}}^{\mathrm{peak}}\propto \left(\chip\right)^{2}$, while close to $R_c$ the agreement is somewhat poorer.} 
\end{figure}

Turning now to the value $R=0.18\approx R_c$, we display in Fig.~\ref{fig:chis_Rc} the complete behaviour of the susceptibility. The $\chip$ values shown in Fig.~\ref{fig:chips} of the main text, which we recall scaled roughly as $\chip \sim N^{1/3}$, were extracted from this data. Finally, in Fig.~\ref{fig:square} we again attempt a parametric plot of the $\chip$ values against the corresponding peak values of the disconnected susceptibility. For the value $R=0.18$ close to $R_c$, we find some deviation from $\chi_{\mathrm{dis}}^{\mathrm{peak}}\propto \left(\chip\right)^{2}$. Namely, $\chi_{\mathrm{dis}}^{\mathrm{peak}}$ appears to be diverging faster than $\sim N^{2/3}$, although given the restricted system size range we cannot conclude whether this actually constitutes a true violation of the law, and leave this as an open question for future work. It is interesting to note that in Fig.~3C of \cite{ozawa_random_2018}, where the same analysis is performed for $3d$ particle simulations, deviations from the $\chi_{\mathrm{dis}}^{\mathrm{peak}}\propto \left(\chip\right)^{2}$ law are also found around the critical preparation temperature. 

\section{\label{app:avalanches_mu_1}Avalanches for $\mu=1$ and $P(x)$}
\begin{figure}
\centering
\includegraphics[width=0.38\textwidth]{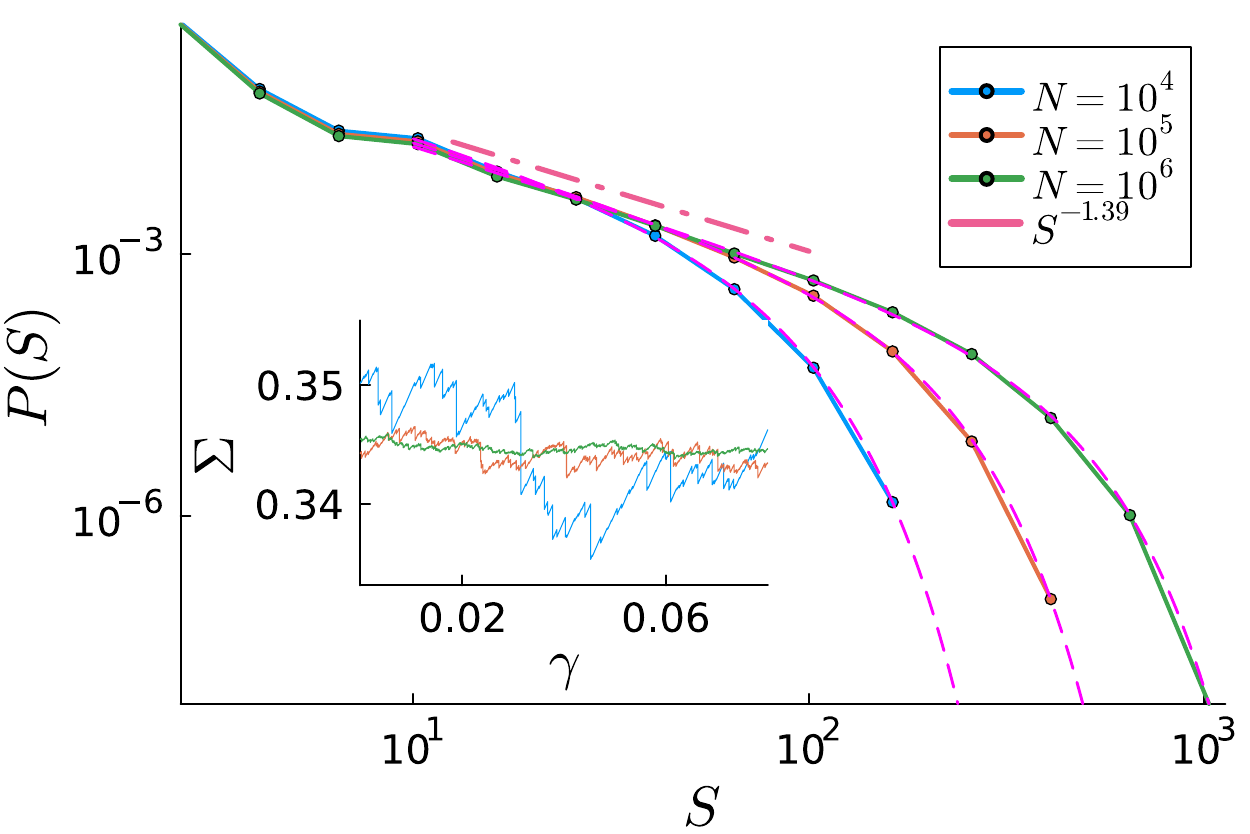}
\caption{\label{fig:ava_A_032}Avalanche distribution in steady state for $\mu=1$ and coupling $A=0.32$, with the upper cutoff as originally defined in~\cite{lin_mean-field_2016}, for three different system sizes. The distributions are again very well fitted by the form (\ref{eq:p_of_S}) (magenta dashed line), from which (for the largest system size $N=10^{6}$) we extract the exponent $\tau$ as $1.395\pm 0.047$ (by fitting to (\ref{eq:p_of_S}) in the range $S\in [10,1026]$). Inset: macroscopic stress versus strain in steady state (as in Fig.~\ref{fig:ava_ss}) for $A=0.32$, from which we extract the avalanche statistics.} 
\end{figure}

We firstly show in Fig.~\ref{fig:ava_A_032} the avalanche distributions measured in the steady state for $\mu=1$, for a lower value of the coupling $A=0.32$ (in the main text, $A=0.47$). We measure a slightly higher avalanche exponent $\tau \approx 1.39$. In the inset, we show the stress versus strain curves in steady state, from which we extract the avalanche statistics.

\begin{figure}
\centering
\includegraphics[width=0.38\textwidth]{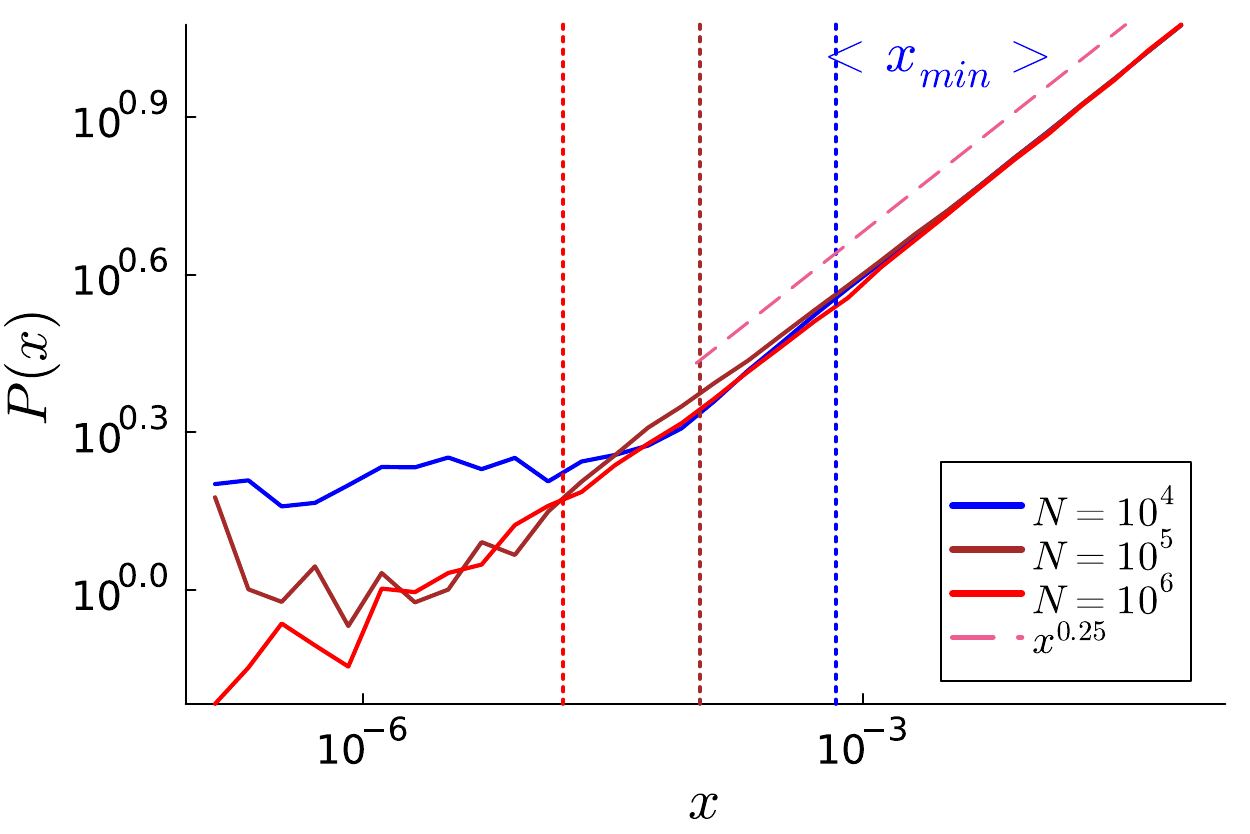}
\caption{\label{fig:p_of_x_A_032}Histogram of local stabilities $\{x_i\}$ after an avalanche in the steady state of the $\mu=1$ model, with $A=0.32$, obtained by averaging over many avalanches. In the limit $N=\infty$, $P(x)$ follows the pseudogap $P(x)\sim x^{\theta}$ with $\theta \approx 0.25$ (see text). For any finite $N$, $P(x)$ instead develops a finite plateau, below a crossover scale that is expected to scale as the smallest stress kicks $\Delta (N)\sim N^{-1}$. Dotted lines indicate the location of the average $\langle x_{\rm min} \rangle$, which clearly lies outside of the plateau (contrast Fig.~\ref{fig:plateau} for HL).} 
\end{figure}

We further support numerically the claim in Sec.~\ref{subsec:extension_PL} regarding the different scaling of the finite size plateau in $P(x)$ for power-law noise ($\mu=1$). We show in Fig.~\ref{fig:p_of_x_A_032} the $P(x)$ statistics extracted for $A=0.32$ in steady state for three different system sizes. We note firstly that the distribution follows the expected pseudogap behaviour $P(x)\sim x^{\theta}$, where $\theta$ can be obtained from the analytical formula derived in \cite{lin_mean-field_2016}
\begin{equation}\label{eq:atan_formula}
    \theta=\frac{1}{\pi}\arctan \left(\frac{\pi A}{v}\right)
\end{equation}
with the ``velocity'' $v=\lim_{\dgamma \to 0} \tY^{-1}$ (see App.~\ref{app:plastic}). In steady state one has from (\ref{eq:stress_eom}) that the inverse scaled yield rate is $\tY^{-1}=\langle \sigma_u \rangle$, i.e.\ the average unstable stress. The fact that one can neglect yields at the opposite threshold thus implies $v \approx 1$. Inserting $A=0.32$ into (\ref{eq:atan_formula}) one obtains $\theta \approx 0.25$, which indeed agrees well with the data in Fig.~\ref{fig:p_of_x_A_032}. 

The two important observations from Fig.~\ref{fig:p_of_x_A_032} are, firstly, that the finite-size plateau is much weaker than for HL (compare Fig.~\ref{fig:plateau}). Naively, one expects the 
plateau to scale as the smallest stress kicks $\sim N^{-1}$, although this is challenging to check explicitly. Secondly, the average $\langle x_{\rm min}\rangle$ is unambiguously outside the plateau and hence within the power-law region, in contrast to HL where both $\langle x_{\rm min}\rangle$ and the plateau scale behave as $\sim N^{-1/2}$.

\section{\label{app:methods}Numerical methods}
We give in this appendix an overview of the various numerical approaches we have adopted to tackle the HL master equation (\ref{eq:hl_adim}). To obtain the finite shear rate numerics presented in Sec.~\ref{sec:first_part} we set up a discretised stress grid $\{\sigma_i\}$, $i=1,\ldots,M$, of $M$ points in a domain $(-l,l)$. The master equation updates are performed in a pseudospectral manner, as in \cite{parley_aging_2020,parley_mean-field_2022-1}, in order to implement the update corresponding to the diffusive term in Fourier space. To this end, we define the corresponding $M$ points $\{k_i\}$ in Fourier space. We choose $l=2$, which for the slow shear rates considered here is wide enough to avoid the effect of periodic images, as the tails of the stress distribution decay quickly beyond the threshold $\sigma=1$.

To evolve the dynamics, we adopt a ``hybrid'' scheme. For the value $R=0.2<R_c$ considered in Sec.~\ref{sec:first_part}, the initial regime is almost purely elastic, as $\tY \ll 1$, and the initial distribution $P(\sigma,\gamma=0)$ is essentially advected towards $\sigma=1$ due to the external shear. In this initial regime, we fix a time step $\md t$ and evolve the dynamics according to (\ref{eq:hl_adim}). However, once $\tY$ reaches a threshold (which we set to $\tY=0.2$), we then switch to an alternative update scheme to resolve the regime where $\tY\gg 1$. This alternative scheme involves considering (\ref{eq:hl_adim}) divided once by the yield rate, and using an adaptive time step $\md t \propto Y^{-1}$. In this way, the prefactor of the diffusive term remains fixed. This adaptive time step scheme may also be interpreted as a fixed plastic strain step $\md \epsilon_p =Y \md t$, as explained further in App.~\ref{app:plastic}.

\section{\label{app:plastic}Plastic strain formulation of self-consistent dynamics}

We discuss in this appendix the plastic strain formulation of the 0th order self-consistent dynamics, which, besides providing the smoothness argument~\cite{popovic_elastoplastic_2018} considered in Sec.~\ref{subsec:TD_limit}, is also useful for the numerical implementation. We recall that this 0th order self-consistent dynamics is defined by the condition
\begin{equation}\label{eq:self}
    c_0(\gamma)= {\left (|Q_0'(1,\gamma)|-Q_0'(-1,\gamma)\right)}^{-1}\left (1-\frac{\partial \Sigma_0}{\partial \gamma}\right)
\end{equation}
along with the equation of motion for $Q_0$ (\ref{eq:Q0_eom}), where as before $c_0(\gamma)=\alpha \lim_{\dgamma \to 0}\tY$. To numerically integrate these equations, we adopt the same strategy as in~\cite{popovic_elastoplastic_2018}, where the dynamics is described in terms of plastic strain $\epsilon_p$. Alternatively, disregarding the clear physical interpretation of this variable, one may think of this merely as an {\it adaptive} strain step. Indeed, the plastic and total strain steps are related by 
\begin{equation}\label{eq:dg_epsilon}
    \md \epsilon_p=\tY(\gamma)\md \gamma
\end{equation}
Adopting a fixed plastic strain step then corresponds to an adaptive (i.e.\ strain-dependent) strain step $\md \gamma \sim \tY^{-1}(\gamma)$. This is useful numerically in order to resolve the regime where $\tY$ becomes very large on approaching $\gamma_c$.

To obtain the formulation in terms of plastic strain one rewrites, 
as in~\cite{popovic_elastoplastic_2018}, 
the equation of motion (\ref{eq:Q0_eom}) for $Q_0(\sigma,\epsilon_p)$ as 
\begin{equation}\label{eq:eom_plastic}
   \partial_{\epsilon_p}Q_0=-v \partial_{\sigma}Q_0+\alpha \partial_{\sigma}^2 Q_0 +\delta (\sigma) 
\end{equation}
with 
the ``velocity'' ~\cite{popovic_elastoplastic_2018}
\begin{equation}\label{eq:velocity}
    v=\lim_{\dgamma \to 0}\frac{\dgamma}{Y}=\lim_{\dgamma \to 0} \tY^{-1}
\end{equation}
The remaining step is to rewrite the self-consistency condition (\ref{eq:self}) in terms of plastic strain. We have on the one hand that \begin{equation}
    \frac{\partial \Sigma_0}{\partial \gamma}=\frac{1}{v}\frac{\partial \Sigma_0}{\partial \epsilon_p}
\end{equation}
Inserting this into (\ref{eq:self}) and solving for $v$ one finds
\begin{equation}\label{eq:self_plastic}
v(\epsilon_p)=\tY^{-1}(\epsilon_p)=\alpha \left (|Q_0'(1,\epsilon_p)|-Q_0'(-1,\epsilon_p)\right)+\frac{\partial \Sigma_0}{\partial \epsilon_p}
\end{equation}
as in~\cite{popovic_elastoplastic_2018}. Equation (\ref{eq:self_plastic}), together with the equation of motion (\ref{eq:eom_plastic}), fully defines the 0th order self-consistent dynamics in terms of plastic strain.

Numerically, one can then fix a plastic strain step (we use $\md \epsilon_p=2 \cdot 10^{-7}$) and set up a stress grid (as detailed in App.~\ref{app:methods}) to evolve the dynamics. From $v=\tY^{-1}(\epsilon_p)$, one is ultimately interested in $\tY(\gamma)$. One therefore has to invert the transformation and find $\gamma(\epsilon_p)$ by integrating in parallel relation (\ref{eq:dg_epsilon})
\begin{equation}
    \md \gamma=v(\epsilon_p)\md \epsilon_p
\end{equation}
starting from the initial values $\gamma_0$ and $\epsilon_{p,0}$. From the inverse transformation one can then study the divergence of $\tY$ through $\tY(\gamma)=v^{-1}\left(\epsilon_p(\gamma)\right)$.

We note that, for a generic initial condition such as the one considered throughout this work (\ref{eq:ansatz}), the self-consistent scheme cannot be adopted directly from the start of the dynamics, given that the quasistatic loading condition (\ref{eq:bc_dQ0}) is not fulfilled. To circumvent this problem, in practice we apply a finite but very small shear rate $\dgamma \pltime=10^{-7}$ during an initial transient (specifically, until a threshold $\tY=0.2$ is reached) to allow for the quasistatic boundary condition (\ref{eq:bc_dQ0}) to develop, before then switching to the self-consistent dynamics.

Finally, we list here the $R$ values considered in Fig.~\ref{fig:effective_exponent}: from black to blue, these are $R=0.308$, $0.306$, $0.3054$, $0.3053$, $0.30522$, $0.305215$; from red to orange: $R=0.30521$, $0.305205$, $0.3052$, $0.3051$, $0.305$, $0.3045$, $0.304$, $0.303$, $0.3025$, $0.302$ and $0.3$.

\section{\label{app:timescales}Additional physical timescales in the HL model}

We comment here on possible additional timescales that may be included in the HL dynamics, as discussed in~\cite{bouchaud_spontaneous_2016}. Once a site ${i}$ yields, which we recall occurs on a typical timescale $\pltime$ whenever $|\sigma_i|>\sigma_c$, it can interpreted as being in a ``fluid'' state: in the original HL dynamics considered in the main text, the site re-jams immediately in a state of zero stress ($\sigma=0$). The authors of \cite{bouchaud_spontaneous_2016} argued that, in fact, a site should remain in its fluid state for a typical finite time $\tau_{\rm fl}$. These fluidised elements are withdrawn from the distribution $P(\sigma,t)$, which becomes unnormalised, and contribute only viscous stress during a finite time. Although the inclusion of a finite $\tau_{\rm fl}$ would not alter our analysis for $\dgamma \ll 1$, as it does not affect the coupling between yield rate and mechanical noise~\cite{bouchaud_spontaneous_2016}, the fact that $\tau_{\rm fl}=0$ in the HL model conceals this important aspect of the physics. Indeed, the brittle yielding regime can be thought of in terms of the appearance of a transient macroscopic fraction of fluidised elements, which in finite dimension will typically be localised in a transient shear band~\cite{singh_brittle_2020}. The work of~\cite{bouchaud_spontaneous_2016} in fact also introduces an additional timescale, associated with the finite stress propagation speed. Within mean-field, this is achieved via a finite lag time in the coupling between diffusion and yield rate~\cite{bouchaud_spontaneous_2016}, so that yield events at a given time do not affect the overall mechanical noise instantaneously. In our analysis of the original HL description, the unique finite physical timescale $\pltime$ is linked to the typical destabilisation time of a mesoscopic block posed at a saddle of its local potential. 
In overdamped systems this is expected to scale as the ratio between a microscopic viscosity and the elastic shear modulus~\cite{nicolas_deformation_2018}~\footnote{Note that there is an additional complication (which we neglect at this point) concerning the shape of the local potential. For cuspy potentials, one indeed expects a uniform destabilisation rate $\pltime^{-1}$ for sites stressed beyond yield, as assumed in the HL model. If the potential is smooth, on the other hand, one expects instead a stress-dependent {\it progressive} rate, as studied in~\cite{ferrero_criticality_2019}.}. Which of the two finite timescales is dominant for a given system may thus depend on microscopic details such as the level of inertia.

\bibliography{main}
\bibliographystyle{apsrev4-2.bst}

\end{document}